\documentclass[fleqn,10pt]{wlscirep}
\usepackage[utf8]{inputenc}
\usepackage[T1]{fontenc}
\usepackage{lineno}
\usepackage{fdsymbol}
\usepackage{amsmath,amssymb,amsfonts}
\usepackage{algorithmic}
\usepackage{tablefootnote}
\usepackage{soul, color}

\usepackage[
style=phys,
citestyle=numeric-comp,
autocite=superscript,
sorting=none
]{biblatex}

\usepackage{helvet}

\newcommand{\res}{\mathrm{R}_{\mathrm{LS}}}
\newcommand{\reslow}{\mathrm{R}_{\mathrm{10s}, 5\%\mathrm{SOC}}}
\newcommand{\reshigh}{\mathrm{R}_{\mathrm{10s}, 90\%\mathrm{SOC}}}
\newcommand{\varq}{\mathrm{Var}(\Delta Q_{100-10}(V))}

\newcommand{\cef}{\mathrm{CE}_{\mathrm{f}}}
\newcommand{\lamne}{\mathrm{LAM}_{\mathrm{NE}}}
\newcommand{\lampe}{\mathrm{LAM}_{\mathrm{PE}}}

\newcommand{\lli}{\mathrm{LLI}}
\newcommand{\Qd}{Q_{\mathrm{d}}}
\newcommand{\Qc}{Q_{\mathrm{c}}}
\newcommand{\QLLI}{Q_{\mathrm{LLI}}}
\newcommand{\mohm}{\mathrm{m}\Omega}
\newcommand{\ymax}{y_\mathrm{max}}

\definecolor{lightblue}{rgb}{0.7,1,1}
\definecolor{lightgray}{rgb}{0.9,0.9,0.9}
\definecolor{lightgreen}{rgb}{0.7,1,0.7}

\newcommand{\rev}[1]{#1}
\newcommand{\reva}[2]{#2}
\newcommand{\revb}[2]{#2}
\newcommand{\revc}[2]{#2}

\newcommand{\beginsupplement}{%
        \setcounter{table}{0}
        \renewcommand{\thetable}{S\arabic{table}}%
        \setcounter{figure}{0}
        \renewcommand{\thefigure}{S\arabic{figure}}%
     }
     
\usepackage{fancyhdr}
\title{Predicting the impact of formation protocols on battery lifetime immediately after manufacturing}

\author[1,*]{Andrew Weng}
\author[1]{Peyman Mohtat}
\author[2]{Peter M. Attia}
\author[1]{Valentin Sulzer}
\author[1]{Suhak Lee}
\author[3]{Greg Less}
\author[1]{Anna Stefanopoulou}
\affil[1]{Department of Mechanical Engineering, University of Michigan, Ann Arbor, MI 48109, USA}
\affil[2]{Department of Materials Science and Engineering, Stanford University, Stanford, CA 94305, USA}
\affil[3]{University of Michigan Battery Lab, Ann Arbor, MI 48105, USA}
\affil[*]{Lead Contact and Corresponding Author (\href{mailto:asweng@umich.edu}{asweng@umich.edu})}

\begin{abstract}

Increasing the speed of battery formation can significantly lower lithium-ion battery manufacturing costs. However, adopting faster formation protocols in practical manufacturing settings is challenging due to a lack of inexpensive, rapid diagnostic signals that can inform possible impacts to long-term battery lifetime. This work identifies the cell resistance measured at low states of charge as an early-life diagnostic feature for screening new formation protocols.  We show that this signal correlates to cycle life and improves the accuracy of data-driven battery lifetime prediction models. The signal is obtainable at the end of the manufacturing line, takes seconds to acquire, and does not require specialized test equipment. We explore a physical connection between this resistance signal and the quantity of lithium consumed during formation, suggesting that the signal may be broadly applicable for evaluating any manufacturing process change that could impact the total lithium consumed during formation.

\end{abstract}
\begin{document}

\flushbottom
\maketitle

\thispagestyle{empty}

\section{Introduction}

With the increasing demand for electric vehicles, global lithium-ion battery manufacturing capacity is quickly approaching the terawatt-hour scale \supercite{AustralianTradeandInvestmentCommission2018, BenchmarkMineralsIntelligence2019, Mackenzie2020}. A key step in battery manufacturing is formation/aging, which has been estimated to account for up to 30\% of total manufacturing costs \supercite{Liu2021a, Nelson2011, Duffner2021, Kuhlmann2018, Wood2015}. The formation/aging process involves charging and discharging hundreds of thousands of cells in environmentally controlled chambers, an expensive process that takes days to weeks to complete but is necessary to improve battery performance and lifetime \supercite{Winter2009, An2016, Wang2018, Peled2017, Goers2011, Lu2014a}. Given the high cost burden, manufacturers are incentivized to develop new formation processes that decrease the total time consumed by formation/aging. A variety of fast formation strategies have been studied in academic literature, which employ some combination of rapid charge-discharge cycles, restricted voltage windows, and optimized temperature \supercite{An2016, An2017, Wood2019, Mao2018, Muller2017, Antonopoulos2018, Zhang2004, Heimes2020a, Ryan2021, Pathan2019, Muller2018a, Rago2019, Lee2004}. Recent studies have shown that formation time can be decreased while preserving battery lifetime \supercite{Wood2019, Rago2019, Ryan2021}, although conclusions remain tenuous due to the limited sample sizes typically used.

In real manufacturing settings, a `one size fits all' formation protocol is unlikely to exist since cell designs with different electrolytes, electrodes, and active materials influence important formation factors such as charging capability, electrode wettability, and solid electrolyte interphase (SEI) reaction pathways. However, cycle life testing often takes months or years to complete, posing a significant barrier to the adoption of new, potentially cost-saving formation protocols. While characterization techniques such as volume change detection \supercite{Wang2007, Bauer2016, Mohtat2020}, impedance spectroscopy \supercite{Zhang2020, An2017}, acoustic spectroscopy \supercite{Bommier2020, Davies2017, Knehr2018, Deng2020} and X-ray tomography \supercite{Pietsch2017, Wood2018} have been proposed for use in manufacturing settings, these methods can be costly to implement since the metrology will need to be deployed at scale in the battery factory. \reva{1}{Diagnostic features that rely only on current-voltage signals that can be obtained using already available cycling equipment \supercite{An2020} and are thus highly attractive.}

In this work, we show that the cell resistance at low states of charge can be used to screen new formation protocols and predict battery lifetime. Our work shows that this signal, measured at the beginning of life, is a stronger predictor of battery lifetime than conventional signals such as Coulombic efficiency. This metric can be measured within seconds and can be integrated directly into the battery manufacturing process with no additional capital costs. This low-SOC resistance metric can thus be deployed in practical manufacturing settings to accelerate the evaluation of new formation protocols. We further demonstrate that the low-SOC resistance decreases as the quantity of lithium lost to the SEI during formation increases. With this physical insight, we propose that this metric, in principle, also be used to diagnose the impact of any manufacturing process that alters the total lithium consumed during formation.

\section{Results and Discussion}

\subsection{Fast Formation Experimental Design}

Two formation protocols have been implemented in this work: a fast formation protocol previously reported by Wood et al. \supercite{Wood2019, An2017}  which completes within 14 hours (Figure \ref{fig:design-of-experiments}b), and a baseline formation protocol (Figure \ref{fig:design-of-experiments}c) which completes in \revb{6}{56} hours. The fast formation protocol \rev{maximizes} the time spent at low negative electrode potentials to promote the creation of a more passivating SEI \supercite{An2017, Attia2021, Kim2011, Zhang2001}.

Forty \rev{nickel manganese cobalt (NMC)}/graphite pouch cells with a nominal capacity of 2.36Ah were built for this study (Table \ref{tab:cell-design}). Half of the cells underwent fast formation, and the remaining cells underwent baseline formation. Cells were further subdivided into `room temperature' and `45\textdegree C' aging groups for cycle life testing. The cycling profile was identical for all cells: 1C charge to 4.2V with a CV hold to 10mA and 1C discharge to 3.0V. Reference performance tests (RPTs) \supercite{Dubarry2020} were inserted throughout the cycle life test, which includes slow (C/20) charge and discharge curves as well as a Hybrid Pulse Power Characterization (HPPC) sequence \supercite{Christoph2015} used to extract the cell internal resistance as a function of SOC.

Our experimental design (Figure \ref{fig:design-of-experiments}a) uses larger samples sizes ($n=10$ per group) compared with those typically reported in the literature, which often use three cells or fewer per group. The increased sample size enables a more statistically rigorous analysis of the impact of different formation protocols on cell characteristics at the beginning and the end of life.

\begin{figure*}[ht]
\includegraphics[width=1\linewidth]{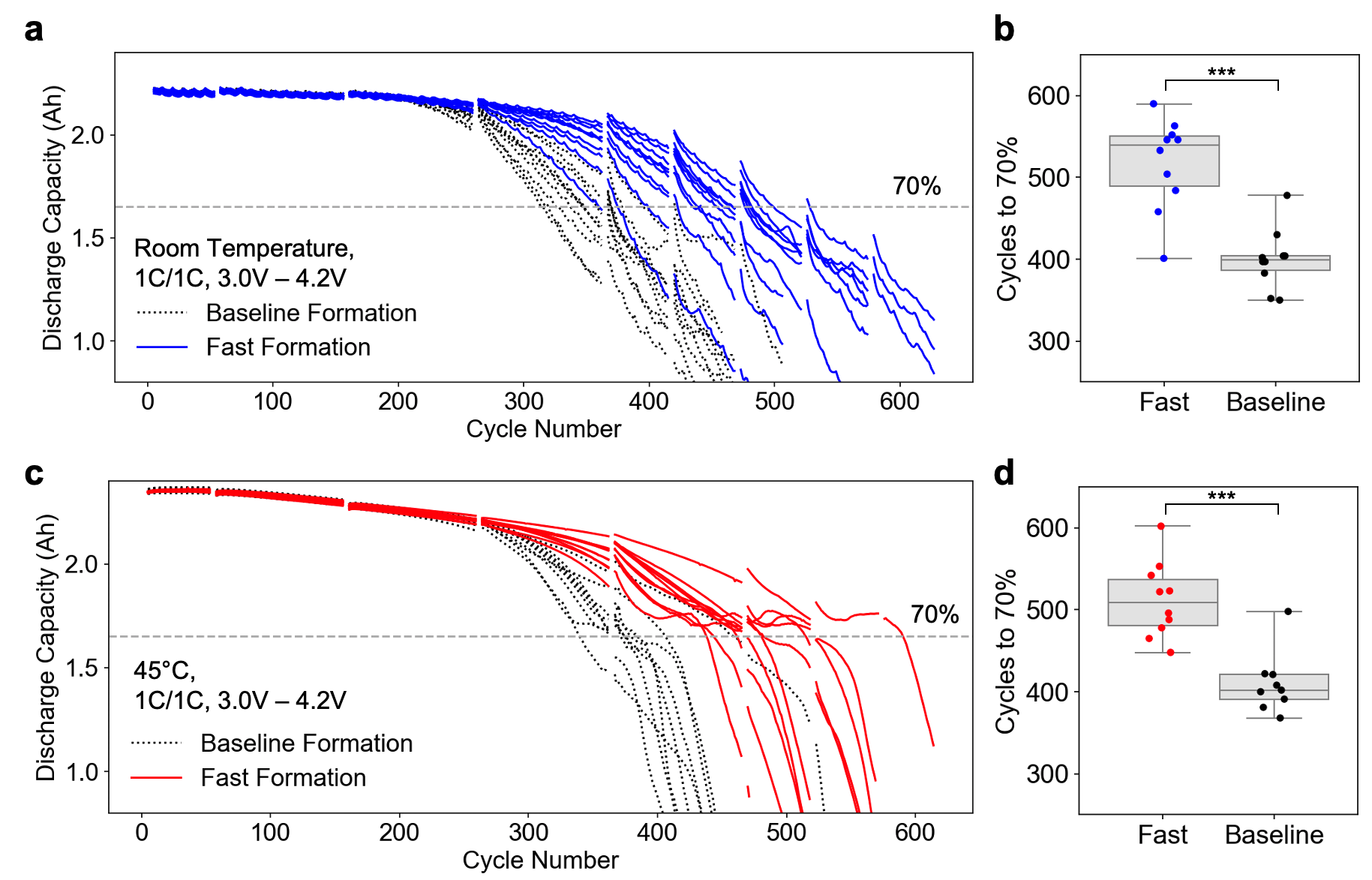}
\caption{\textbf{Cycle Life Test Results} \\
(a,c) Discharge capacity for individual cells measured during the 1C/1C aging test at (a) room temperature and (c) 45\textdegree C. Gaps in the curves correspond to the embedded reference performance test (RPT) cycles. (b,d) End-of-life capacity retention distributions, defined as when the cell discharge capacity reaches 70\% of initial capacity. `***' - statistically significant with $p$-value $< 0.001$.}
\label{fig:aging-test-result-combined}
\end{figure*}

\subsection{Fast Formation Cells Had Longer Cycle Life}
 
Fast formation cells had higher average lifetimes than the baseline formation cells under the cycle life test, as shown in Figure \ref{fig:aging-test-result-combined}. \rev{The degradation rate of fast formation cells initially track the baseline formation cells closely under both temperatures tested (Panels a, c)}. However, after 250 cycles, all cells begin to lose capacity rapidly. \rev{The fast formation cells sustained over 100 cycles longer before reaching the end of life, defined as when cells reach 70\% of their initial measured capacity (Panels b, d)}. This result is highly statistically significant ($p$-value < 0.001). The general result that fast formation improved \rev{lifetime} performance holds across multiple performance metrics, including Coulombic efficiency (Figure \ref{fig:aging-test-ce}) and voltage efficiency (Figure \ref{fig:aging-test-voltage-efficiency}), \revb{2}{as well as when plotted against equivalent cycles (Figure \ref{fig:fig_aging_test_result_combined_ahah})}. Together, these results support the growing body of evidence that well-designed fast formation protocols can improve cycle life \supercite{An2017, Ryan2021, Attia2021}.

\subsection{Finding Diagnostic Signals at the Beginning of Life}

Given the demonstrated impact of formation protocol on battery cycle life, \rev{we turn to investigating} methods to quantify the impact of fast formation on the initial cell state. Differences in the initial cell state \rev{(e.g. lithium consumed during formation)} may offer clues as to how fast formation could have improved cycle life. We focused our work on studying signals directly obtainable from full cell current-voltage data, which offer the lowest barrier-to-entry for deployment in real manufacturing settings.

\subsubsection{Conventional Metrics of Formation Efficiency}

Figures \ref{fig:initial-cell-state}\rev{(a-c)} show standard measures of formation efficiency extracted from the formation cycling data. The discharge capacity, $\Qd$, was measured at the end of each formation protocol during a C/10 discharge step from 4.2V to 3.0V. $\Qd$ corresponds to the capacity of cyclable lithium excluding the \rev{contribution from} lithium irreversibly lost to the SEI during formation. Fast formation decreased $\Qd$ by 0.3\%, a small but statistically significant difference ($p = 0.01$). The charge capacity, $\Qc$, was taken during the initial charge cycle, and includes both the capacity of cyclable lithium as well as the capacity of lithium lost irreversibly to the SEI. The quantity of lithium inventory lost to the SEI can be calculated as $\QLLI = \Qc - \Qd$ \rev{(Panel b)}. Note that while the two formation protocols differed in the initial charging rate, $\Qc$ remains a fair comparison metric since both charge protocols ended on a potentiostatic hold at 4.2V until the current dropped below C/100. \rev{Fast formation increased $\QLLI$ by 23 mAh ($p = 0.03$)}. Finally, we also included another common evaluation metric, the formation Coulombic efficiency, defined as $\cef = \Qd / \Qc$ \rev{(panel c), which shows that fast formation decreased $\cef$ by 0.8\% ($p=0.02$)}. Measured values are summarized in Table \ref{tab:initial-state-comparison}. Together, the results show that fast formation marginally increased the amount of lithium consumed during formation. A $p$-value of less than 0.05 in all cases indicate that the measured differences, while small, are statistically significant to a least a 95\% confidence level. 

\subsubsection{Low-SOC Resistance}

\begin{figure*}[ht]
\includegraphics[width=1\linewidth]{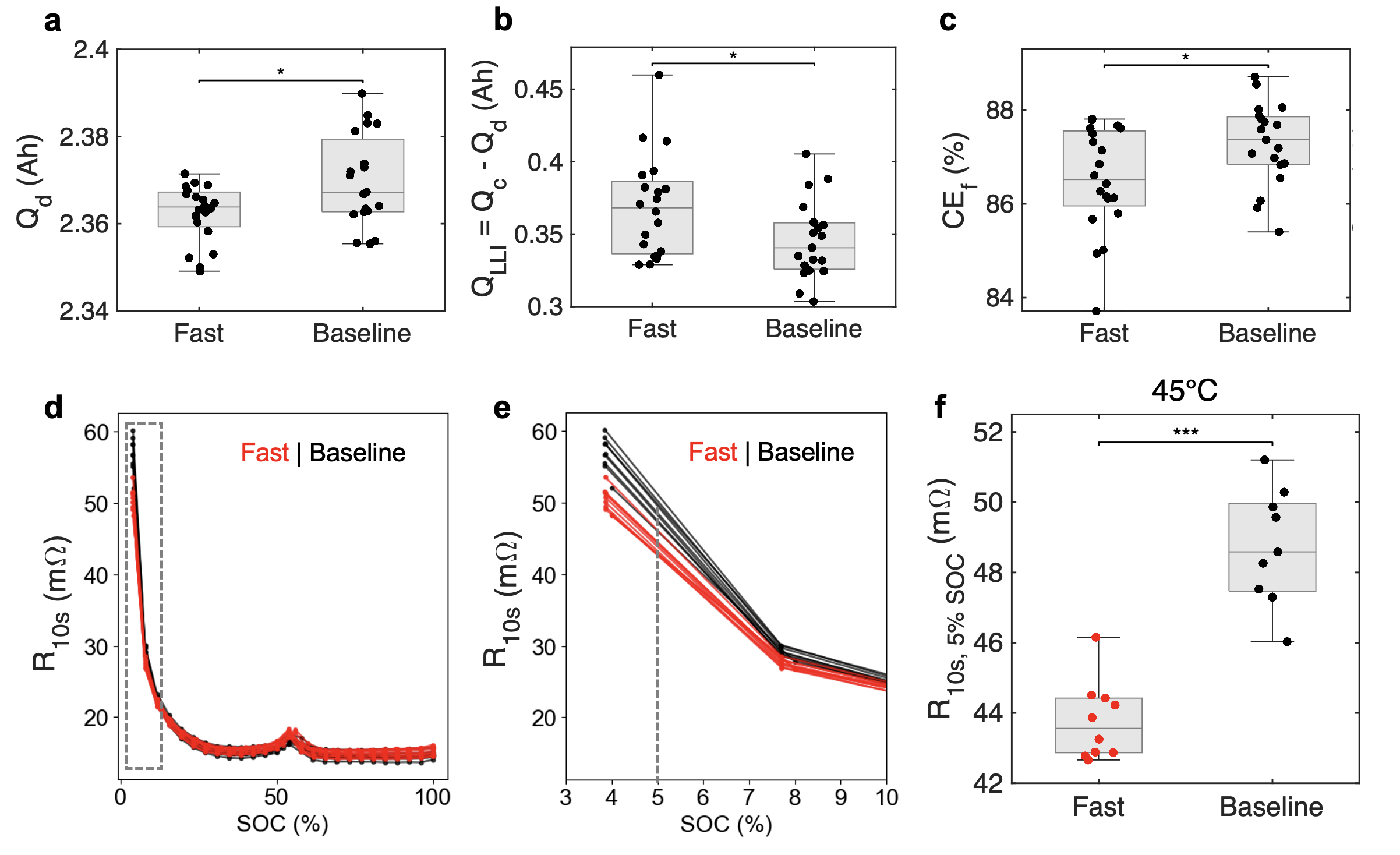}
\caption{\textbf{Diagnostic Signals for Differences in the Initial Cell State} \\
(a) Final discharge capacity, (b) capacity of lithium inventory lost during formation, and (c) formation Coulombic efficiency, measured from the formation protocol. (d) 10-second resistance obtained from the Hybrid Pulse Power Characterization test prior to the start of the cycle life test. (e) Magnification of the 10-second resistance at low SOCs. (f) Distribution of 10-second resistance at 5\% SOC comparing between the two formation protocols. (d-f) are sourced from the initial reference performance test from the 45\textdegree C cycle life test (see Figure \ref{fig:initial-dcr-both-temps} for the results at room temperature cycle life test). `*' - statistically significant with $p$-value $< 0.05$. `***' - statistically significant with $p$-value $< 0.001$. }
\label{fig:initial-cell-state}
\end{figure*}

Following formation, the cell internal resistance was measured using the Hybrid Power Pulse Characterization (HPPC) technique \supercite{Christoph2015} prior to the start of the cycle life test. During this test, a series of \revb{3}{10-second,} 1C discharge pulses were applied to the cell at varying SOCs\rev{,} and the resistance is calculated using Ohm’s law (Figure \ref{fig:hppc-example}). The 10-second resistance, R$_\mathrm{10s}$, was plotted against SOC for all cells cycled at 45\textdegree C (Figure \ref{fig:initial-cell-state}d). R$_\mathrm{10s}$ generally remained flat at mid-to high SOCs. The peak at 55\% SOC corresponds to the stage 2 solid-solution regime of the graphite negative electrode \supercite{Dahn1991}. R$_\mathrm{10s}$ rose sharply below 10\% SOC. Focusing on the low-SOC region (Figure \ref{fig:initial-cell-state}e), we observed that R$_\mathrm{10s}$ measured at 4\% and 8\% SOC were lower for fast formation cells compared to that of baseline formation cells. This result was highly statistically significant, with a $p$-value less than 0.001 (Figure \ref{fig:initial-cell-state}f). A similar result held when R$_\mathrm{10s}$ was measured at room temperature (Figure \ref{fig:initial-dcr-both-temps}). At mid to high SOCs, differences in R$_{\mathrm{10s}}$ between fast formation and baseline formation cells were generally not statistically significant (Figure \ref{fig:initial-dcr-both-temps}). Thus, differences in resistance between the two formation protocols appeared uniquely at low SOCs. All initial cell state metrics are summarized as part of Table \ref{tab:initial-state-comparison}.

\begin{table}[h!]
    \begin{center}
        \begin{tabular}{r | c c c c c c c c} 
          \toprule
             Metric & Unit & Temperature & Baseline Formation & Fast Formation & $\Delta$ (abs) & $\Delta$ (\%) & $p$-value \\
          \midrule
          $\Qd$ & mAh & Room temp & 2370 (11) & 2362 (7) & -8 & -0.3\% & 0.01 \\
          
          $\QLLI$ ($\Qc - \Qd$) & mAh & Room temp & 346 (27) & 369 (35) & +23 & +6.6\% & 0.03 \\
          
          $\cef$ & \% & Room temp & 87.3 (0.9) & 86.5 (1.1) & -0.8 & -0.9\% & 0.02 \\
          
          $R_{\mathrm{10s,5\%SOC}}$ ($\res$) & $\mohm$ & Room temp & 139.7 (2.9) & 130.0 (2.3) & -9.7 & -6.9\% & < 0.001 \\
          
          $R_{\mathrm{10s,5\%SOC}}$ ($\res$) & $\mohm$ & 45\textdegree C & 48.7 (1.6) & 43.8 (1.1) & -4.9 & -10.0\% & < 0.001 \\
          
          $R_{\mathrm{10s,90\%SOC}}$ & $\mohm$ & Room temp & 23.6 (0.1) & 23.9 (1.0) & +0.3 & +1.3\% & 0.28 \\
          
          $R_{\mathrm{10s,90\%SOC}}$ & $\mohm$ & 45\textdegree C & 14.5 (0.4) & 14.9 (0.5) & +0.4 & +2.8\% & 0.10 \\
          \bottomrule 
          
        \end{tabular}
    \end{center}
    \caption{\textbf{Comparison of Initial Cell State Metrics} \\ 
    Values are reported as mean (standard deviation). $\Qd$, $\QLLI$, and $\cef$ are extracted directly from the formation test protocol. R$_{\mathrm{10s}}$ metrics are extracted from the initial reference performance test at the beginning of the cycle life test profile.}
    \label{tab:initial-state-comparison}
  
\end{table}

To study the robustness of the \rev{low-SOC resistance signal}, we varied the SOC set-point between 4\% and 10\% and also computed the resistance under 1-second and 5-second pulse durations. In all cases, the resistance metric provided a high degree of contrast between the two different formation protocols (Figures \ref{fig:initial-dcr-effect-of-soc} and \ref{fig:initial-dcr-dffect-of-duration}). The lowest SOC measured in our dataset was 4\% SOC.

The remainder of the paper will focus on the resistance measured at 5\% SOC and with a 10-second pulse duration. From hereon, this metric will be referred to as the `low-SOC resistance', $\res$.

\subsection{Low-SOC Resistance as a Diagnostic Signal: A Data-Driven Perspective}

\subsubsection{Low-SOC Resistance Correlates to Cycle Life}

\begin{figure*}[ht]
\includegraphics[width=1\linewidth]{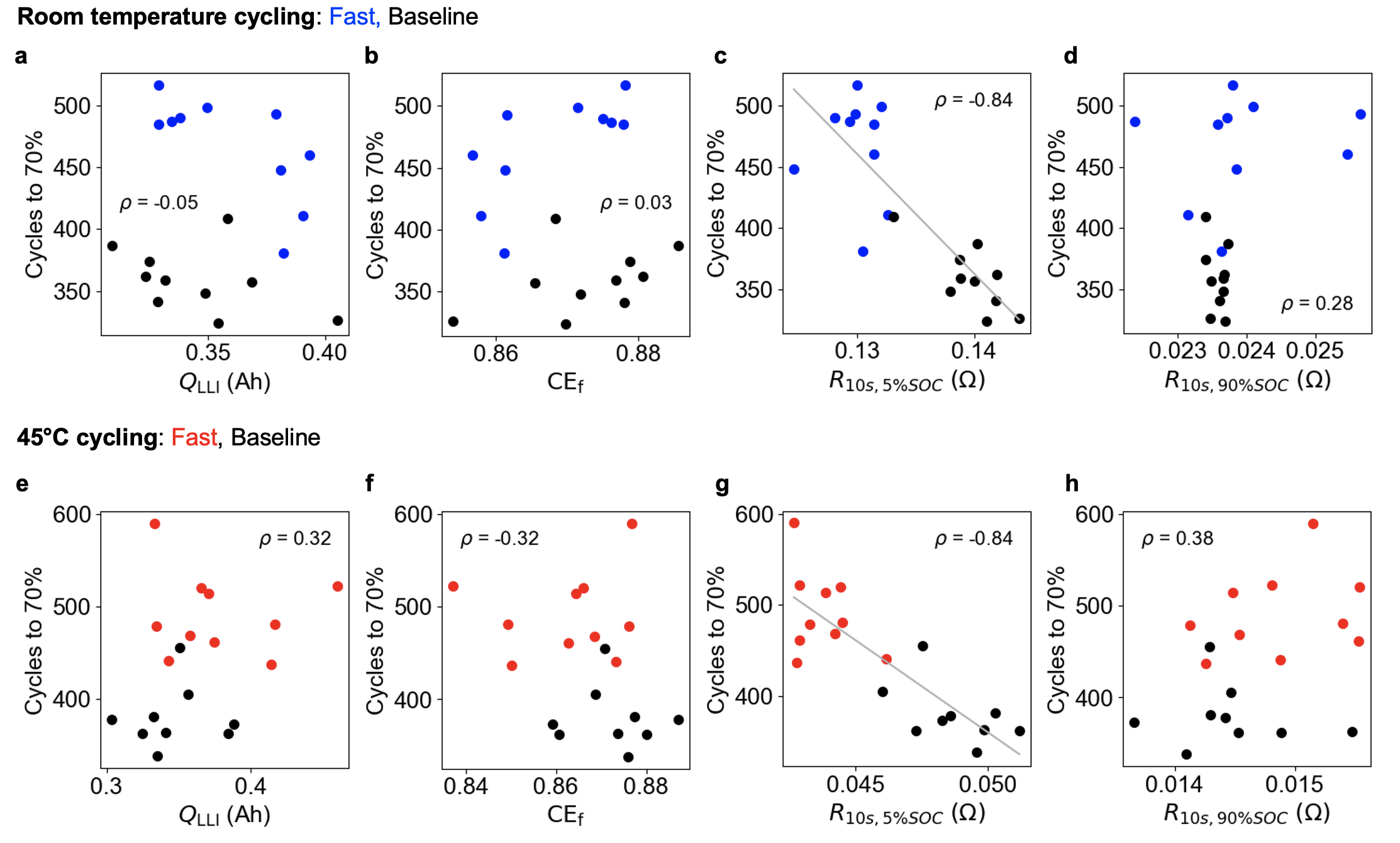}
\caption{\textbf{Correlation Between Early-Life Diagnostic Signals and Cycle Life} \\(a-d) Correlations under room temperature cycling. (e-h) correlations under 45\textdegree C cycling. Cycle life is defined as cycles to 70\% of initial capacity. $\QLLI$ and $\cef$ are taken directly from the formation test. $\reslow$ ($\res$) and $\reshigh$ are measured at the beginning of the cycle life test and thus share the same temperature as the cycle life test.}
\label{fig:correlation}
\end{figure*}

To evaluate the merit of low-SOC resistance ($\res$) as a diagnostic feature, we explored the correlations between the initial cell metrics (Figure \ref{fig:initial-cell-state}) and the cycle life, defined as cycles to 70\% of the initial capacity. The results are shown in Figure \ref{fig:correlation}. Out of all metrics studied, $\res$ was the only signal with a meaningful correlation to cycle life, with a correlation coefficient of $\rho = -0.84$. Other metrics such as $\Qd$ and $\cef$ were poorly correlated to cycle life ($|\rho| < 0.5$). We attribute the weakness of these correlations to the poor signal-to-noise inherent in cell capacity measurements in the absence of high-precision cycling \supercite{Smith2011a, Fathi2014}\revb{5}{, a topic we explore in detail later.}. The resistance measured at high SOCs also did not correlate to cycle life. From these results, we observe that the low-SOC signal uniquely holds information related to cycle life. These results have been reproduced for different end-of-life definitions ranging between 50\% and 80\% (Figures \ref{fig:correlations-different-eol-definitions-rt}, \ref{fig:correlations-different-eol-definitions-ht}), \revb{4}{as well as for charge pulses (Figure \ref{fig:correlations_charge_vs_discharge}}).

\subsubsection{Low-SOC Resistance Predicts Cycle Life}

To understand if $\res$ can be used to improve battery lifetime prediction, we trained univariate prediction models with regularized linear regression models inspired by Severson et al. \supercite{Severson2019}. 
The performance of the predictive models are summarized in Table \ref{tab:predictive-model}. A dummy regressor\rev{,} which predicts the mean of the training set and requires no \rev{cycling} data\rev{,} was included as a benchmark. For room temperature cycling, the model trained using $\res$ achieved the lowest test error of 6.9\% compared to 13.3\% for the dummy regressor. A similar result held under 45\textdegree C cycling. To compare, we also included the $\varq$ metric introduced by Severson et al. \supercite{Severson2019}, defined as the variance in the discharge capacity versus voltage curve between cycle 10 and cycle 100. When applied to our dataset, this metric did not yield a significant improvement over the dummy regressor. This result suggests that $\res$ is a stronger predictor of battery lifetime than $\varq$.

We repeated this study with multivariate regularized linear regressions: one using the three capacity-based features from formation ($\QLLI$, $\cef$, and $\Qd$) and another using the previous three formation features plus $\res$. Using only the features from formation, no improvement over the dummy regressor was achieved. By including $\res$ in the feature set, however, the test error was improved. Yet, the test error achieved did not exceed the test error of the univariate model using $\res$ alone. This result suggests that the chosen set of formation features does not provide useful information about cycle life beyond what is provided by $\res$. This result is counter-intuitive considering the important role that lithium consumption plays in determining battery lifetime \supercite{Winter2009, An2016, Wang2018, Peled2017, Goers2011, Lu2014a}, which should be reflected in the formation features such as $\QLLI$ and $\cef$. We \rev{hypothesize} that the reason for the poor model performance using formation signals is not because these formation signals lack physical meaning. Rather, due to the absence of high-precision cycling and temperature control, the useful information within these signals may be masked by the noise in the data (e.g. due to current integration errors, temperature variations over the course of 10+ hours of formation, etc.) $\res$ appears to be able to overcome these limitations. We explore the connection between $\res$ and the other formation metrics in detail later.

\begin{table}[]
    \centering
    \begin{tabular}{l c c c c c}
        \toprule
         Model & Data needed & \multicolumn{2}{c}{Room temp} & \multicolumn{2}{c}{45\textdegree C} \\
         & & Train & Test & Train & Test \\
        \midrule
        Dummy regressor & none & 13.3 (1.0) & 14.4 (4.0) & 14.0 (0.9) & 15.1 (3.6) \\
        \midrule
        $\res$ & 3 cycles & 6.9 (0.5) & 8.0 (2.8) & 6.5 (0.6) & 7.4 (2.9) \\
        $\QLLI$ & formation & 12.2 (1.2) & 14.0 (4.6) & 14.1 (0.8) & 15.2 (4.4) \\
        $\cef$ & formation & 12.2 (1.2) & 13.8 (4.5) & 14.1 (0.7) & 15.1 (4.3) \\
        $\Qd$ & formation &12.0 (1.2) & 13.6 (5.0) & 13.5 (0.8) & 15.0 (4.0) \\
        $\varq$ & 100 cycles & 11.6 (1.7) & 14.4 (5.2) & 10.3 (1.1) & 11.5 (4.7) \\
        \midrule
        $\QLLI$ + $\cef$ + $\Qd$ & formation & 12.8 (1.3) & 14.5 (5.1) & 13.4 (1.1) & 14.1 (4.0) \\
        $\QLLI$ + $\cef$ + $\Qd$ + $\res$ & 3 cycles & 7.2 (1.1) & 9.4 (4.0) & 6.5 (1.0) & 7.4 (2.9) \\
        \bottomrule
    \end{tabular}
    \caption{\textbf{Training and Testing Errors for Different Lifetime Prediction Models} \\
    Values represent means (standard deviations). The dummy regressor model uses no features and simply returns the mean of the training set, and hence is the baseline against which to judge the performance of other features. All remaining models use a Ridge regression with nested cross-validation to determine the optimal regularization strength (see Experimental Procedures).}
    \label{tab:predictive-model}
\end{table}

As defined in this work, the model trained using $\res$ required just \rev{three} cycles of lifetime testing, i.e., one diagnostic cycle. (The two preceding cycles consisted of slow-rate charge-discharge cycles as part of the reference performance test inserted at the beginning of the cycle life test.) By comparison, $\varq$ requires 100 cycles of lifetime testing. For future implementations, $\res$ can be incorporated directly into the formation protocol, further decreasing the required measurement time. The total amount of data required to exercise each predictive model is summarized in Table \ref{tab:predictive-model}. 

Overall, the correlation and prediction results suggest that $\res$ may be useful for advancing broad-scale efforts to improve cycle life prediction using small and readily-obtainable datasets at the beginning of life. While the results are promising, they are also limited, since only two types of formation protocols have been studied here. To understand the extent to which $\res$ can generalize to other applications (e.g. chemistries, use cases, cell designs) and to understand the relation between $\res$ and the other formation signals, the rest of the paper will focus on providing a physical interpretation of $\res$. A mechanistic understanding of $\res$ will provide the necessary context required to evaluate the general scope of applicability and limitations of the method.

\subsection{Low-SOC Resistance as a Diagnostic Signal: A Mechanistic Perspective}

Understanding the physical interpretation of diagnostic signals can help assess whether prediction frameworks leveraging such signals can generalize to new systems. \revb{10}{In principle, different formation protocols, manufacturing process changes, and cell design changes could all lead to changes in lithium consumption and active material losses during formation.} Towards this end, we will first review the commonly accepted theory of SEI passivation and showed how our observations of $\QLLI$ and $\cef$ support this theory. Next, we will show that our observations of $\res$ are consistent with this theory but provide a stronger and more easily measurable signal compared to conventional measures.

\subsubsection{Benefits of Fast Formation on Cycle Life} 

Lithium intercalation at \rev{negative electrode} potentials higher than 0.25V-0.5V vs Li/Li$^+$ is generally associated with the formation of a porous, poorly-passivated SEI film \supercite{Peled2017, Zhang2001, Edstrom2006, Lu2011a, Lu2014a}. Conversely, lithium intercalation at negative electrode potentials below 0.25V-0.5V has been found to promote the formation of a more conductive and passivating SEI film \supercite{Attia2021, Zhang2001}. Attia et al. \supercite{Attia2021} showed that the reduction of ethylene carbonate (EC) at negative electrode potentials above 0.5V vs Li/Li$^+$ is non-passivating. This negative electrode potential corresponds to a full cell voltage of below 3.5V, neglecting overpotential contributions. Hence, an ideal formation protocol would minimize the time spent charging below 3.5V while maximizing the time spent above 3.5V. The fast formation protocol we tested \supercite{An2017} achieves this \rev{objective by rapidly charging the cell to above 3.9V at a 1C rate and subsequently cycling the cell between 3.9V and 4.2V, thus decreasing the time associated with the non-passivating EC reduction reaction.
%The protocol  to promote the continued formation of \rev{the} passivating SEI film.
Focusing on the initial charge cycle, fast formation cells spent only 2 minutes below 3.5V and 12.9 hours above 3.5V, while baseline formation cells spent 30 minutes below 3.5V and 9.4 hours above 3.5V. Fast formation decreased the time spent below 3.5V by 28 minutes.}
%, resulting in a decrease in the amount of lithium lost to form the non-passivating SEI. Simultaneously, fast formation increased the time spent above 3.5V by 3.5 hours, resulting in an increase in the amount of lithium lost to form the passivating SEI. 
Fast formation resulted in a net increase in total lithium consumed during formation, $\Delta\QLLI$, by 23 mAh (Table \ref{tab:initial-state-comparison}). This increase is attributed to the \rev{additional} lithium consumed to form the passivating SEI.

While fast formation cells consumed more lithium during formation and thus exhibited lower $\cef$ (or, equivalently, higher $\QLLI$), these cells lasted longer on the cycle life test. \rev{While a lower initial coulombic efficiency is conventionally associated with poor cycle life performance \supercite{Smith2011a, Burns2013}, the opposite was true in our study since the additional lithium consumed during fast formation was associated with the creation of a more passivating SEI.} A more passivating SEI can, for example, lower the rate of electrolyte reduction reactions associated with the formation of solid products that decrease the negative electrode porosity which could subsequently increase the propensity for lithium plating during charge \supercite{Reniers2019, Yang2017}. A more passivating SEI could therefore play a role in delaying the `knee-point' observed in the cycle test data. Our result reinforces the notion that passivation of the SEI during the first cycle plays an important role in determining battery cycle life.

\subsubsection{Lithium Loss Dominates Overall Cell Capacity Loss Over Cycling}

\revb{7}{We performed a voltage fitting analysis \supercite{Bloom2005, Smith2011a, Dubarry2012, Dubarry2017d, Lee2020a} to confirm that the main failure mode in our cells is the loss of lithium inventory (LLI) over cycle life (Figures \ref{fig:aging_test_signals_high_temp}, \ref{fig:aging_test_signals_low_temp}). We found that LLI can fully account for the thermodynamic (i.e. C/20) cell capacity loss over life. The knee-point in LLI over cycle life coincides with the knee-point in the capacity loss. All cells also experienced an increase in the loss of active material in the negative electrode ($\lamne$) after the knee-point, which could indicate the occurrence of porosity decrease and/or electrolyte depletion as a result of a less passivating SEI, as discussed previously. The increased $\lamne$ after the knee-point was less prominent in the fast formation cells, suggesting that the more passivating SEI generated from fast formation could be playing a role in delaying the knee-point to improve lifetime. Finally, all cells experienced a knee-point in the capacity fade rate irrespective of whether the discharge capacity is measured at higher (C/3) or lower (C/20) C-rates (Figure \ref{fig:aging_test_diagnostic_signals_2}), indicating that kinetic limitations cannot fully account for the observed knee-point in the cycle life data. The origin of the capacity loss therefore has a strong thermodynamic component which can be attributed to the loss of lithium inventory. This analysis further supports the theory that consuming more lithium at low negative electrode potentials during formation can create a passivating SEI that is beneficial to cycle life \supercite{Attia2021}}. 

\subsubsection{Low-SOC Resistance is Attributed to Kinetic Limitations in the Positive Electrode}

\revb{10}{To explore possible physical connections between $\res$ and the impact of fast formation on cycle life, we first develop a physical interpretation of the low-SOC resistance. We focus our discussion on the resistance contributions from the positive and negative electrode. While other cell components (e.g. current collectors, tabs, and electrolyte) also contribute to the total cell resistance, they are not known to depend on SOC and hence cannot explain the rising resistance measured at low SOCs.}

\revb{10}{Positive electrode diffusion limitations generally play a significant role in the low-SOC cell resistance in NMC/graphite systems. The solid-state diffusion coefficient in NMC materials has been measured to decrease by more than one order of magnitude at high states of lithiation \supercite{Yang2012}, a phenomenon attributed to the depletion of divacancies needed to support diffusion as the electrode becomes fully lithiated \supercite{Zhou2019, Liu2021c}. Using half-cell HPPC measurements, we experimentally verified that the positive electrode dominates the low-SOC resistance. In the coin cell form factor, the 10-second resistance of graphite/Li stayed below 100m$\Omega$ as the graphite approached full delithiation, while the 10-second resistance of NMC/Li exceeded 1000m$\Omega$ as the NMC approached full lithiation (Figure \ref{fig:coin_cell_hppc}). This finding is consistent with previous empirical studies on NMC/graphite systems \supercite{An2017a, Abraham2005, Wu2000}. In particular, An et al. \supercite{An2017a} used a three-electrode pouch cell configuration to show that, for an NMC/graphite system, the positive electrode accounts for nearly all of the measured full cell resistance at all SOCs.}

\revb{10}{Charge transfer kinetics at either electrode could also play a role at determining total cell resistance. The charge transfer process at either electrode can be modeled using the Butler-Volmer equation \supercite{Plett2015}:}

\begin{equation}
    j = k_0 c_e^{1-\alpha} (c_{s,\mathrm{max}} - c_{s,e})^{1-\alpha} c_{s,e}^\alpha \Big(\exp\Big(\frac{(1-\alpha)F}{RT}\eta\Big) - \exp\Big(-\frac{\alpha F}{RT}\eta\Big)\Big). 
\end{equation}

\revb{10}{In this equation, $j$ is the reaction flux, the exponential terms describe the overpotential dependence of the forward and backward reactions, and the exponential prefactor terms together describe the exchange current density. $c_{s,\mathrm{max}}$ is the theoretical maximum allowable lithium concentration in the solid phase, $c_{s,e}$ is the surface concentration of lithium, and $k_0$ is the reaction rate constant. The exchange current density approaches zero as the electrode becomes either fully lithiated or fully delithiated. Indeed, our coin cell data shows that as the graphite negative electrode approaches full delithiation, the measured resistance rises steeply (\ref{fig:coin_cell_hppc}i). However, the magnitude of this charge transfer effect remains small compared to the contribution from the diffusion-limited NMC positive electrode  (\ref{fig:coin_cell_hppc}h) at high states of lithiation.}

\revb{10}{In summary, we attribute the low-SOC resistance to kinetic limitations in the positive electrode. This result was experimentally verified using coin cell measurements of electrode resistances and is consistent with literature findings \supercite{An2017a, Abraham2005, Wu2000}. The kinetic limitation arises from a combination of diffusion and charge transfer limitations in the positive electrode. For NMC/graphite systems, diffusion limitations (i.e. `kinetic hindrance' \supercite{Liu2021c}) is a major component of the rapid rise in measured resistance at low SOCs.}

\subsubsection{Lithium Consumption Leads to an Apparent Decrease in Low-SOC Resistance}

\begin{figure*}[ht]
\includegraphics[width=1\linewidth]{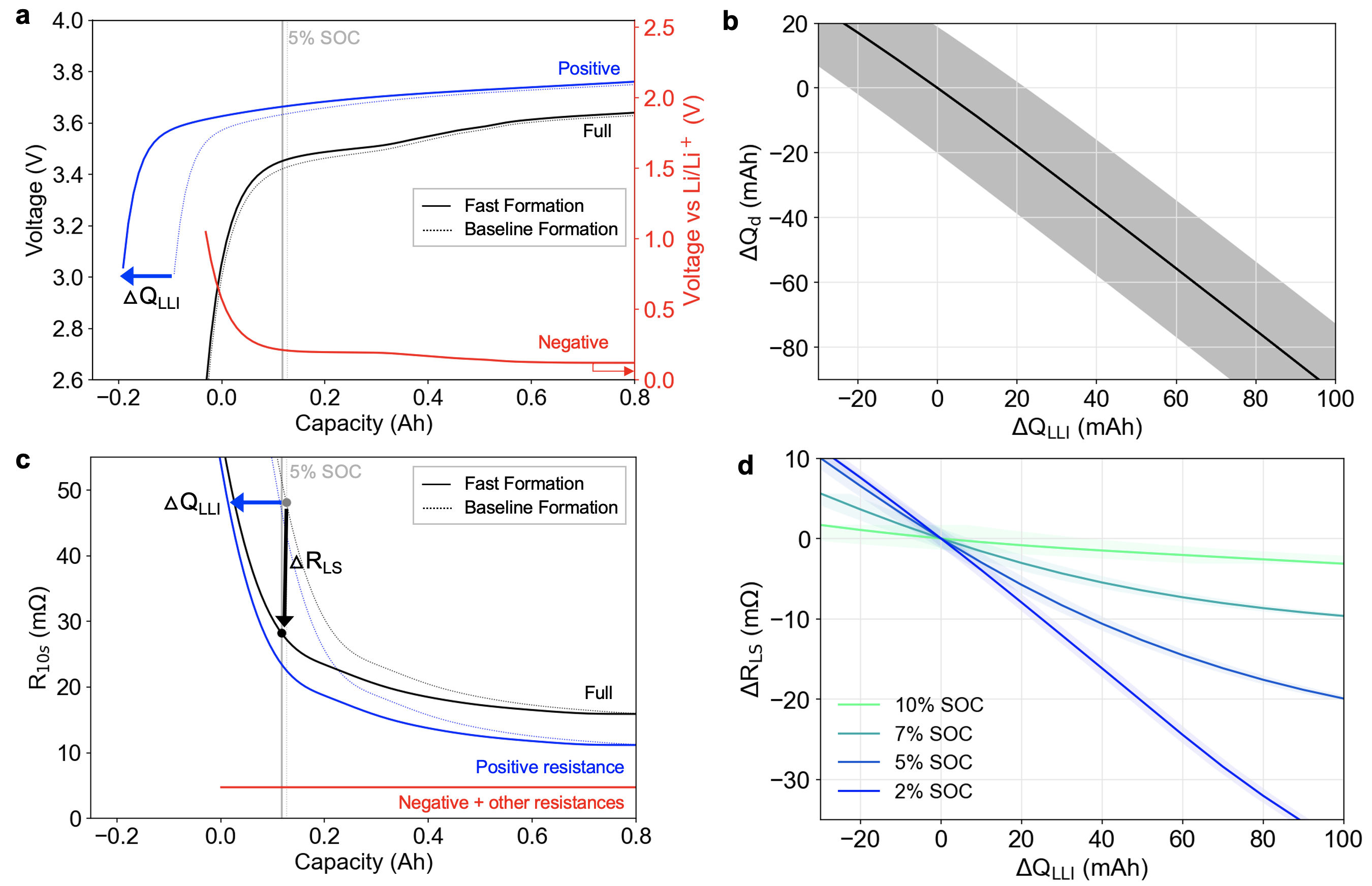}
\caption{\textbf{Electrode Stoichiometry Model Illustrating the Impact of Lithium Consumption on Low-SOC Resistance} \\
(a) Relative alignment of the positive and negative equilibrium potential curves after baseline formation and fast formation. (b) Effect of increasing lithium consumption ($\Delta\QLLI$) on the measured discharge capacity. (c) The corresponding cell resistance curves, where the measured full cell resistance (black lines) have been broken down into positive electrode charge resistance (blue) and all other resistances (red). (d) The effect of increasing $\Delta\QLLI$ on the measured low-SOC resistance for varying SOC set-points. Bands in (b,d) indicate estimates of the measurement error bound using conventional battery cycling equipment, clarifying $R_{LS}$'s improved accuracy in representing changes in $Q_{LLI}$ when compared to $Q_d$ (see Supplementary Materials).}
\label{fig:initial-cell-state-mechanism}
\end{figure*}

Fast formation decreased the measured low-SOC resistance ($\res$). \rev{From our previous analysis, fast formation also increased the lithium consumed during formation ($\QLLI$) to create a more passivating SEI. To explain the connection between these two quantities, we employ a simple electrode stoichiometry model which describes both the thermodynamic potentials and kinetic limitations of both electrodes. Figure  \ref{fig:initial-cell-state-mechanism}a shows the relative alignment of the positive and negative equilibrium potential curves after baseline formation and fast formation. The origin of the capacity axis corresponds to 0\% SOC (3.0V) after baseline formation. The gap between the positive and negative potential curve endpoints is attributed to the lithium lost to the SEI during formation, or $\QLLI$ \supercite{Dubarry2012, Smith2011a}. By comparison, the curves prior to formation do not have a gap, corresponding to $\QLLI = 0$ (Figure \ref{fig:initial-cell-state-before-formation}). We shift the positive electrode curve to the left by some amount $\Delta\QLLI$ to emulate the impact of additional lithium consumed during fast formation. Here, $\Delta\QLLI$ has been set to an exaggerated value of 100mAh for graphical clarity. An alternative graphic is provided in Figure \ref{fig:initial-cell-state-mechanism-precise}, which sets $\Delta \QLLI$ = 23mAh to coincide with the measured difference between baseline formation and fast formation.}

\rev{Figure \ref{fig:initial-cell-state-mechanism}c shows the corresponding full cell 10-second resistance measured from the HPPC test.} \revb{9}{The full cell resistance is partitioned to model a scenario in which the positive electrode dominates the low-SOC resistance, consistent with previous findings.} \rev{The resistance curve of the positive electrode must also translate to the left by the same amount $\Delta\QLLI$ due to the increased lithium consumed during fast formation. From the reference frame of the full cell, the measured low-SOC resistance will decrease by $\Delta\res$. In this manner, $\res$ can decrease without any real change in positive electrode kinetic properties. The decrease in $\res$ reflects the shifting of the positive electrode stoichiometry window as lithium is consumed.}

Two additional observations support the connection between $\Delta\QLLI$ and $\Delta\res$. First, $\res$ appears to be positively correlated to $\cef$ and negatively correlated to $\QLLI$ (Figure \ref{fig:initial-cell-state-correlations})\rev{, a result which is consistent with theory and predicted by the electrode stoichiometry model.} The strengths of the correlations are generally weak, with correlation coefficients, \rev{$|\rho|$, ranging between 0.2 and 0.5}. We attribute the weakness of the correlations to the poor signal-to-noise of the capacity measurements using typical battery cycling equipment, which may compound at room temperature where the temperature is not strictly controlled (Figure \ref{fig:aging-test-temperature}). \rev{Second, we note that the resistance around 90\% SOC is insensitive to small changes in SOC, so changes in resistance at 90\% SOC provides a measure of true} resistance changes rather than \rev{apparent changes due to electrode stoichiometry shifts (Figure \ref{fig:initial-cell-state}d). Fast formation did not significantly increase the resistance at 90\% SOC (Figure \ref{fig:initial-dcr-both-temps}), so the changes in $\res$ is not likely to be due to material changes in the cell resistance (e.g. due to resistive surface films). This observation further supports the hypothesis that changes in $\res$ are due to electrode stoichiometry window shifts in the presence of lithium consumption.}

\subsubsection{Low-SOC Resistance Improves the Observability of Lithium Loss During Formation}

\revb{5}{Figure \ref{fig:initial-cell-state-mechanism}b shows that the sensitivity of the measured cell discharge capacity ($\Delta\Qd$) to the lithium consumed ($\Delta\QLLI$) is 0.9 mAh/mAh. The error in measuring $\Qd$ is 20 mAh due to current integration inaccuracies using ordinary cycling equipment. Hence, using $\Qd$ to estimate $\QLLI$ leads to a measurement error of 22 mAh. Since the total difference in lithium consumed between fast formation and baseline formation is 23mAh, measurement noise may prevent $\Delta\Qd$ from effectively resolving this difference. In our experiments, we relied on large sample sizes ($n=10$ per group) to resolve the small difference in lithium consumption between the two formation protocols.}

\revb{5}{Figure \ref{fig:initial-cell-state-mechanism}d shows that the sensitivity of the low-SOC resistance ($\Delta\res$) to $\Delta\QLLI$ is 0.22 m$\Omega$ / mAh when measured at 5\% SOC. The error in measuring $\res$ is 0.88 m$\Omega$ due to the voltage and current precision for calculating resistance using Ohm's law using ordinary cycling equipment. Hence, using $\res$ measured at 5\% SOC to estimate $\QLLI$ leads to a measurement error of 4 mAh, a five-fold improvement over using $\Qd$. Figure \ref{fig:initial-cell-state-mechanism}d further shows that the sensitivity of $\res$ is improved at lower SOCs. For example, $\res$ measured at 2\% SOC leads to a measurement error of 2.5 mAh. Any SOC set-point lower than 7\% SOC makes $\res$ a more precise measure of $\QLLI$ compared to $\Qd$. See Supplementary Materials for a detailed derivation of the measurement errors.}

\subsection{Generalizability}

\revb{10}{So far, we have explored the sensitivity of $\res$ to lithium lost during formation for an NMC/graphite system. By understanding the benefits of fast formation \supercite{Attia2021}, we rationalized why $\res$ is predictive of cycle life for our system. Here, we discuss the application of $\res$ towards understanding other degradation modes, chemistries, and use cases. This discussion sets the stage for understanding how $\res$ may be incorporated into generalizable lifetime prediction and diagnostic frameworks.}

\subsubsection{$\res$ Can Detect Active Material Losses}

\revb{10}{In principle, some small quantity of positive and negative active material could be lost during formation, i.e. due to expansion and contraction of the electrodes during initial lithiation and delithiation. In the positive electrode, lithiation-induced stresses can induce particle fracturing in the metal oxide particles \supercite{Watanabe2014, Yan2017, Zhang2020b}, leading to capacity loss. In the negative electrode, while graphite cracking is unlikely to occur under most applications \supercite{Takahashi2015}, insufficient binder adhesion or electrolyte wetting \supercite{Kupper2018} could create local islands of isolated graphite particles, leading to active material loss.}

\revb{10}{We develop a simple mechanistic electrode stoichiometry model to examine the influence of active material losses in both the positive and negative electrodes. Our model differentiates between loss of active material in the lithiated phase versus the delithiated phase \supercite{Dubarry2012}. For the positive electrode, loss of active material in the delithiated phase is represented by shrinking the positive electrode equilibrium potential curve with the point of minimum stoichiometry fixed (i.e. shrinking from the bottom, Figure \ref{fig:generalizability_lam_pe}a), while loss of active material in the lithiated phase is represented by shrinking the positive electrode equilibrium potential curve with the point of maximum stoichiometry fixed (i.e. shrinking from the top, Figure \ref{fig:generalizability_lam_pe}d).  $\res$ was found to increase with loss of positive active material, but only in the delithiated phase (Figures \ref{fig:generalizability_lam_pe}b). By contrast, active material lost in the lithiated phase bears a negligible effect on $\res$ (Figures \ref{fig:generalizability_lam_pe}d,e). This result can be understood graphically by considering the influence of the positive curve shifts on the positive electrode stoichiometry at low SOCs. In the case of loss of active material in the lithiated phase, the positive electrode stoichiometry at low SOCs does not significantly change, whereas in the delithiated case, the maximum positive electrode stoichiometry increases, causing $\res$ to increase. Note that $\Qd$ has the opposite sensitivity: $\Qd$ is sensitive to loss of active material in the lithiated state only. Hence, $\res$ and $\Qd$ complement each other in the study of positive electrode active material loss mechanisms. A similar analysis can be done on the negative electrode (Figure \ref{fig:generalizability_lam_ne})}.

\revb{10}{Figure \ref{fig:generalizability_lam_comparison} compares the sensitivity of $\res$ and $\Qd$ to the four different modeled cases of active material losses. The results highlight that the measured value of $\res$ is determined by multiple degradation factors, including both lithium inventory loss and active active material losses. It would therefore be impractical to use $\res$ to identify any dominant degradation mode without some \textit{a priori} understanding of the system through additional characterization and analysis. For diagnostic purposes, we recommend that $\res$ be used within the context of a broader set of non-destructive techniques to enrich the understanding of degradation mechanisms. From a data-driven prediction perspective, however, the sensitivity of $\res$ to active material losses in addition to lithium inventory loss may make it a more robust indicator for multiple degradation modes. In general, $\res$ may need to be coupled with other signals to improve the observability of distinct degradation modes.}

\subsubsection{When is $\res$ Sensitive to Lithium Loss?}

\revb{10}{We have so far focused on an NMC111/graphite system where kinetic limitations in the positive electrode dominates $\res$, a result which holds for nickel-rich cathode chemistries such as nickel cobalt aluminum (NCA) and higher nickel content NMC materials \supercite{Liu2021c, Zhou2019}. In general, electrode design factors such as particle size \supercite{Taleghani2017} and surface modifications \supercite{Mohanty2016} could impact the relative contribution of each electrode to $\res$. To study how such changes could modify the sensitivity of $\res$ to changes in $\QLLI$, we performed a sensitivity study using our electrode stoichiometry model by varying the proportion of the total cell resistance attributed to the positive electrode. The results (Figures \ref{fig:generalizability_cathode_dominance_1}, \ref{fig:generalizability_cathode_dominance_2}) show that $\res$ becomes ineffective at quantifying $\QLLI$ if the positive electrode contributes to less than 50\% of the total cell resistance at low SOCs. This result suggests that the utility of $\res$ as a diagnostic signal for $\QLLI$ diminishes for systems where the positive electrode is not the main contributor to $\res$.}

\subsubsection{When Can $\res$ Predict Cycle Life?}

\revb{10}{Our cycle life correlation study was presented in the context of the study of fast formation. To understand whether $\res$ can predict cycle life for other use cases (i.e. chemistries and aging conditions), we start by reviewing why $\res$ was predictive of cycle life for fast formation. Figure \ref{fig:mechanism} outlines the connection between fast formation and cycle life. In brief, fast formation spent more time above 3.5V, creating a higher quantity of SEI that is more passivating \supercite{Attia2021}. The passivating SEI improved cycle life by protecting the negative electrode against side reactions over life. The low-SOC resistance, $\res$, provided an estimate of the amount of lithium consumption during formation, $\QLLI$, and thus served as a proxy for both the amount of passivating SEI formed and the cycle life of the cell. This physical description rationalizes the predictive power of $\res$ within the context of the degradation pathway (fast formation) and chemistry (NMC/graphite) explored in this study.

\begin{figure*}[ht]
\includegraphics[width=1\linewidth]{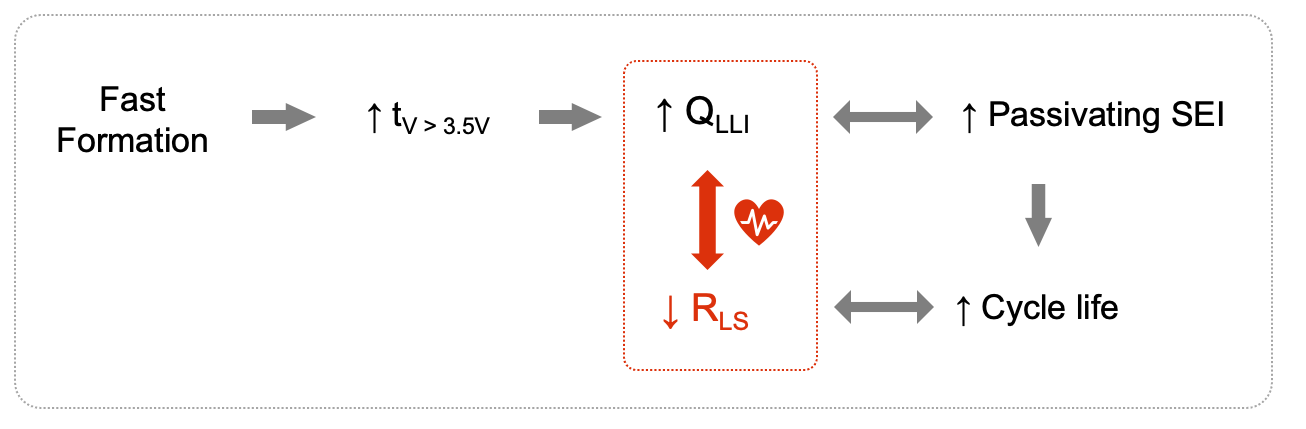}
\centering
\caption{\textbf{Connection Between the Fast Formation Degradation Pathway and the Low-SOC Resistance Early Life Diagnostic Signal} \\ Inner box: the relationship between low-SOC resistance ($\res$) and lithium consumed during formation ($\QLLI$) is general. Outer box: the relationship between low-SOC resistance ($\res$) and cycle life applies specifically to fast formation, where higher $\QLLI$ signaled the creation of a more passivating SEI \supercite{Attia2021} which improved cycle life. The relationship between $\res$ and cycle life may differ for other use cases.}
\label{fig:mechanism}
\end{figure*}

To gain confidence that $\res$ can predict cycle life for other use cases, the relationship between lithium loss ($\QLLI$) and cycle life must first be understood. For our study, the knowledge that increased $\QLLI$ signals a more passivating SEI was necessary for rationalizing why higher $\QLLI$ after formation could be beneficial to cycle life. For other use cases, the opposite may be true. For example, low first-cycle efficiencies for silicon-containing anodes \supercite{Jin2017} or lithium metal anodes \supercite{Xiao2020a} generally indicate poor negative electrode passivation which leads to poor cycle life. Under such use cases, $\res$ may still be predictive of cycle life, but the relationship may become inverted.}

\subsection{Unique Properties of $\res$}

Here, we highlight several unique properties of using the low-SOC resistance ($\res$) as an early-life diagnostic signal. First, since the positive electrode \rev{kinetics becomes poorer} as the electrode becomes fully lithiated, the sensitivity of $\res$ to lithium loss ($\QLLI$) improves as the measurement SOC decreases (Figure \ref{fig:initial-cell-state-mechanism}d). The results from this study used $\res$ measured at 5\% SOC. For future work, the sensitivity to $\QLLI$ may be further improved by taking the measurement at even lower SOCs. Second, \rev{$\res$ can be used to extract information about $\QLLI$ within seconds and therefore can be deployed in manufacturing settings without decreasing production speed. By contrast, conventional measurements of $\QLLI$ relying on Coulomb counting require full charge-discharge cycles during formation which could take hours to days to complete. Since measuring $\res$ does not require full cycles, $\res$ is also suitable for diagnosing differences in lithium consumption between formation protocols with different charge and discharge conditions.} Finally, $\res$ becomes stronger the earlier in life it is measured. As the cell ages, continual loss of lithium inventory will cause the highly sloped region of the \rev{positive electrode resistance} curve to become inaccessible during the normal full cell voltage operating window. Typically, diagnostic features become less predictive of cycle life the earlier in life the feature is sampled \supercite{Severson2019}. $\res$ is expected to have the opposite relationship: the earlier in life $\res$ is sampled, the more sensitive it will be to changes in $\QLLI$.

\subsection{Diagnosing State of Health Beyond Cycle Life: Practical Considerations}

Our discussion has so far focused on evaluating the merits of $\res$ for diagnosing cycle life. However, in real manufacturing settings, cycle life is only one of many considerations for adopting new formation protocols. Here, we introduce two such considerations: (1) impact to gas buildup over life, and (2) impact to aging variability over life. In our analysis, $\res$ could not be used to learn the impact of fast formation on gas buildup or aging variability. \rev{Here, we give an overview of these observations.}

\subsubsection{Gas Buildup Over Life}

Swollen cells in a battery pack can compromise pack integrity and pose safety hazards to first-responders for electric vehicle fire accidents \supercite{Sun2020a}. Understanding the impact of formation protocols on cell swelling is therefore just as important as understanding the impact on cycle life for practical purposes.

Fast formation caused a significant degree of swelling at the end of life for cells cycled at 45\textdegree C (Figures \ref{fig:swelling-comparison}, \ref{fig:swelling-all}). At this temperature, 9 of 10 fast formation cells showed visible signs of swelling, compared to only 2 of 10 for baseline formation. None of the cells cycled at room temperature showed any appreciable degree of swelling. All swollen pouch cells were compliant and compressible, indicating that gas is occupying the space inside the pouch bags. Since the cells were de-gassed after formation, the gases present excludes the gas generated during formation and represent only the accumulation of gas over the course of the cycle life test. The absence of gas during room temperature cycling indicates that the gas evolution is thermally activated. \revc{iii}{More experimental work is needed to determine the origin of gas evolution over cycle life due to fast formation. We provide speculation into the origin of gas evolution as part of the Supplementary Materials.}

Our study found no correlations between $\res$ and the gas amount as measured by pouch thickness. We attribute the lack of correlation primarily to the fact that the cell age was not well-controlled at the time of the pouch thickness measurement: cells stopped cycling anywhere between 0\% and 50\% capacity retention. Future studies will be needed to confirm the relationship between $\res$ and gas build-up.

\subsubsection{Aging Variability}

Adopting a new formation protocol in practice also requires a close understanding of the impact of new formation protocols on cell aging variability over life. Cells with non-uniform capacity fade could take longer to balance in a pack and cause a deterioration of energy available at the pack-level \supercite{Liu2019b}. Pack imbalance issues could lead to consumer products being retired earlier, compounding existing battery recycling challenges \supercite{Harper2019}. Non-uniform cell degradation will also be more difficult to re-purpose into new modules \supercite{Engel2019, Rumpf2018, Rasheed2020}, creating higher barriers for pack reuse.

\rev{The inter-quartile range (IQR) of cycle lives for fast formation cells was higher than that of baseline formation cells (Figures \ref{fig:aging-test-result-combined}b,d). The same result held under both room temperature and 45\textdegree C cycling, as well as across different end-of-life definitions (Figure \ref{fig:aging-variability})}, suggesting that fast formation increased aging variability. A key question is whether fast formation created more heterogeneous aging behavior which caused the higher variability in aging, or if the higher variability is due to the cells lasting longer. To answer this question, we employed the modified signed-likelihood ratio test \supercite{Krishnamoorthy2014} to check for equality of the coefficients of variation, defined as the ratio between the standard deviation and the mean cycle life. The resulting $p$-values were greater than 0.05 in all cases. Therefore, with the available data, \rev{we} cannot conclude that fast formation increased the variation in aging beyond the effect of improving cycle life. While a relationship between formation protocol and aging variability may still generally exist, this difference \rev{could not} be determined with our sample \rev{sizes ($n=10$ cells per group)}. This result motivates the continued usage of larger samples sizes for future studies on the impact of formation protocol on aging variability.

\section{Conclusion}

In this work, we demonstrated that low-SOC resistance ($\res$) correlates to cycle life across two different battery formation protocols. As a predictive feature, $\res$ provided higher prediction accuracy compared to conventional measures of formation quality such as Coulombic efficiency as well as state-of-the art predictive features based on changes in discharge voltage curves. $\res$ is measurable at the end of the manufacturing line using ordinary battery test equipment and can be measured within seconds. Changes in $\res$ are attributed to differences in the amount of lithium consumed to the SEI during formation, where a decrease in $\res$ indicates that more lithium is consumed. \rev{The sensitivity of $\res$ to lithium consumption is due to the presence of kinetic limitations in the positive electrode causing the total cell resistance to increase at low SOCs. For this reason, $\res$ provides a particularly strong signature in nickel-rich positive electrode systems where kinetic hindrance plays a strong role in limiting lithium transport towards high states of lithiation.} Since the physical interpretation of $\res$ is general, $\res$ can be broadly applicable for screening any manufacturing process that impact the amount of lithium consumed during battery formation. As a whole, our results hold promise for decreasing lithium-ion battery formation time and cost while improving lifetime, as well as identifying rapid diagnostic signals for screening new manufacturing processes and cell designs based on cycle life.

\section{Experimental Procedures}

\subsection{Resource Availability}

\subsubsection{Lead Contact}

Further information and requests for resources and materials should be directed to and will be fulfilled by Andrew Weng (\href{mailto:asweng@umich.edu}{asweng@umich.edu}).

\subsubsection{Materials Availability}

All materials are commercially available, with the exception of the CMC binder material used in the anode formulation which is proprietary. 

\subsubsection{Data and Code Availability}

Data and code used in this study are available at \href{https://doi.org/10.7302/pa3f-4w30}{https://doi.org/10.7302/pa3f-4w30}. The source code can be accessed at \href{https://doi.org/10.5281/zenodo.5525258
}{https://doi.org/10.5281/zenodo.5525258
}.

\subsection{Cell Build Process}

The cathode was comprised of 94:3:3 NMC 111 (TODA North America), C65 conductive additive (Timcal), and PVDF (Kureha 7208). The slurry was mixed in a step-wise manner, starting with a dry solids homogenization, wetting with NMP, and then addition of the PVDF resin. The slurry was allowed to mix overnight under static vacuum with agitation from both the double helix blades (30 rpm) and the high-speed disperser blade (1600 rpm). The final slurry was gravity filtered through a 125 $\mu$m paint filter before coating on a roll-to-roll coating machine (Creative \& Innovative Systems). The electrode was coated using the reverse comma method at 2 m/min. The final double-sided loading was 34.45 mg/cm$^2$. 

The anode was comprised of 97:0:(1.5/1.5) graphite (Hitachi MAG-E3), no conductive additive, and equal parts CMC (proprietary) and SBR (Zeon BM-451B). The graphite and pre-dispersed CMC were mixed prior to further let-down with de-ionized water and overnight dispersion under static vacuum and double helix blade agitation (40 rpm). Prior to coating, the SBR was added and mixed in with helical blade agitation for 15 minutes under active vacuum. The final slurry was gravity filtered through a 125 $\mu$m paint filter before coating on a roll-to-roll coating machine (Creative \& Innovative Systems). The electrode was coated using the reverse comma technique at 1.5 m/min. The final double-sided loading was 15.7 mg/cm$^2$.

Both anode and cathode were calendared at room temperature to approximately 30\% porosity prior to being transferred to a -40°C dew point dry room for final cell assembly and electrolyte filling. The cells, comprising 7 cathodes and 8 anodes, were z-fold stacked, ultrasonically welded, and sealed into formed pouch material (mPlus). The assembled cells were placed in a vacuum oven at 50°C overnight to fully dry prior to electrolyte addition. Approximately 10.5 g of electrolyte (1.0M LiPF$_6$ in 3:7 EC:EMC v/v + 2wt\% VC from Soulbrain) was manually added to each cell prior to the initial vacuum seal (50 Torr, 5 sec). The total mass of all components of the battery is 56.6 $\pm$ 0.3g.

The now-wetted cells were each placed under compression between fiberglass plates held in place using spring-loaded bolts. The compression fixtures are designed to allow the gas pouch to protrude and freely expand in the event of gas generation during formation. All cells were allowed to fully wet for 24 hours prior to beginning the formation process.

After formation, the cells were removed from the pressure fixtures, returned to the -40°C dew point dry room and degassed. The degassing process was completed in an mPlus degassing machine, automatically piercing the gas pouch, drawing out any generated gas during the final vacuum seal (50 Torr, 5 sec) and then placing the final seal on the cell. Cells are manually trimmed to their final dimensions before being returned to their pressure fixtures.

The pouch cell architecture is summarized in Figure \ref{fig:battery-architecture}.

\subsection{Formation Protocols}

Figure \ref{fig:design-of-experiments}b describes the two different formation protocols used in this study. The fast formation protocol borrows from the `Ultra-fast formation protocol' reported in An et al. \supercite{An2017} and Wood et al. \supercite{Wood2019} In this protocol, the cell is brought to 3.9V using a 1C (2.36Ah) charge, followed by five consecutive charge-discharge cycles between 3.9V and 4.2V at C/5, and finally ending on a 1C discharge to 2.5V. Each charge step terminates on a CV hold until the current falls below C/100. A C/10 charge and C/10 discharge cycle was appended at the end of the test to measure the post-formation cell discharge capacity. A 6-hour step was included in between the C/10 charge-discharge steps to monitor the voltage decay. The formation sequence takes 14 hours to complete after excluding time taken for diagnostic steps. 

A baseline formation protocol was also implemented which serves as the control for comparing against the performance of fast formation. This protocol consists of three consecutive C/10 charge-discharge cycles between 3.0V and 4.2V. A 6-hour rest was also added between the final C/10 charge-discharge step to monitor the voltage decay signal. The total formation time \revb{6}{was 56} hours after excluding the diagnostic steps. Formation was conducted at room temperature for all cells and across both formation protocols.

All formation cycling was conducted on a Maccor Series 4000 cycler (0-5V, 30$\mu$A - 1A, auto-ranging). Following formation, one cell (\#9) was excluded from this study due to tab weld issues. Consequently, the sample count for the `baseline formation, 45\textdegree C' cycling group was decreased to 9. The remaining groups had sample counts of 10.

The mean cell energy measured at a 1C discharge rate from 4.2V to 3.0V at room temperature is 8.13 Wh. Full cell level volumetric stack energy density is estimated to be 365 Wh/L based on a volume of 69mm x 101mm x 71 mm x 3.2 mm, and the gravimetric stack energy density is estimated to be 144 Wh/kg based on a total cell mass of 56.6g.

\subsection{Cycle Life Testing}

Following completion of formation cycling, cells were placed in spring-loaded compression fixtures to maintain a uniform stack pressure. Half of the cells from each formation protocol were placed in a thermal chamber (Espec) with a measured temperature of $44.2 \pm 0.1$°C. The remaining cells were left at room temperature and were exposed to varying temperatures throughout the day ($24.5 \pm 0.6$°C). Long-term cycle life testing was conducted on a Maccor Series 4000 cycler (0-5V, 10A, auto-ranging). The cycle life test protocol was identical for all cells and consisted of 1C (2.37A), CC charge to 4.2V with a CV hold to 10mA and 1C discharges to 3.0V. At every 50 to 100 cycles, the test was interrupted so that a Reference Performance Test (RPT) could be performed \supercite{Dubarry2020}. The RPT consists of a C/3 charge-discharge cycle, a C/20 charge-discharge cycle, followed by the Hybrid Pulse Power Characterization (HPPC) protocol \supercite{Christoph2015}. The HPPC test is used to extract 10-second discharge resistance ($R_{10s}$) as a function of SOC (Figure \ref{fig:hppc-example}). Every cell was cycled until the discharge capacity was less than 1.18 Ah, corresponding to less than 50\% capacity remaining. The total test time varied between 3 to 4 months and the total cycles achieved ranged between 400 and 600 cycles. Cycle test metrics are shown in Figures \ref{fig:aging-test-average-voltage}, \ref{fig:aging-test-ce}, \ref{fig:aging-test-voltage-efficiency}, \ref{fig:aging-test-discharge-energy}.

\subsection{Statistical Significance Testing}

The standard Student's \textit{t}-test for two samples was used throughout this study to check if differences in measured outcomes between the two different formation protocols were statistically significant. The $p$-value was used to quantify the level of marginal significance within the statistical hypothesis test and represents the probability that the null hypothesis is true. A $p$-value less than 0.05 was used to reject the null hypothesis that the population means are equal. All measured outcomes were assumed to be normally distributed. Box-and-whisker plots are also used throughout the paper to summarize distributions of outcomes. Boxes denote the inter-quartile range (IQR) and whiskers show the minimum and maximum values in the set. No outlier detection methods are employed here due to the small sample sizes ($n<10$). \rev{Finally, the Pearson correlation coefficient, $-1 \leq \rho \leq 1$, was used to determine the significance of correlations between initial state variables and lifetime output variables. $|\rho| > 0.5$ is taken to indicate a statistically meaningful correlation.}

\subsection{Predictive Lifetime Model}

Due to the small number of data points available, the model prediction results are sensitive to which cells are chosen for validation. Therefore, we used nested cross-validation \supercite{Krstajic2014} to evaluate the regularized linear regression model on all the data without over-fitting. The nested cross-validation algorithm is as follows: first, we separated the data into 20\% `validation' and 80\% `train/test'. Then, we performed four-fold cross-validation on the `train/test' data to find the optimal regularization strength for Ridge regression, $\alpha^*$, using grid search. Finally, we trained the Ridge regression algorithm with regularization strength $\alpha^*$, using all of the train/test data, and evaluated the error on the validation data. We repeated this process for 1000 random train-test/validation splits and reported the mean and standard deviation of the mean percent error for each run,

\begin{equation}
    \text{MPE} [\%] = \frac{1}{N}\sum_{k=1}^{N}\frac{y^\text{pred}_k-y^\text{true}_k}{y^\text{true}_k}.
\end{equation}

Each run can select a different optimal regularization strength $\alpha^*$.

\subsection{Electrode Stoichiometry Model}

\rev{To construct the stoichiometry model shown in Figure \ref{fig:initial-cell-state-mechanism}a, a full cell near-equilibrium potential curve was first extracted using the C/20 charge cycle from the reference performance test (RPT)}. A randomly selected cell from the 45\textdegree C cycling group was selected for this data extraction. Positive and negative electrode near-equilibrium potential curves were adapted from Mohtat et al. \supercite{Mohtat2020}. The electrode-specific utilization windows are determined by fitting the positive and negative electrode potential curves to match the full cell curve by solving a least squares optimization problem as outlined in Lee et al. \supercite{Lee2020a}. The resulting positive and negative electrode alignment minimized the squared error of the modeled versus the measured full cell voltage. The fast formation curve equilibrium potential curve was constructed by shifting the positive electrode curve horizontally and re-computing the full cell voltage curve.

\rev{The full cell resistance curves in Figure \ref{fig:initial-cell-state-mechanism}(c) sourced data from the HPPC sequence as part of the same RPT used to obtain the equilibrium potential curve shown in Figure \ref{fig:initial-cell-state-mechanism}(a). A cubic spline fit was used to create smooth resistance curves. (A model generated using a linear fit is provided in Figure \ref{fig:initial-cell-state-mechanism-precise}). To break down the resistance contribution into `positive resistance' and `negative + other resistances', a baseline reference resistance $R_\mathrm{ref}$ was first defined as the minimum measured full cell resistance below 1Ah. The `negative + other resistances` was then assigned a value of ($1 - f_\mathrm{pos}) \cdot R_\mathrm{ref}$. The remaining resistance was then assigned to the positive electrode. $f_\mathrm{pos}$ was set to 0.7 to model a generic NMC/graphite system \supercite{Wu2000, An2017, Abraham2005}.}

\subsection{Voltage Fitting Algorithm}

\revb{7}{Methods for estimating electrode-specific state-of-health metrics using half-cell reference curves has been previously reported \supercite{Bloom2005, Dubarry2017d, Dubarry2012}. Here, we applied an automated voltage fitting approach based on work by Lee et al. \supercite{Lee2020a} to extract electrode capacity losses LAM$_\mathrm{PE}$ and LAM$_\mathrm{NE}$, as well as lithium inventory loss (LLI) for both fresh and aged cells. The input data consisted of C/20 charge curves measured at each RPT. An example set of C/20 charge curves over age is shown in Figure \ref{fig:dvdq_aging_example}.}

\revb{7}{The method to extract electrode-specific state of health indicators LLI, LAM$_\mathrm{PE}$ and LAM$_\mathrm{NE}$ is adapted from \supercite{Lee2020a}. Positive and negative near-equilibrium potential curves were adapted from Mohtat et al. \supercite{Mohtat2020}. The positive and negative electrode potential curves are obtained during delithiation and lithiation, respectively, which correspond to charging in the full cell. The curves were obtained at the C/20 rate and serve as proxies for the true equilibrium potential curves. The same equilibrium potential curves were used to model data at both test temperatures.}

\revb{7}{To prevent over-fitting, the positive electrode stoichiometry at 100\% SOC ($y_{100}$) was fixed to 0.03 at every instance for this analysis. Fixing this value yielded smoother and more physical degradation trajectories over cycle life. Figure \ref{fig:dvdq_fitting_example} shows an example of voltage fitting results for a single cell. The degradation metrics, including Loss of Lithium Inventory (LLI) and loss of active material (LAM) were computed in the usual manner (see Lee et al. \supercite{Lee2020a} for more details).}

\subsection{Hybrid Power Pulse Characterization of Half Cells}

\revb{10}{Coin cell half cells were built with LFP, NMC111, and graphite as the working electrode and lithium metal as the counter electrode. The NMC material used were identical to that used in the pouch cells for the formation experiments (TODA North America). The graphite material used differed from the ones used in the pouch cells. The coin cell construction consisted of 2032 form factor components including a wavespring and spacer. The electrolyte used was 1M LiPF$_6$ with EC/EMC. The lithium counter electrode was 16 mm in diameter, the separator was 19 mm in diameter, and the working electrodes were 14mm in diameter. Working electrodes were measured to be approximately 60 $\mu$m thick and the lithium counter-electrodes were approximately 750 $\mu$m thick. Working electrodes were single side coated. Calculated theoretical capacities for the NMC111, LFP, and graphite cells were 2.0 mAh, 2.9 mAh, and 4.6 mAh, respectively. }

\revb{10}{The Hybrid Pulse Power Characterization (HPPC) protocol was adapted for the coin cells. Potential ranges were modified depending on the working electrode. The currents used in the pulses were also scaled down to 0.4 mA for all cells (Figure \ref{fig:coin_cell_hppc_example}). The measured resistance drop includes a large Ohmic contribution due to the presence of the lithium metal counter electrode. However, since this counter electrode was present in all cells, differences in measured, SOC-dependent resistances between the different cells remain meaningful. All coin cells were pre-conditioned using at least three slow charge-discharge cycles prior to starting the HPPC sequence.}

\section*{Acknowledgements}

This work was supported by the National Science Foundation, Grant Number 176224, and the University of Michigan Battery Laboratory. The authors acknowledge Voltaiq \href{https://voltaiq.com}{(www.voltaiq.com)} for providing software that made it easy to remotely monitor cycle life test data during the COVID-19 pandemic in 2020. The authors also thank Joseph Gallegos for building the coin cells for the study.

\section*{Author Contributions}

Conceptualization: A.W. and A.S.; Methodology: A.W., P.M., G.L., P.M.A., V.S., and S.L.; Investigation: A.W. and G.L.; Data Curation: A.W; Software: A.W; Visualization: A.W; Formal Analysis: V.S; Writing - Original Draft: A.W; Writing - Review \& Editing: A.W., P.M.A, A.S. Funding Acquisition: A.S.

\section*{Declaration of Interests}

Andrew Weng and Peter M. Attia are employees of Tesla, Inc. A patent application relating to this work has been filed. The authors declare no other competing interests.

\section*{Glossary of Terms}

\begin{tabular}{ l l }

 CC & constant current \\ 
 CE & coulombic efficiency \\
 CE$_\mathrm{f}$ & formation coulombic efficiency \\
 CMC & carbon methyl cellulose \\
 CV & constant voltage \\
 EC & ethylene carbonate \\
 EMC & ethyl methyl carbonate \\
 HPPC & hybrid pulse power characterization \\
 IQR & inter-quartile range \\
 $\lamne$ & loss of active material in the negative electrode \\
 $\lampe$ & loss of active material in the positive electrode \\
 LiPF$_6$ & lithium hexafluorophosphate \\
 $\lli$ & loss of lithium inventory \\
 NMC & Nickel manganese cobalt \\
 NMP & n-methyl-2-pyrrolidone \\
 PVDF & polyvinylidene fluoride \\
 $\Qc$ & first cycle charge capacity \\
 $\Qd$ & post-formation C/10 discharge capacity \\
 $\QLLI$ & capacity of lithium inventory lost during formation = $\Qc - \Qd$ \\
 R$_{10\mathrm{s}}$ & 10-second discharge resistance \\
 $\res$ & low-SOC resistance \\
 RPT & reference performance test  \\ 
 SBR & styrene butadiene rubber \\
 SEI & solid electrolyte interphase \\
 SOC & state of charge \\
 VC & vinylene carbonate \\
 $\ymax$ & maximum positive electrode stoichiometry \\
 
\end{tabular}

%%% REFERENCES %%%

\vspace{12pt}

\printbibliography

\clearpage

%%% APPENDICES %%%
\beginsupplement

\section*{Supplemental Material}

\begin{refsection}

\begin{figure}[ht]
\centering
\includegraphics[width=1\linewidth]{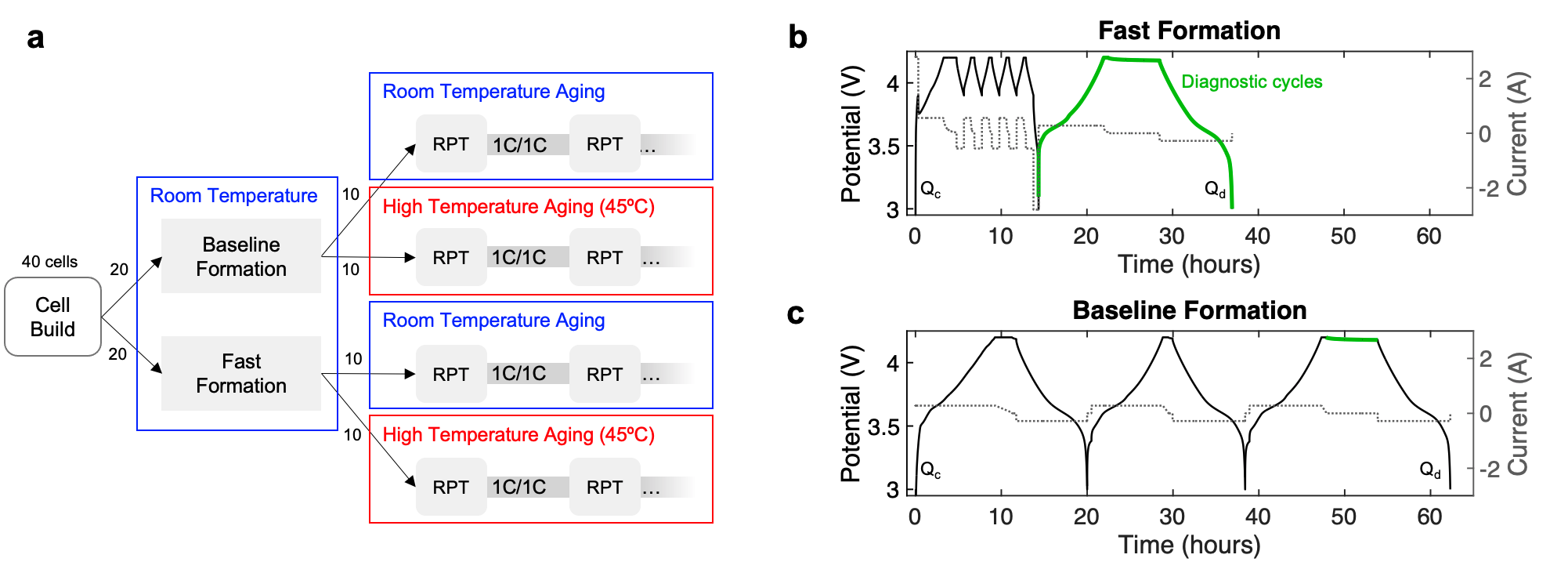}
\caption{\textbf{Experimental Design} \\ (a) Distribution of cells across two formation protocols and two aging temperatures. The aging test consists of 1C charge/discharge cycles between 3.0V and 4.2V, with reference performance tests (RPTs) inserted periodically into the test. (b, c) Voltage and current versus time profiles for (b) fast formation and (c) baseline formation. Green lines show diagnostic steps inserted into the formation protocols for the purpose of this study only. The diagnostic cycles are not considered to be part of the formation protocol. Ignoring the diagnostic cycles, the fast formation protocol lasts 14 hours and the baseline formation protocol lasts 56 hours.}
\label{fig:design-of-experiments}
\end{figure}

\clearpage

\begin{figure*}[ht]
\centering
\includegraphics[width=1\linewidth]{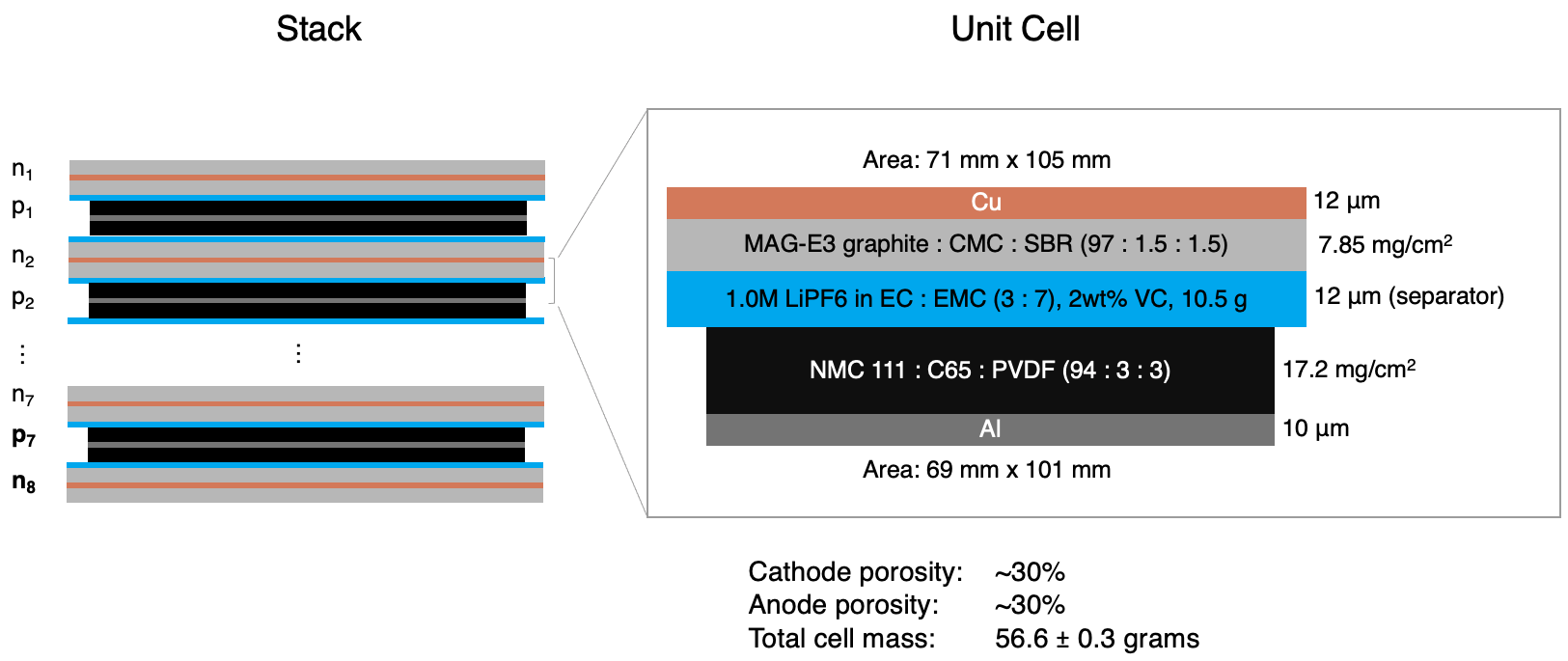}
\caption{\textbf{Pouch Cell Architecture} \\ Left: cell stack definition. Right: unit cell definition. The same cell architecture is used for all cells in this study.}
\label{fig:battery-architecture}
\end{figure*}

\begin{figure*}[ht]
\centering
\includegraphics[width=0.7\linewidth]{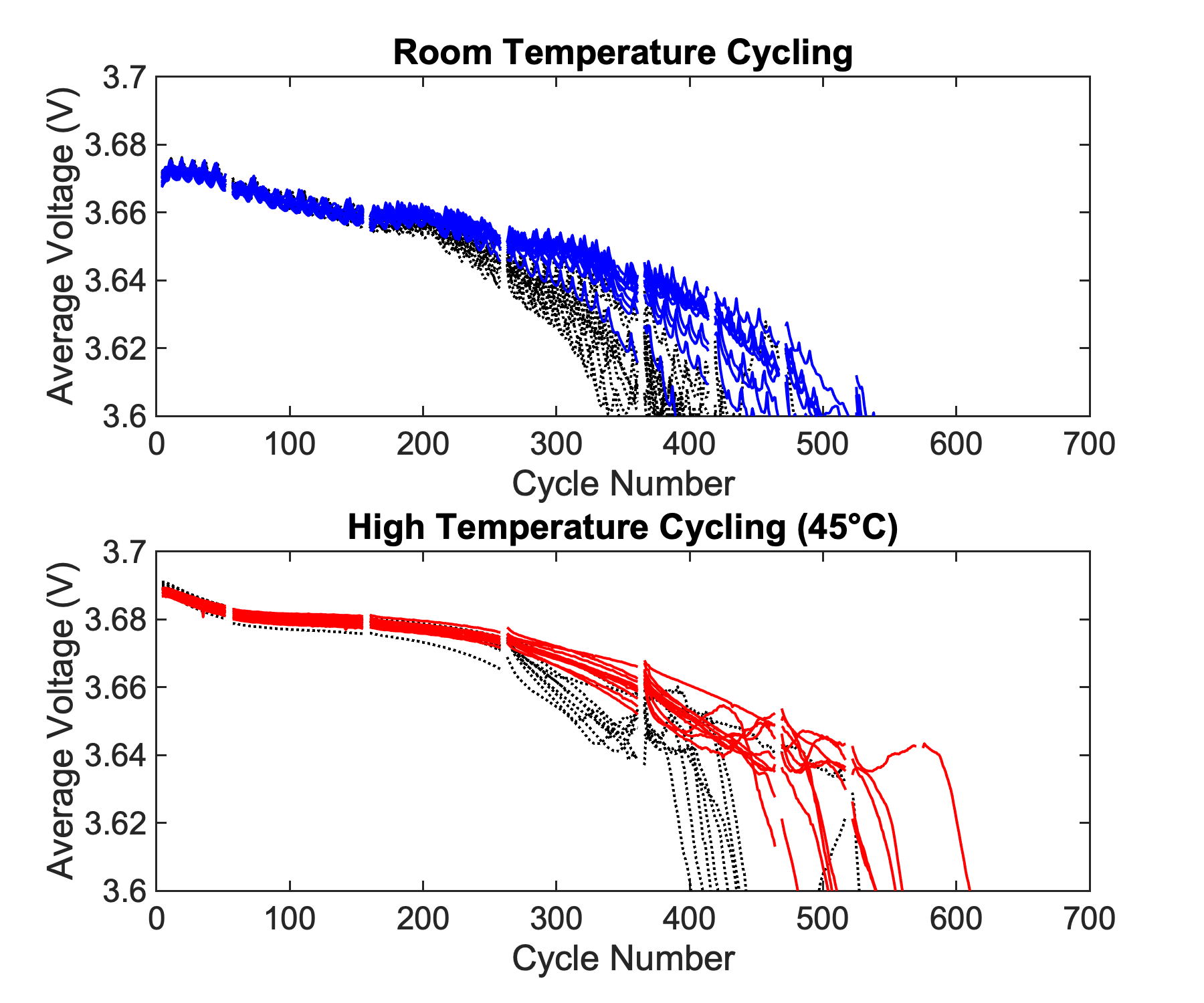}
\caption{\textbf{Mean Discharge Voltage Over Cycle Number}}
\label{fig:aging-test-average-voltage}
\end{figure*}

\begin{figure*}[ht]
\centering
\includegraphics[width=0.7\linewidth]{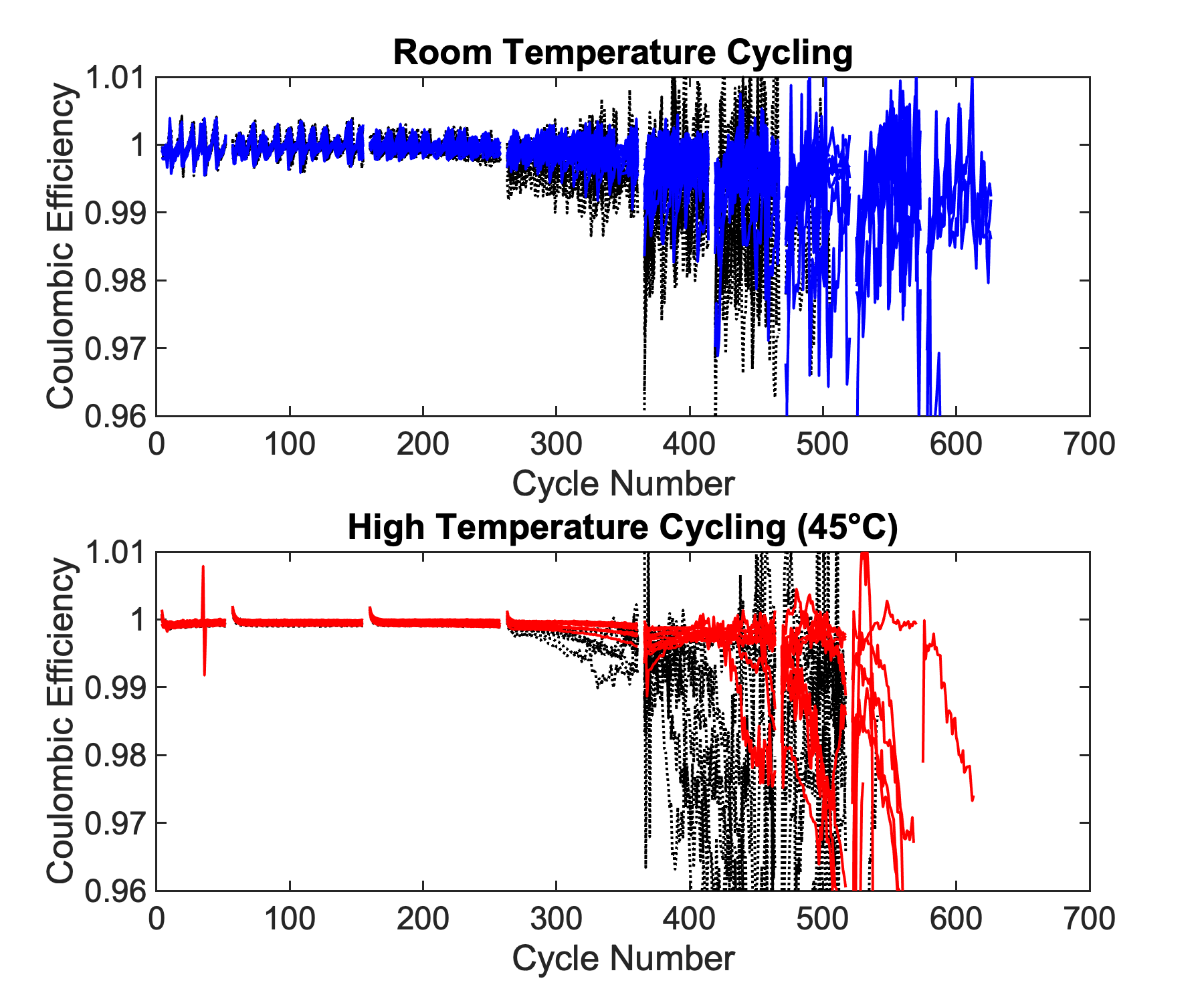}
\caption{\textbf{Coulombic Efficiency Over Cycle Number}}
\label{fig:aging-test-ce}
\end{figure*}

\begin{figure*}[ht]
\centering
\includegraphics[width=0.7\linewidth]{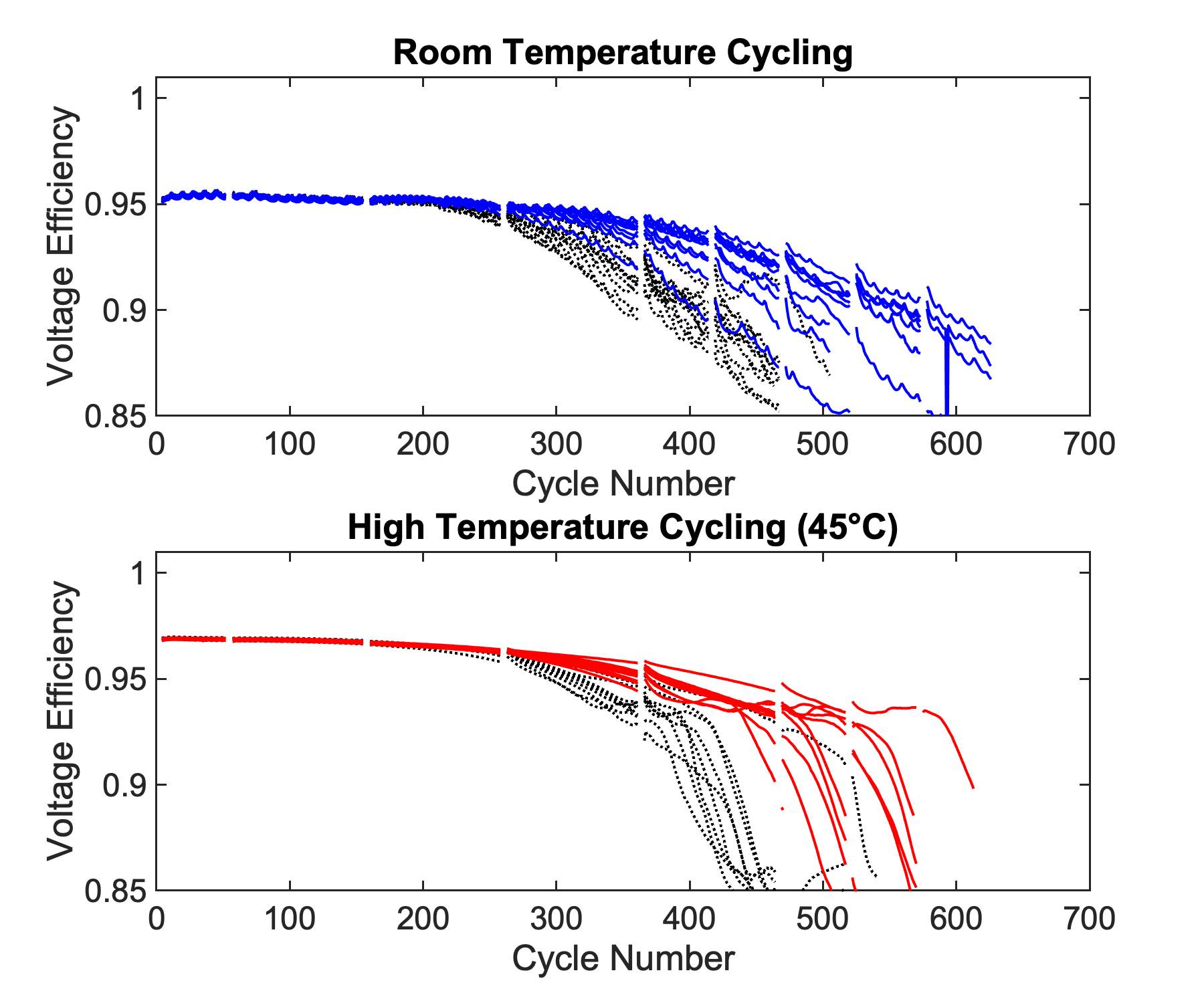}
\caption{\textbf{Voltage Efficiency Over Cycle Number}}
\label{fig:aging-test-voltage-efficiency}
\end{figure*}

\begin{figure*}[ht]
\centering
\includegraphics[width=0.7\linewidth]{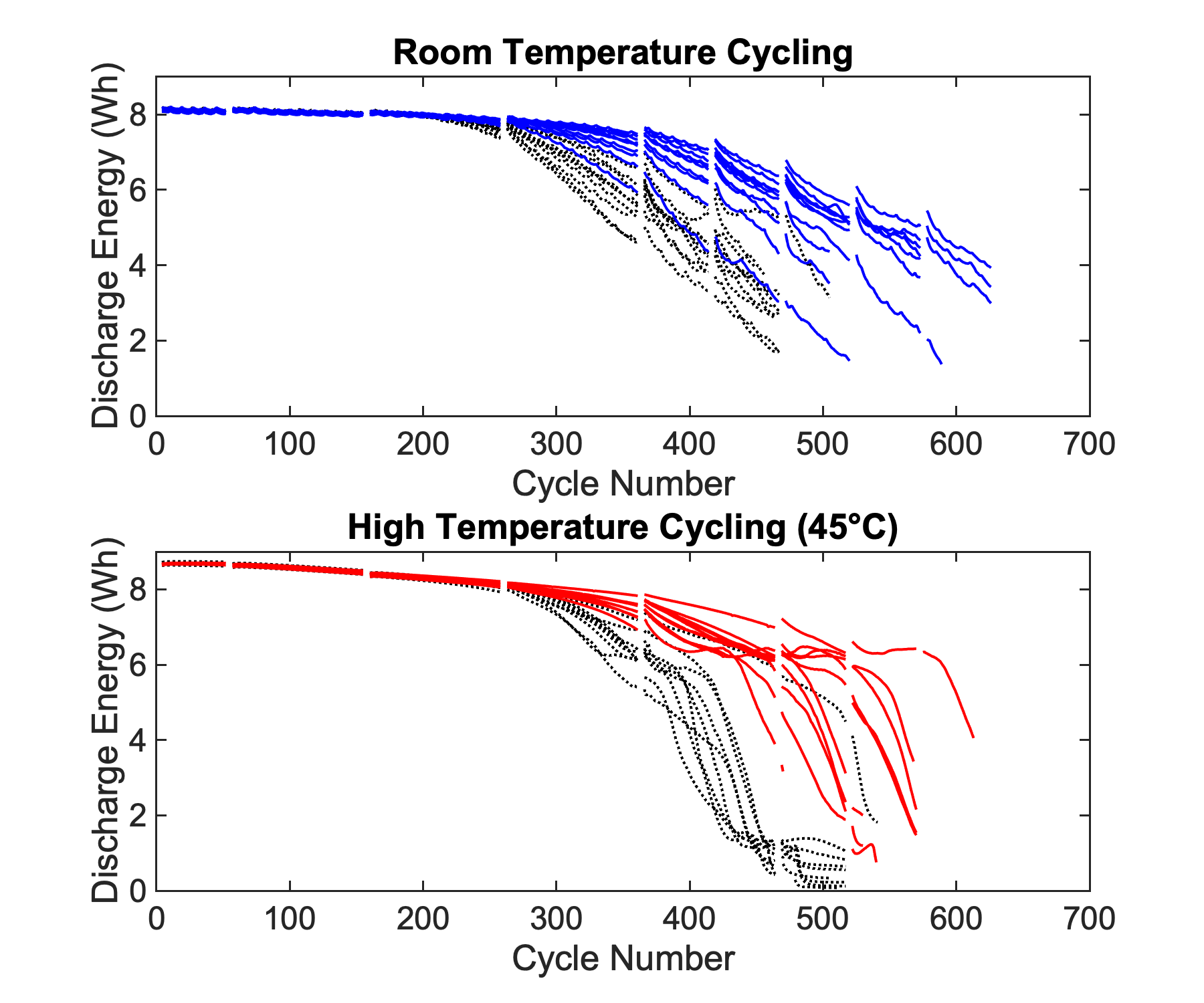}
\caption{\textbf{Discharge Energy Over Cycle Number}}
\label{fig:aging-test-discharge-energy}
\end{figure*}

\begin{figure*}[ht]
\centering
\includegraphics[width=1.0\linewidth]{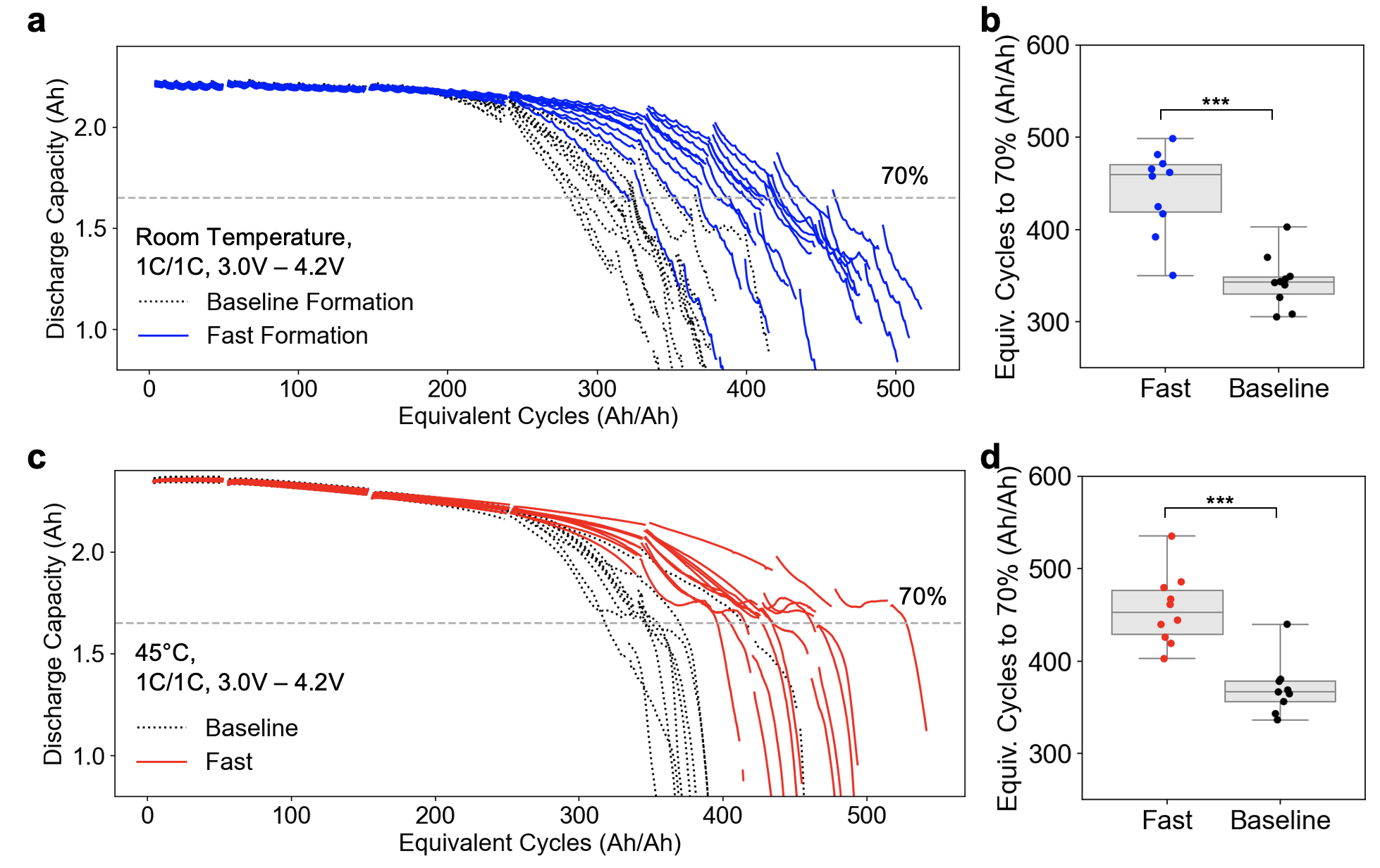}
\caption{\textbf{Cycle Life Test Results Plotted With Equivalent Cycles (Ah/Ah) as the Throughput Basis} \\ The total throughput is calculated by taking the cumulative discharge capacity and dividing by the cell nominal capacity (2.36 Ah). (a,c) Discharge capacity for the room temperature (a) and 45\textdegree C (c) tests. (b,d) Distribution of outcomes at the end of life for the room temperature (b) and 45\textdegree C (d) tests. End of life is defined as equivalent cycles to reach 70\% of initial capacity.}
\label{fig:fig_aging_test_result_combined_ahah}
\end{figure*}

\begin{figure*}[ht]
\centering
\includegraphics[width=1\linewidth]{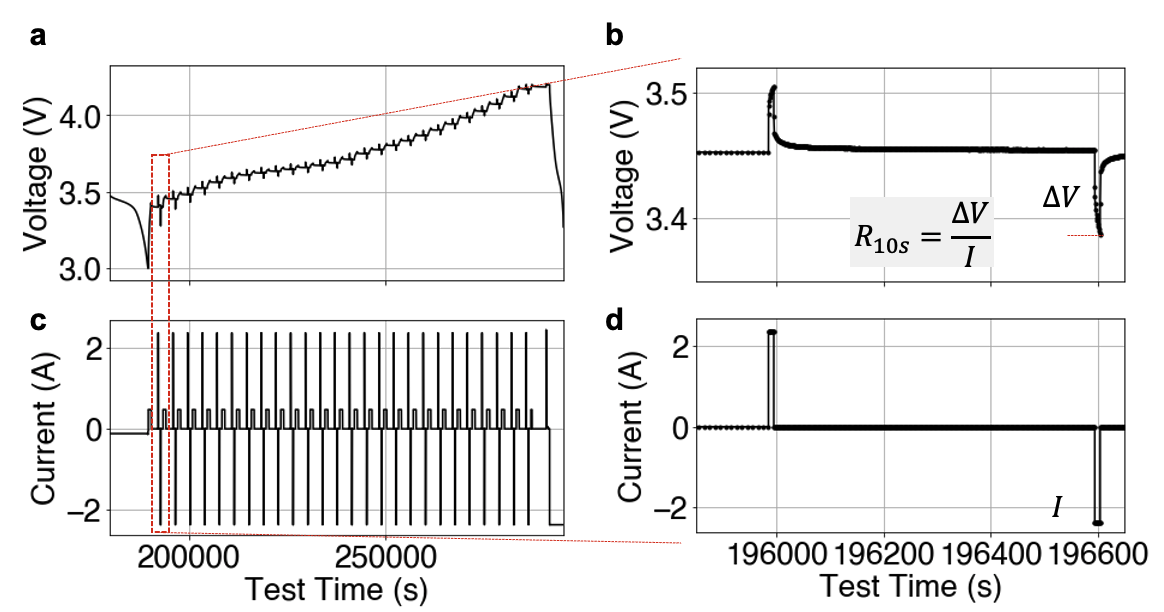}
\caption{\textbf{Example Hybrid Power Pulse Characterization (HPPC) Sequence} \\ HPPC is used to extract the 1C, 10-second charge and discharge resistances across different SOCs. (a,c) current and voltage traces from the HPPC protocol. (b,d) magnified view of the first charge and discharge pulse from the pulse sequence, which occurs at $\sim$4\% SOC. The HPPC pulses are included as part of every reference performance test (RPT).} 
\label{fig:hppc-example}
\end{figure*}

\begin{table}[h!]
  \begin{center}
    
    \begin{tabular}{r|l|c} 
      \toprule
       & Dimensions   & 72 mm x 110 mm \\
       & Stack (pos/neg)  & 7/8 \\
      \midrule
      \textbf{Positive} & Chemistry & NMC111 \\
                        & Composition & NMC111 : C65 : PVDF \\
                        & Ratio       & 94 : 3 : 3 \\
                        & Loading (double-sided)  & 34.45 mg/cm$^2$ \\
                        & Collector thickness & 12 um \\
      \midrule
      \textbf{Negative} & Chemistry & Graphite \\ 
                        & Composition & MAG-E3 : CMC : SBR \\
                        & Ratio       & 97 : 1.5 : 1.5 \\
                        & Loading (double-sided) & 15.7 mg/cm$^2$ \\
                        & Collector thickness & 10 um \\
      \midrule
      \textbf{Electrolyte} & Salt & 1.0 M LiPF$_6$ \\
                           & Solvent & EC : EMC \\
                           &         & (3 : 7) \\
                           & Additive & 2 wt\% VC, 4g/Ah \\
                           & Mass & 10.55 g \\
      \midrule
      \textbf{Separator} & Supplier & Entek \\
                         & Thickness & 12 um  \\
      \bottomrule
    \end{tabular}
    \caption{\textbf{Cell Design Parameters}}
    \label{tab:cell-design}
  \end{center}
\end{table}

\begin{figure*}[ht]
\centering
\includegraphics[width=1\linewidth]{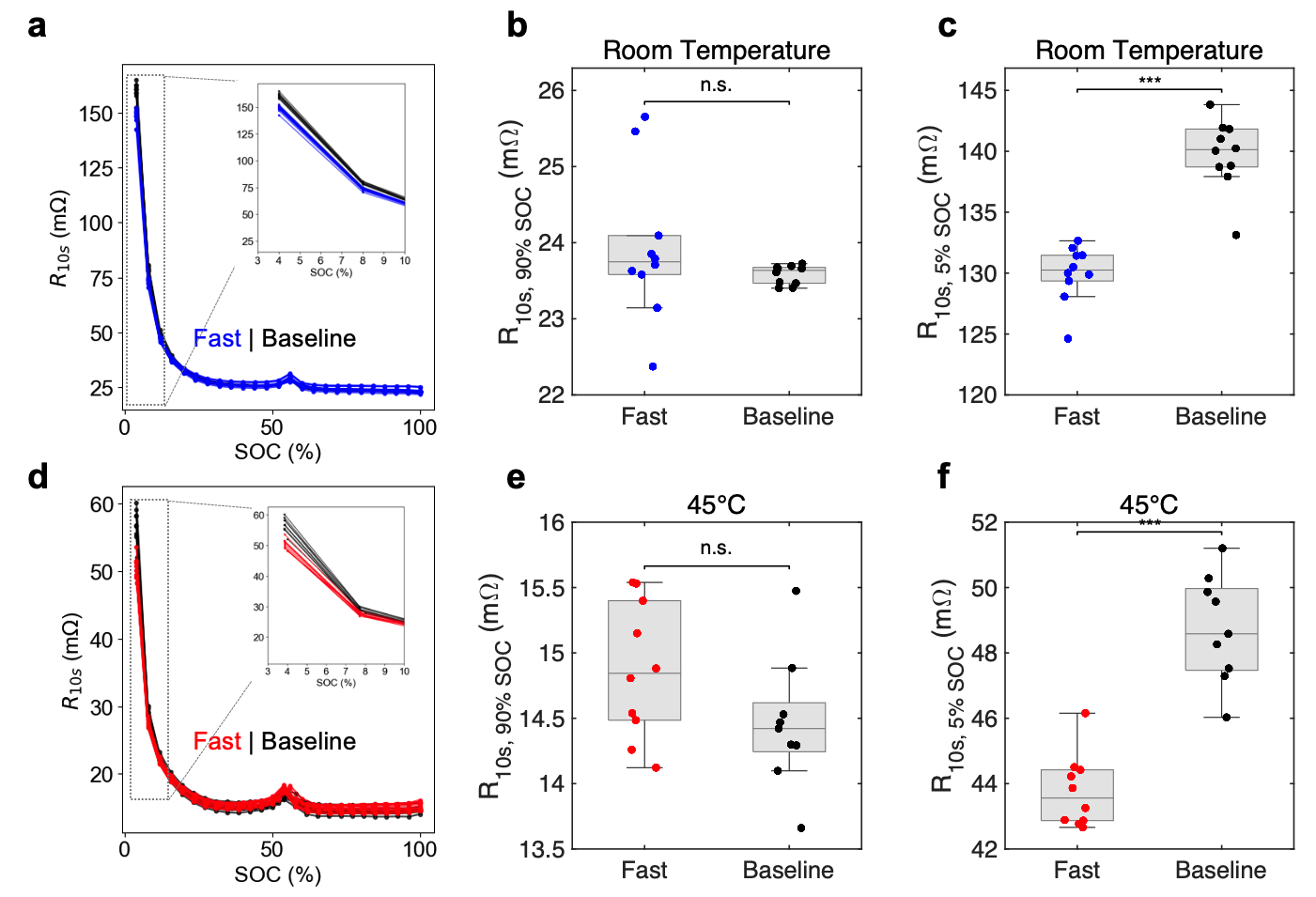}
\caption{\textbf{Initial Cell Resistance Comparison} \\ Distribution of resistance measured at (a-c) room temperature and (d-f) 45\textdegree C. Resistance is calculated from the Hybrid Power Pulse Characterization (HPPC) test from the first RPT. `n.s.': not statistically significant. `***': statistically significant with $p$-value < 0.001.}
\label{fig:initial-dcr-both-temps}
\end{figure*}

\begin{figure*}[ht]
\centering
\includegraphics[width=1\linewidth]{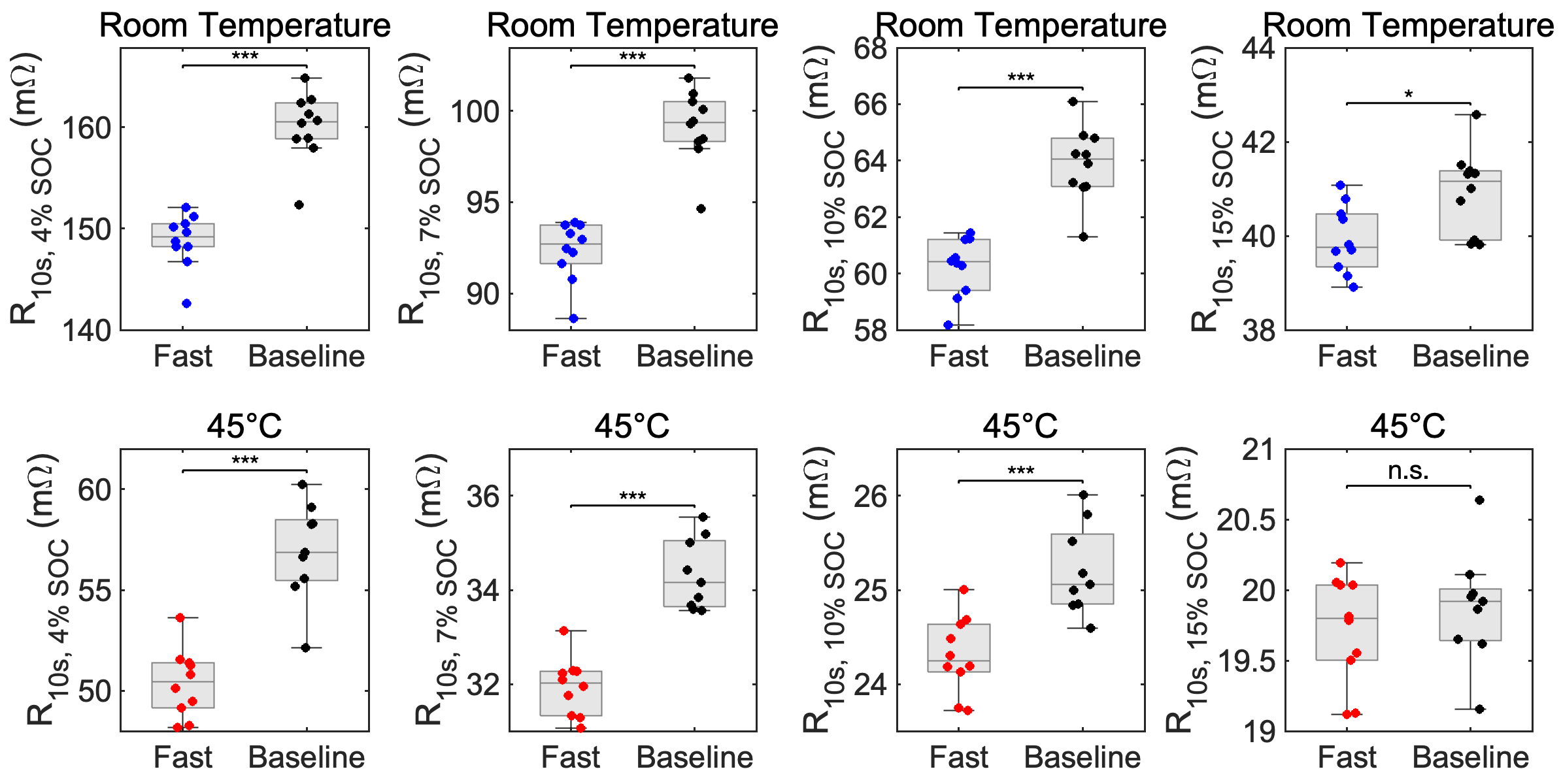}
\caption{\textbf{Effect of Measurement SOC on the Cell Resistance Measured from HPPC} \\ Top row: room temperature. Bottom row: 45\textdegree C. Values are obtained from the initial Reference Performance Test (RPT) run prior to the start of cycling. `n.s.': not statistically significant. `***': statistically significant with $p$-value < 0.001. `*': statistically significant with $p$-value < 0.05.}
\label{fig:initial-dcr-effect-of-soc}
\end{figure*}

\begin{figure*}[ht]
\centering
\includegraphics[width=0.75\linewidth]{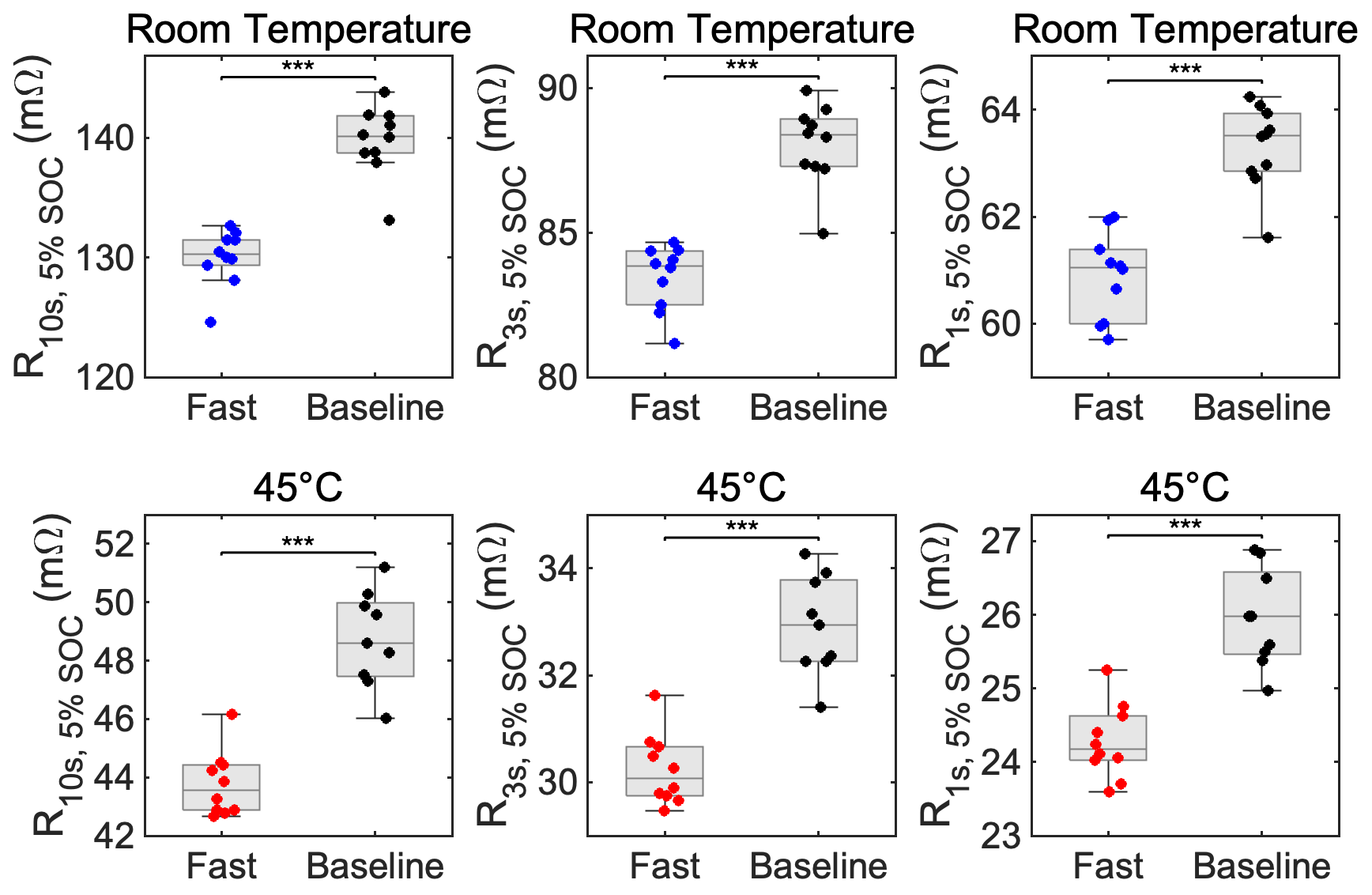}
\caption{\textbf{Effect of Pulse Duration on the Cell Resistance Measured From HPPC} \\ Top row: room temperature. Bottom row: 45\textdegree C. Values are obtained from the initial Reference Performance Test (RPT) run prior to the start of cycling. `***': statistically significant with $p$-value < 0.001.}
\label{fig:initial-dcr-dffect-of-duration}
\end{figure*}

\begin{figure*}[ht]
\centering
\includegraphics[width=0.7\linewidth]{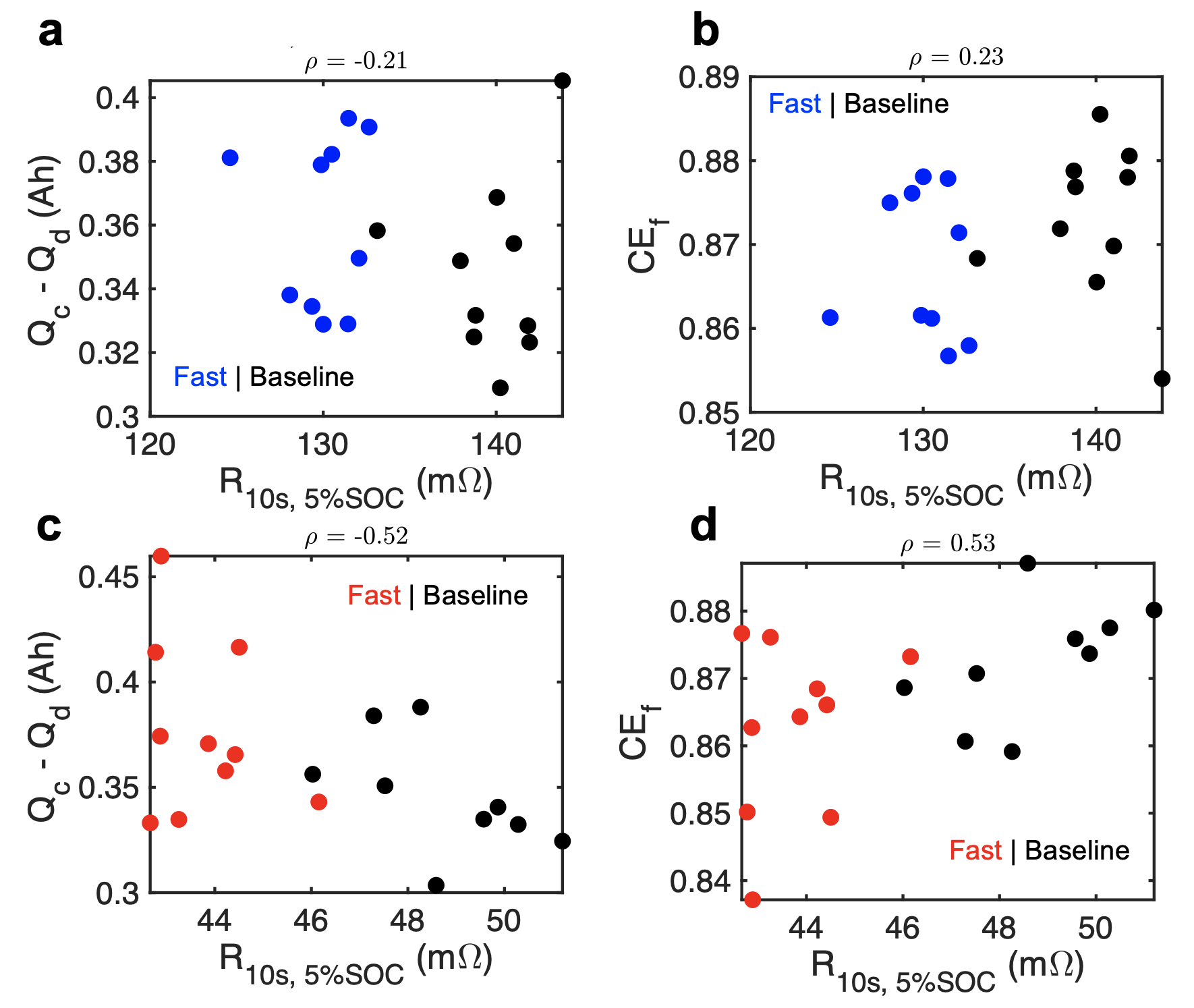}
\caption{\textbf{Correlation Between $\res$ and Conventional Metrics of Lithium Consumption During Formation} \\ Conventional metrics include lithium consumed during formation ($\QLLI = \Qc - \Qd$) and Coulombic efficiency ($\cef$). (a,b) Correlation with $\res$ measured at room temperature. (c,d) Correlation with $\res$ measured at 45\textdegree C. $\QLLI = \Qc - \Qd$ and $\cef$ are always measured at room temperature since these signals are derived directly from the formation protocol which ran at room temperature. $\res$ was measured at the temperature of the cycling test.}
\label{fig:initial-cell-state-correlations}
\end{figure*}

\begin{figure*}[ht]
\centering
\includegraphics[width=1\linewidth]{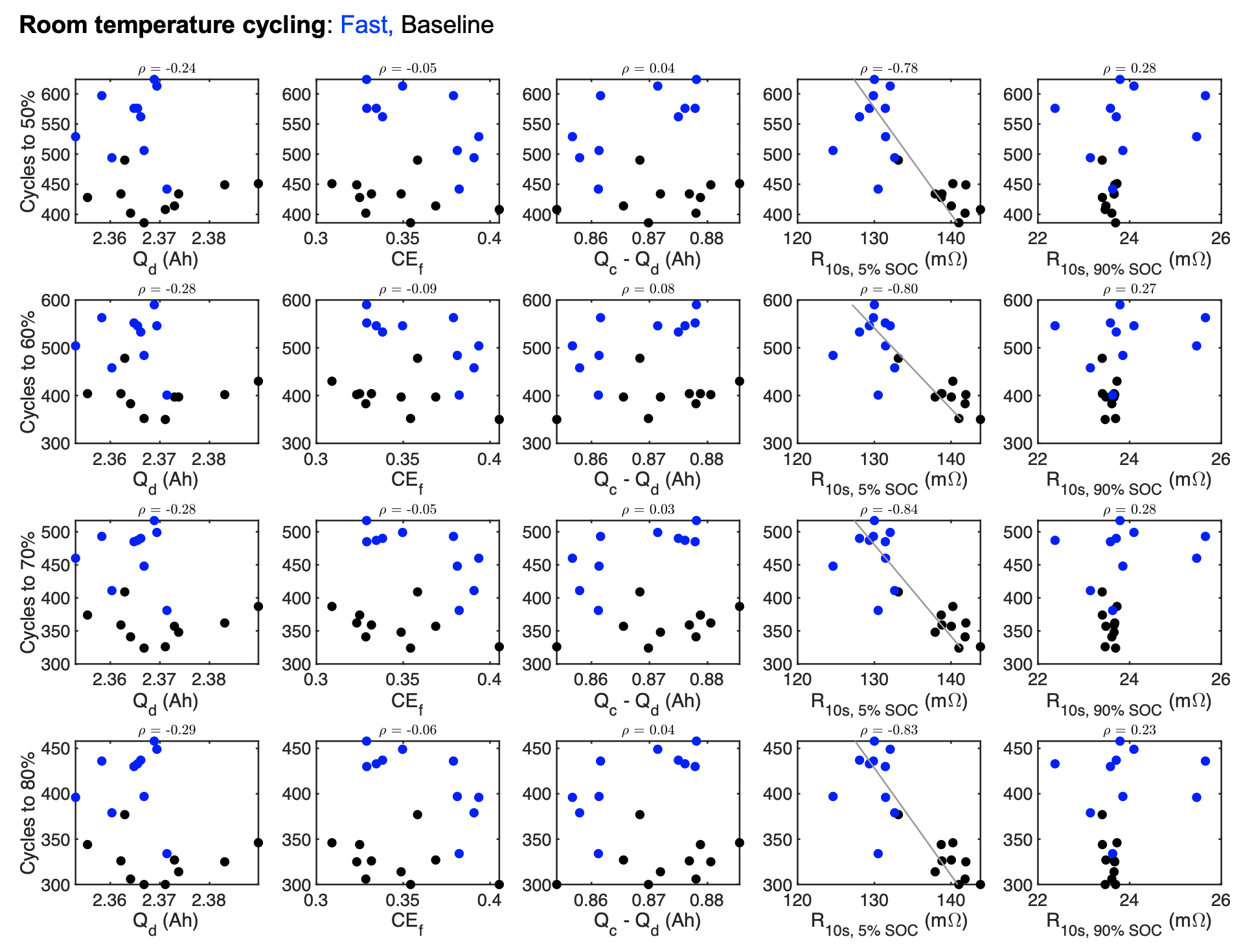}
\caption{\textbf{Correlation Between Initial Cell Signals and Cycle Life for Room Temperature Cycling} \\ End of life is defined as cycles to $x$\% of initial capacity, where $x = \{50, 60, 70, 80\}$. Formation signals ($\QLLI = \Qc-\Qd$ and $\cef$) are always measured at room temperature. $\reslow$ and $\reshigh$ are measured at the same temperature as the cycle life test.}
\label{fig:correlations-different-eol-definitions-rt}
\end{figure*}

\begin{figure*}[ht]
\centering
\includegraphics[width=1\linewidth]{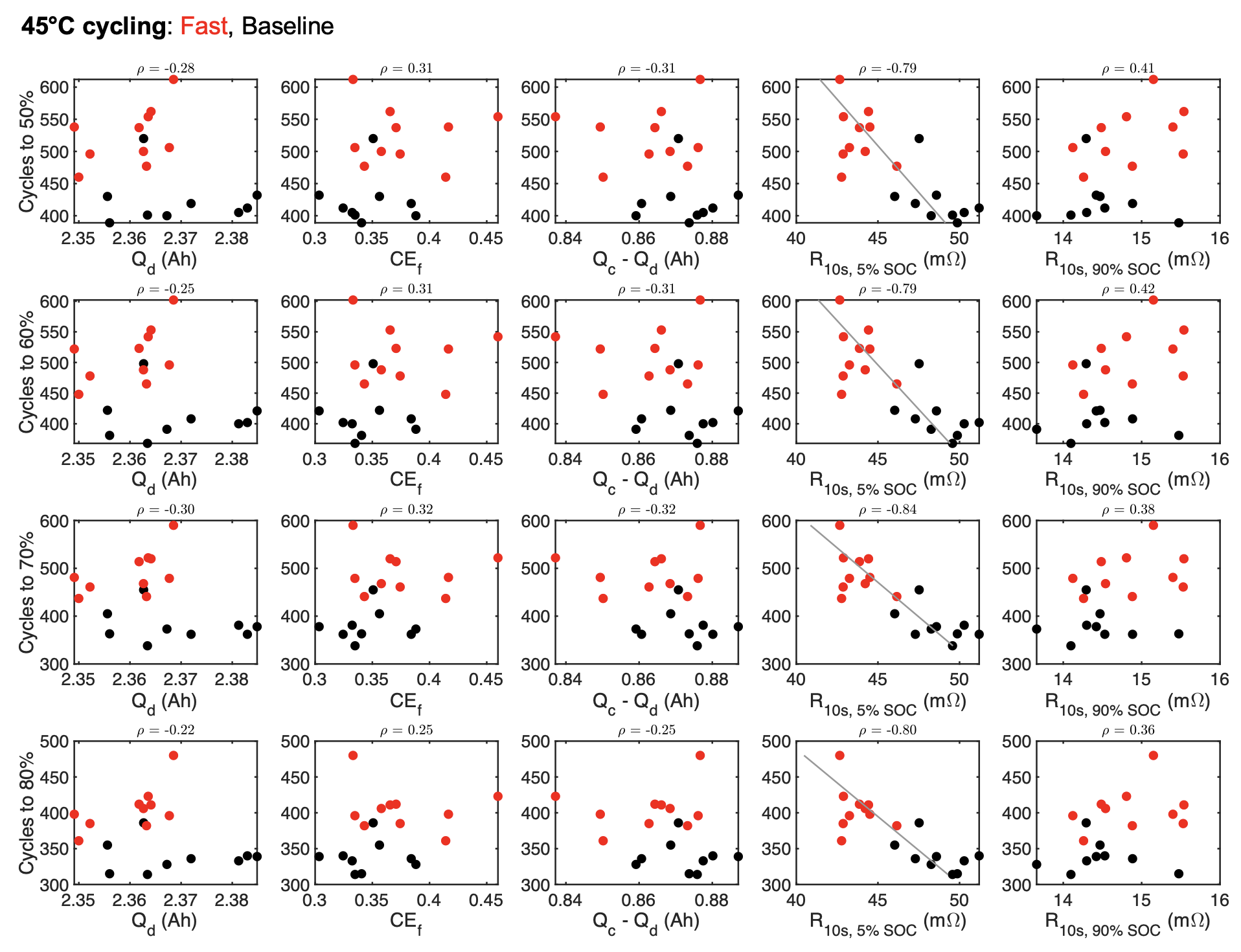}
\caption{\textbf{Correlation Between Initial Cell State Signals and Cycle Life for 45\textdegree C Cycling} \\ End of life is defined as cycles to $x$\% of initial capacity, where $x = \{50, 60, 70, 80\}$. Formation signals ($\QLLI = \Qc-\Qd$ and $\cef$) are always measured at room temperature. $\reslow$ and $\reshigh$ are measured at the same temperature as the cycle life test.}
\label{fig:correlations-different-eol-definitions-ht}
\end{figure*}

\begin{figure*}[ht]
\centering
\includegraphics[width=1\linewidth]{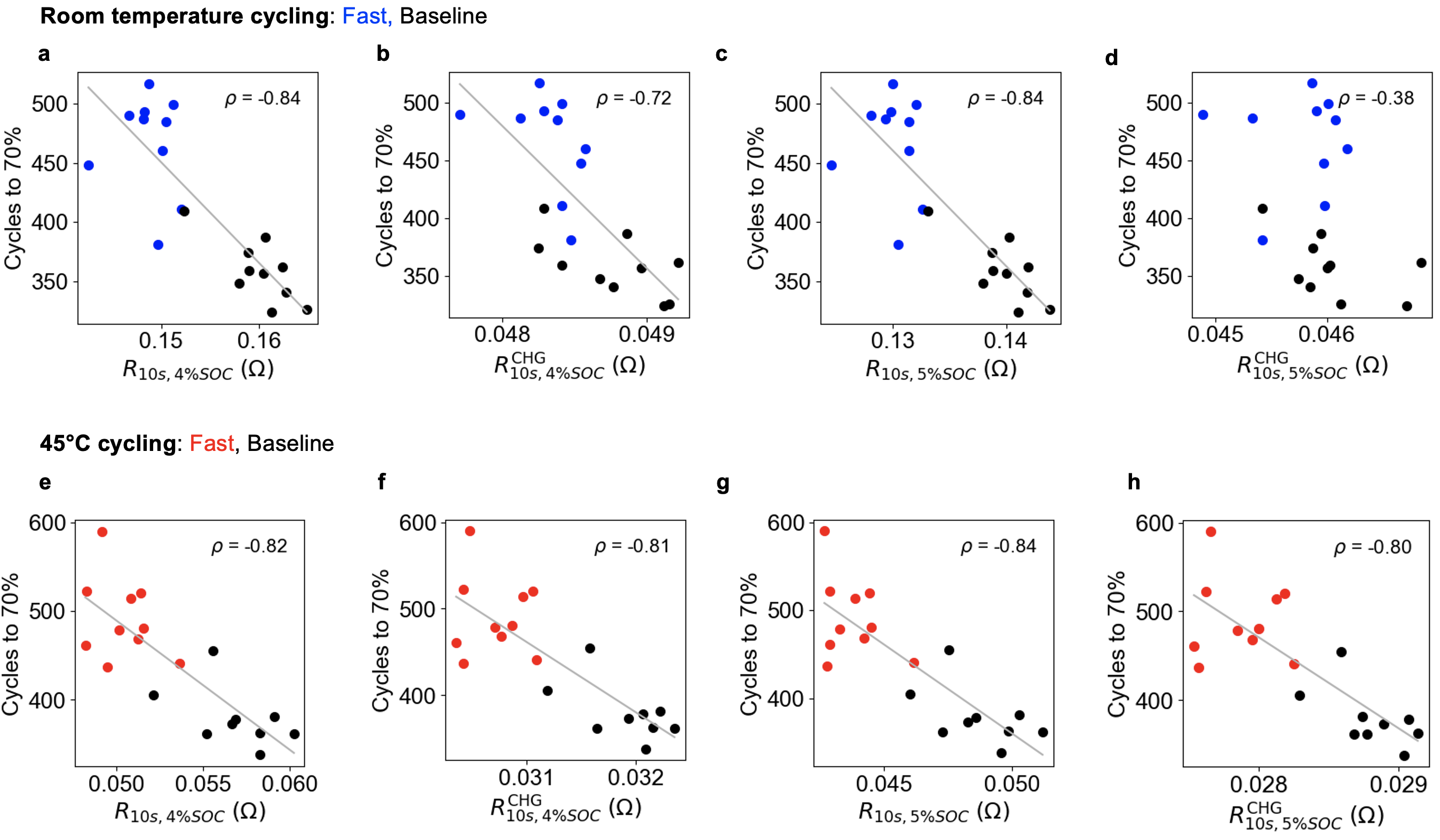}
\caption{\textbf{Correlation Between Low-SOC Resistance and Cycle Life: Charge Versus Discharge Pulses} \\ Resistances are calculated from either discharge (a,c,e,g) or charge (b,d,f,h) pulses from the Hybrid Pulse Power Characterization (HPPC) sequence. The comparison is provided for resistances evaluated at both 4\% SOC (a,b,e,f) and 5\% SOC (c,d,g,h).}
\label{fig:correlations_charge_vs_discharge}
\end{figure*}

%%% VOLTAGE FITTING ANALYSIS RESULTS %%%

\begin{figure*}[ht]
\centering
\includegraphics[width=1.0\linewidth]{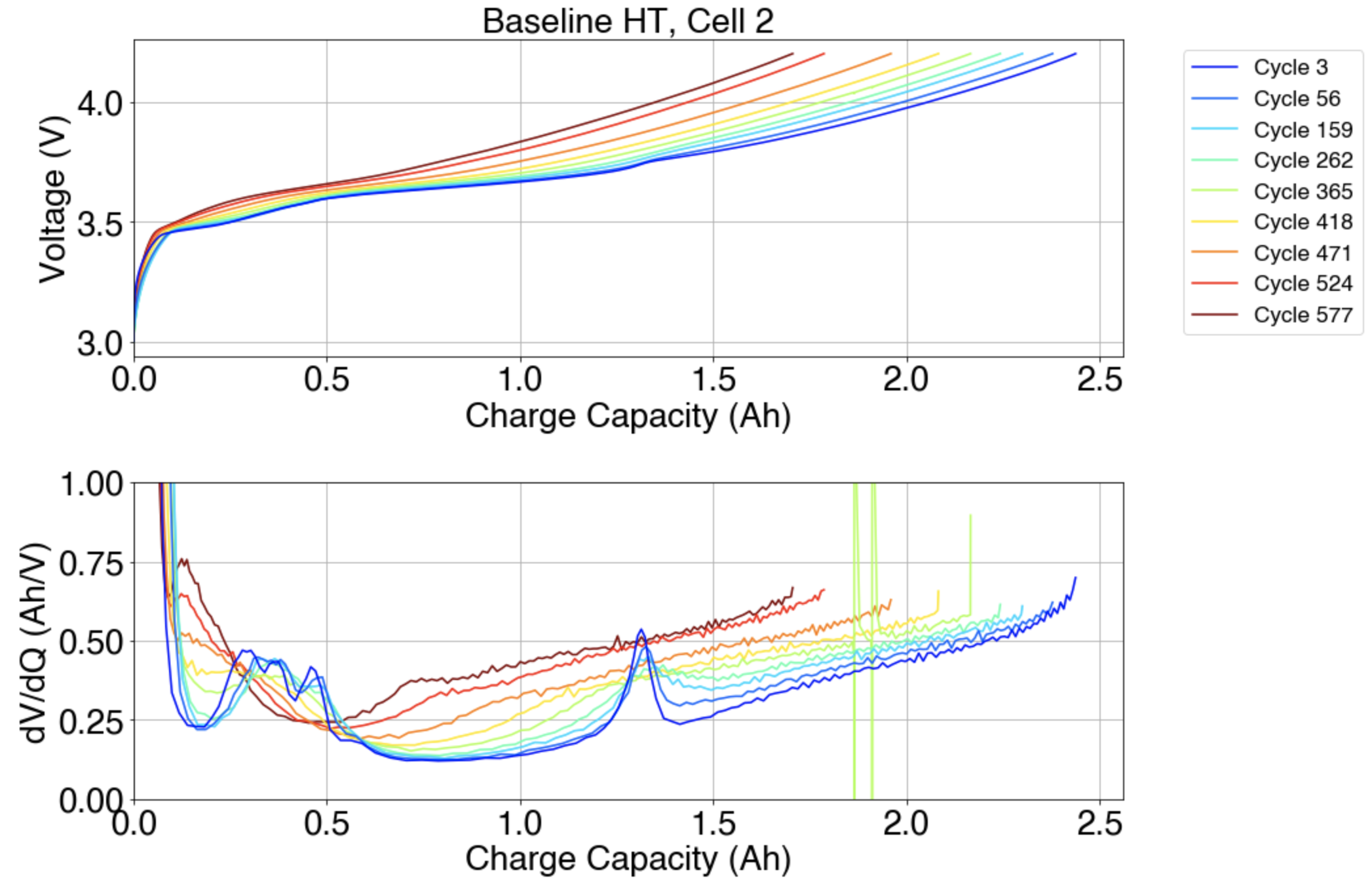}
\caption{\textbf{Example Set of C/20 Charge Curves Used for Voltage Fitting Analysis} \\The C/20 charge curves are embedded as part of the reference performance tests (RPTs).}
\label{fig:dvdq_aging_example}
\end{figure*}

\begin{figure*}[ht]
\centering
\includegraphics[width=1.0\linewidth]{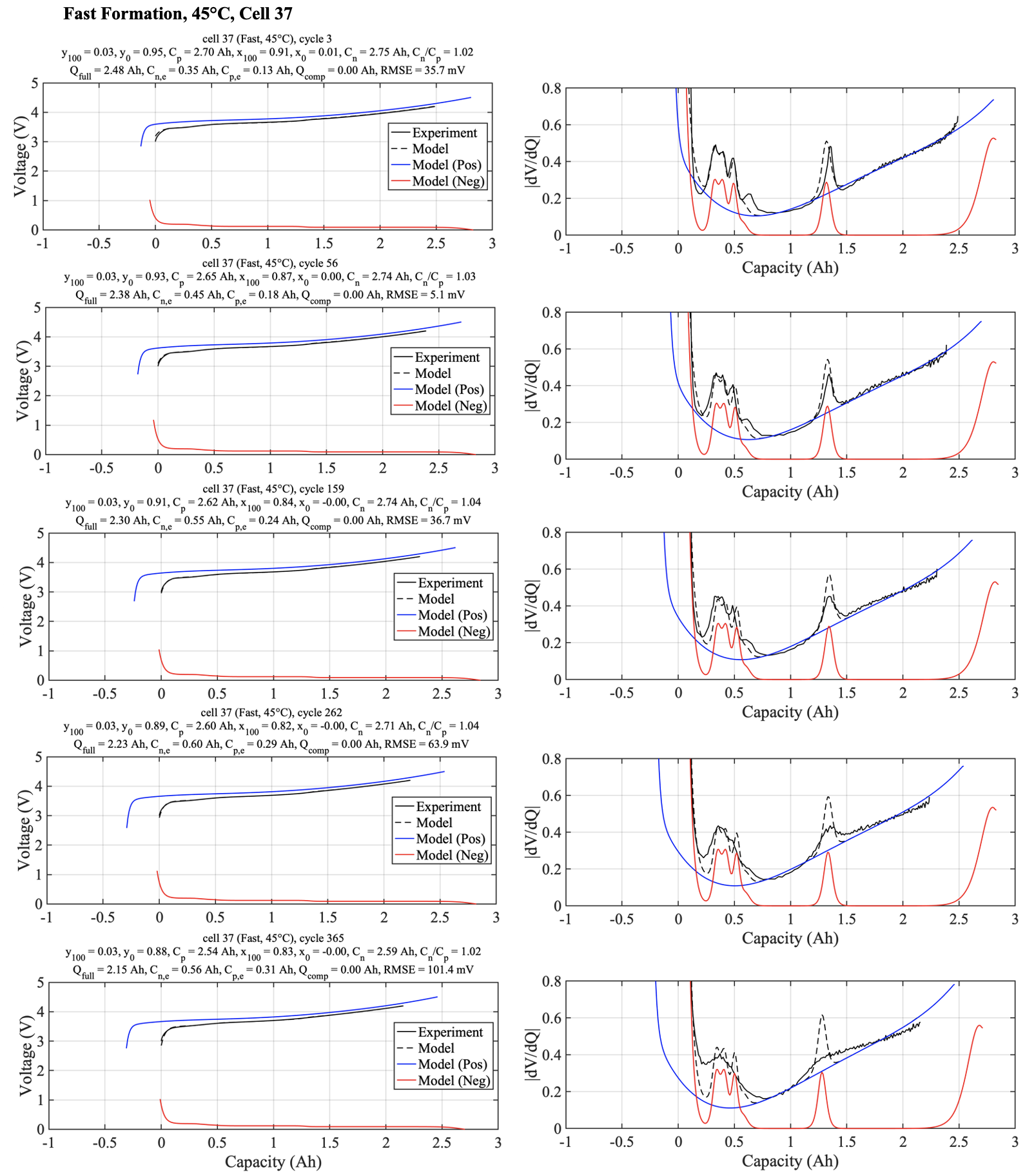}
\caption{\textbf{Example of Voltage Fitting Results Obtained on a Single Cell Over Aging}\\ Left column: voltage feature alignment. Right: differential voltage feature alignment. Top row: fresh cell. Bottom row: after 365 cycles have elapsed. Note that the voltage fitting algorithm only consumes voltage data.}
\label{fig:dvdq_fitting_example}
\end{figure*}

\begin{figure*}[ht]
\centering
\includegraphics[width=1\linewidth]{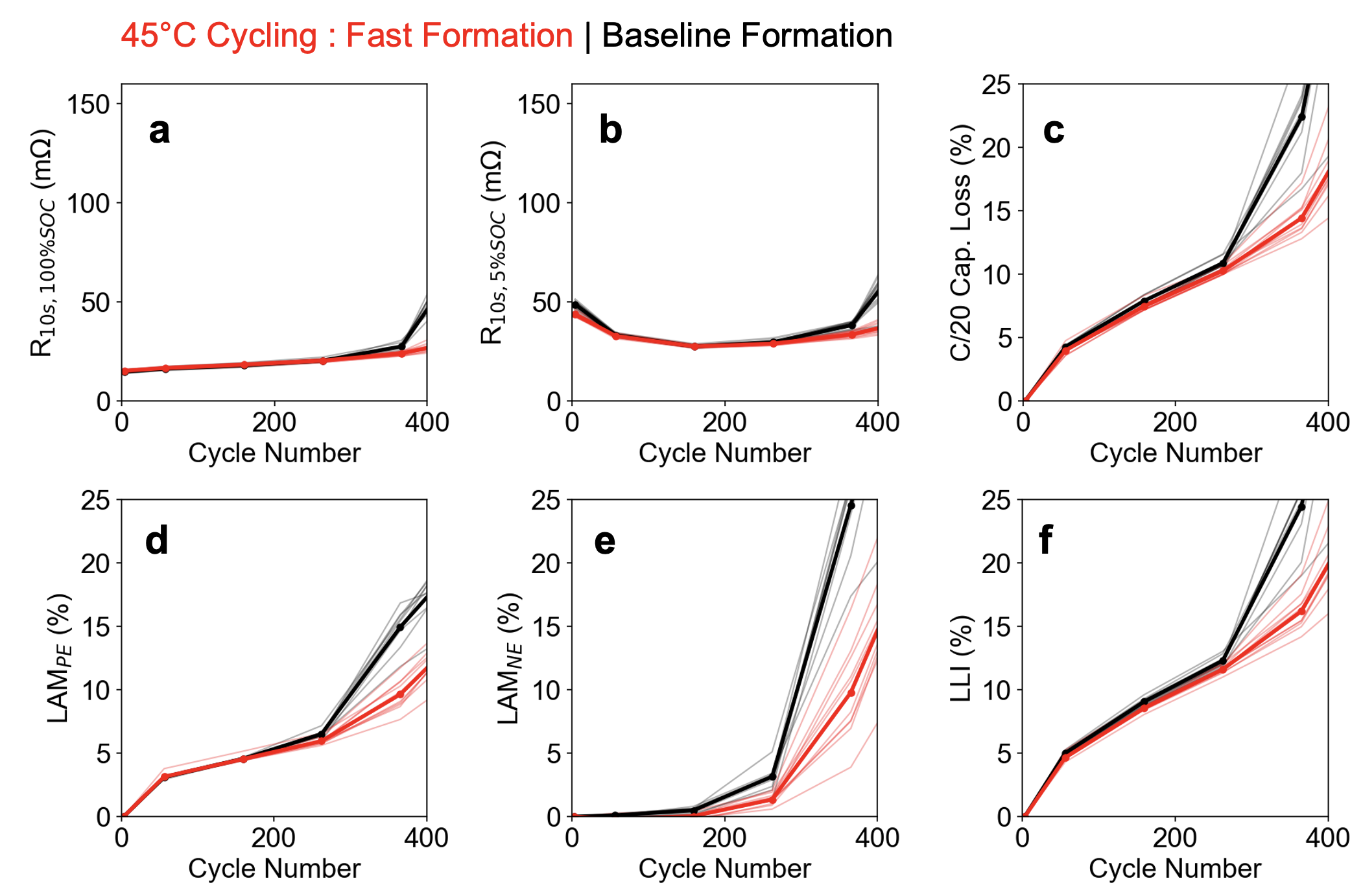}
\caption{\textbf{Evolution of Diagnostic Signals Over the course of the 45\textdegree C Cycle Life Test} \\All metrics shown are calculated using data collected from the Reference Performance Tests (RPTs) run over the course of the cycle life test. Metrics include (a) 10-second resistance calculated at 4\% SOC, (b) 10-second resistance calculated at 100\% SOC, (c) C/20 discharge capacity loss, (d) loss of active material in the positive electrode, (e) loss of active material in the negative electrode, and (f) loss of lithium inventory. Loss of active material and loss of lithium inventory are calculated using the voltage fitting techniques reported by Lee et al. \supercite{Lee2020a} (e.g. Figure \ref{fig:dvdq_fitting_example}). Thick lines show mean values averaged from each group of baseline and fast formation cells. Thin lines show results from individual cells.}
\label{fig:aging_test_signals_high_temp}
\end{figure*}

\begin{figure*}[ht]
\centering
\includegraphics[width=1\linewidth]{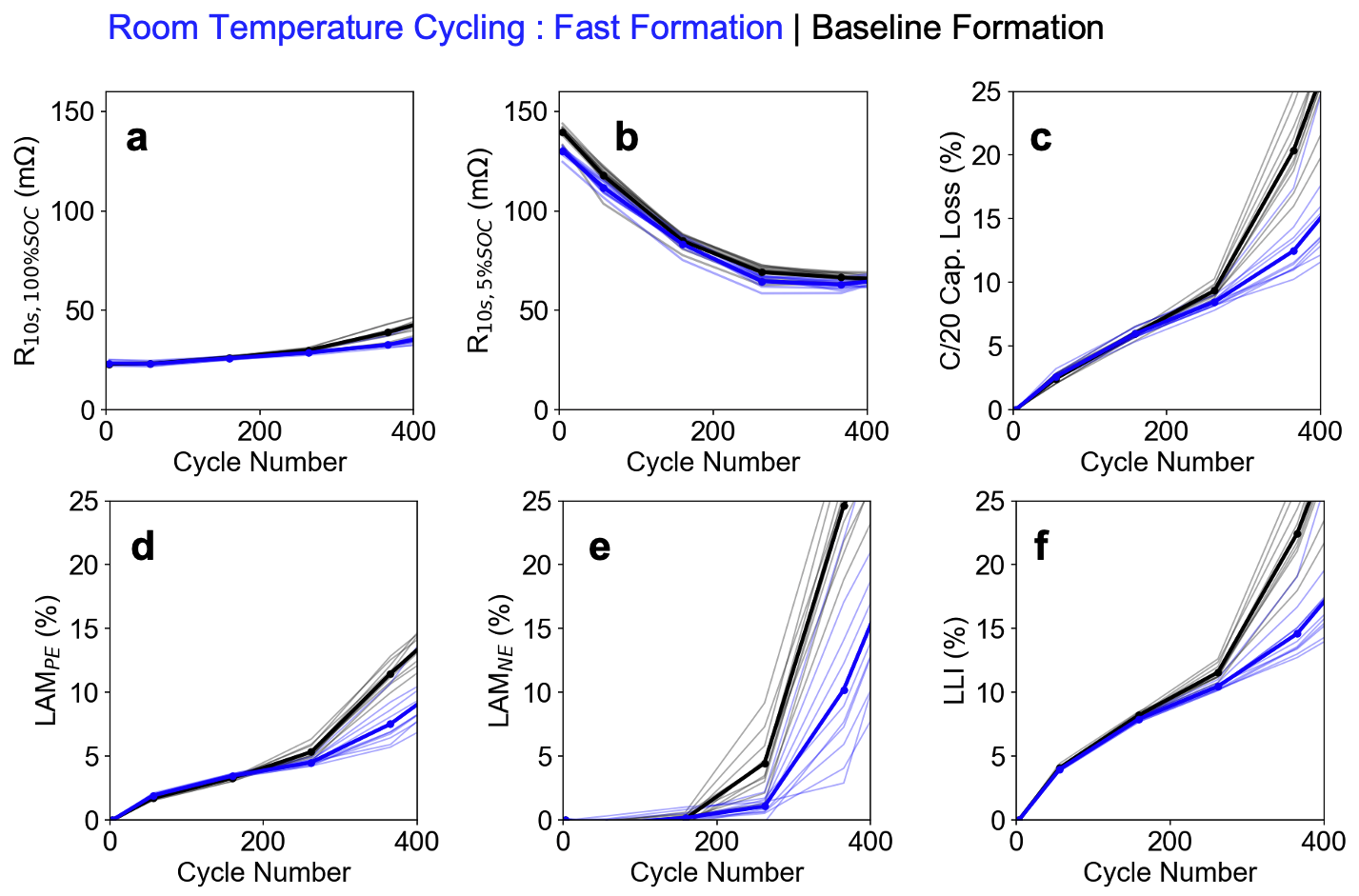}
\caption{\textbf{Evolution of Diagnostic Signals Over the Course of the Room Temperature Cycle Life Test} \\All metrics shown are calculated using data collected from the Reference Performance Tests (RPTs) run over the course of the cycle life test. Metrics include (a) 10-second resistance calculated at 4\% SOC, (b) 10-second resistance calculated at 100\% SOC, (c) C/20 discharge capacity loss, (d) loss of active material in the positive electrode, (e) loss of active material in the negative electrode, and (f) loss of lithium inventory. Loss of active material and loss of lithium inventory are calculated using the voltage fitting techniques reported by Lee et al. \supercite{Lee2020a} (e.g. Figure \ref{fig:dvdq_fitting_example}). Thick lines show mean values averaged from each group of baseline and fast formation cells. Thin lines show results from individual cells.}
\label{fig:aging_test_signals_low_temp}
\end{figure*}

\begin{figure*}[ht]
\centering
\includegraphics[width=0.8\linewidth]{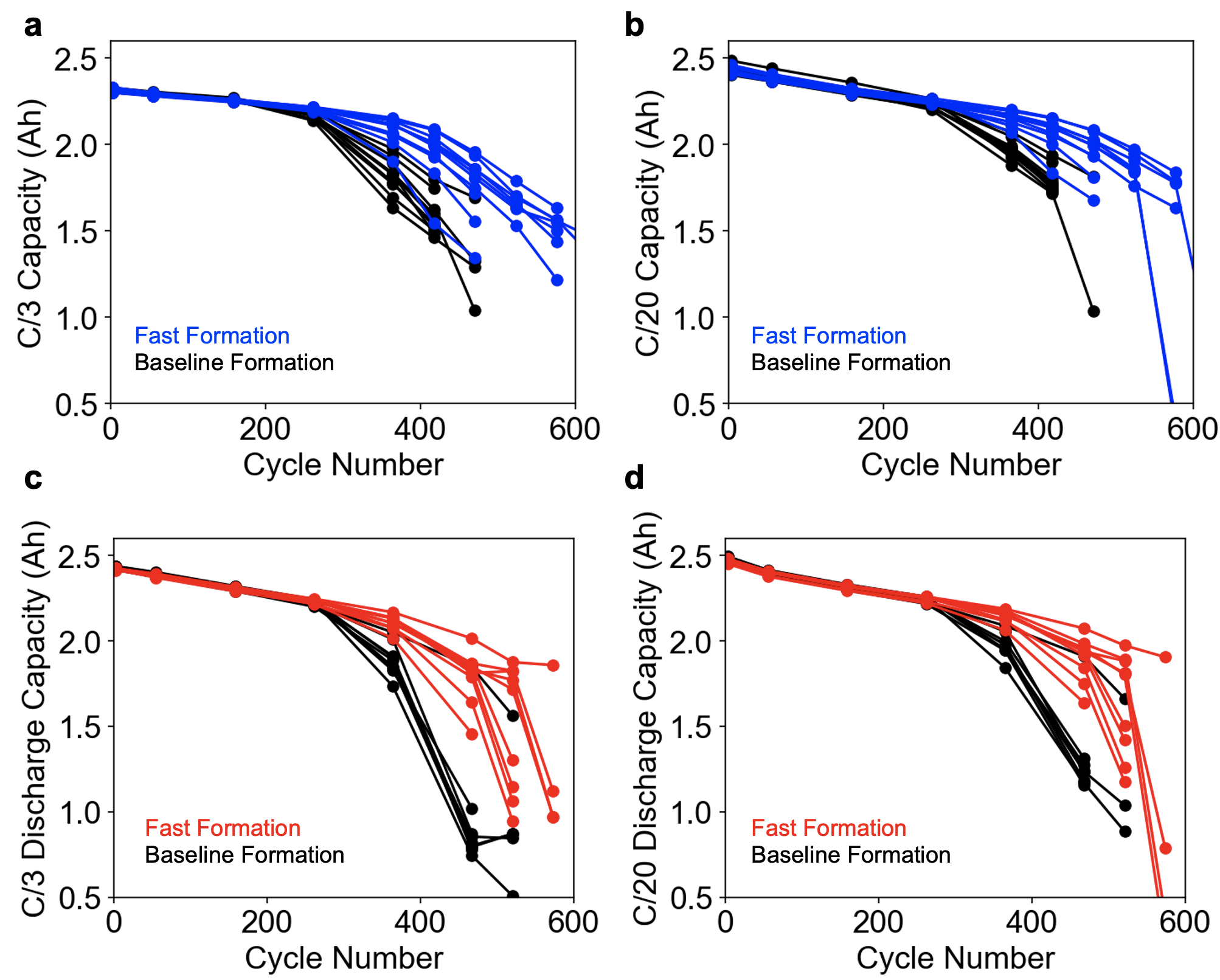}
\caption{\textbf{Comparison of Discharge Capacity Versus Cycle Number at Different C-Rates}\\(a,b) Cycle life tests run at room temperature. (c,d) Cycle life tests run at 45\textdegree C. (a,c) C/3 discharge capacity. (b,d) C/20 discharge capacity. All discharge capacities here are measured from the Reference Performance Tests (RPTs).}
\label{fig:aging_test_diagnostic_signals_2}
\end{figure*}

%%%% COIN CELL HPPC

\begin{figure*}[ht]
\centering
\includegraphics[width=0.7\linewidth]{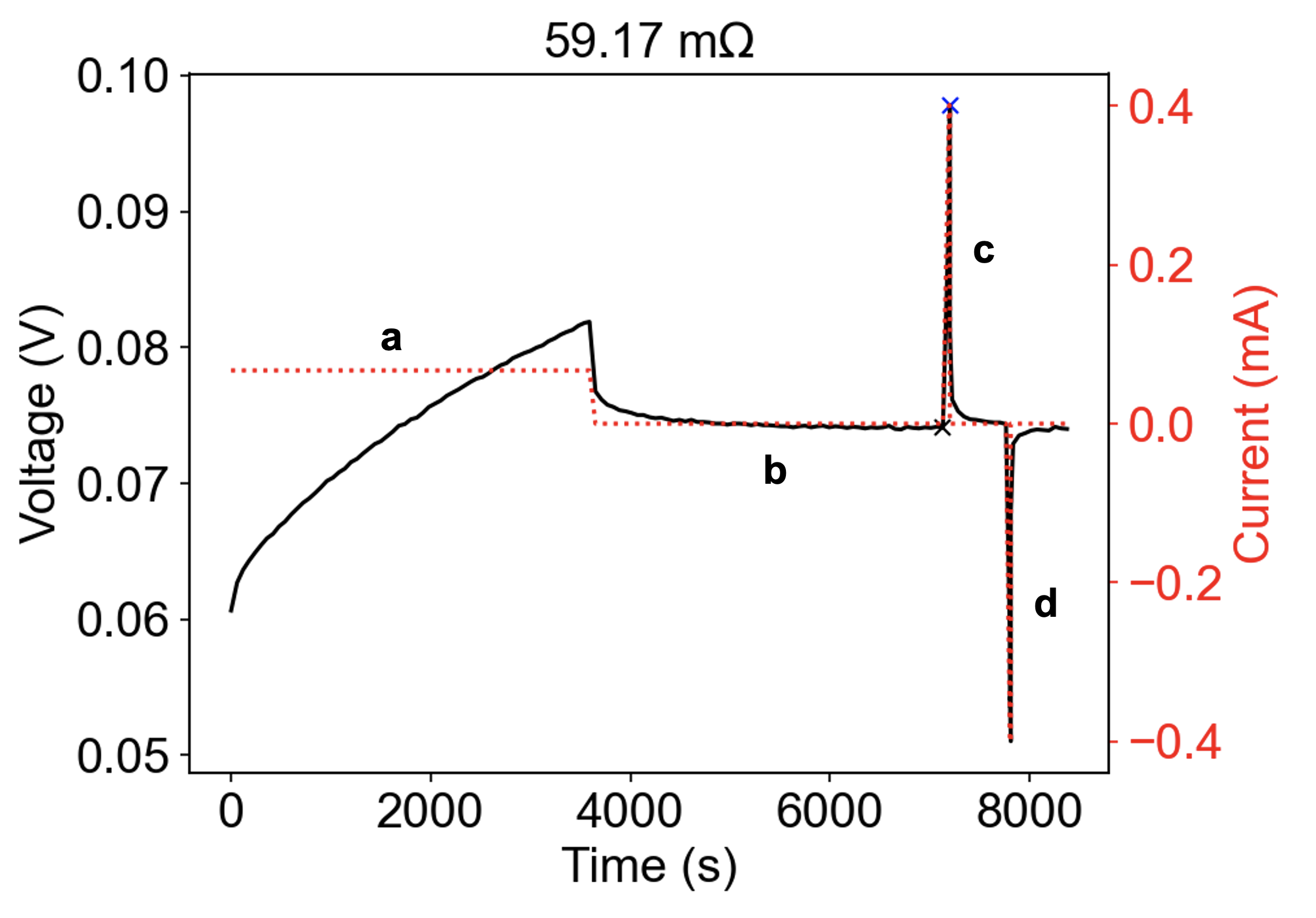}
\caption{\textbf{Example Pulse Profile for a Graphite/Li Coin Cell Half Cell} \\ The pulses are obtained as part of the coin cell HPPC test. The coin cell HPPC profile consists of the following repeated sequence: (a) background charge, (b) rest, (c) 10-second charge pulse, and (d) 10-second discharge pulse. The markers (x) indicate the potential difference taken to compute the 10-second resistance using Ohm's law. The potential measured at the end of the rest period (b) is assumed to be in near-equilibrium. The current used during the pulses is 0.4mA for all cells.}
\label{fig:coin_cell_hppc_example}
\end{figure*}

\begin{figure*}[ht]
\centering
\includegraphics[width=1.0\linewidth]{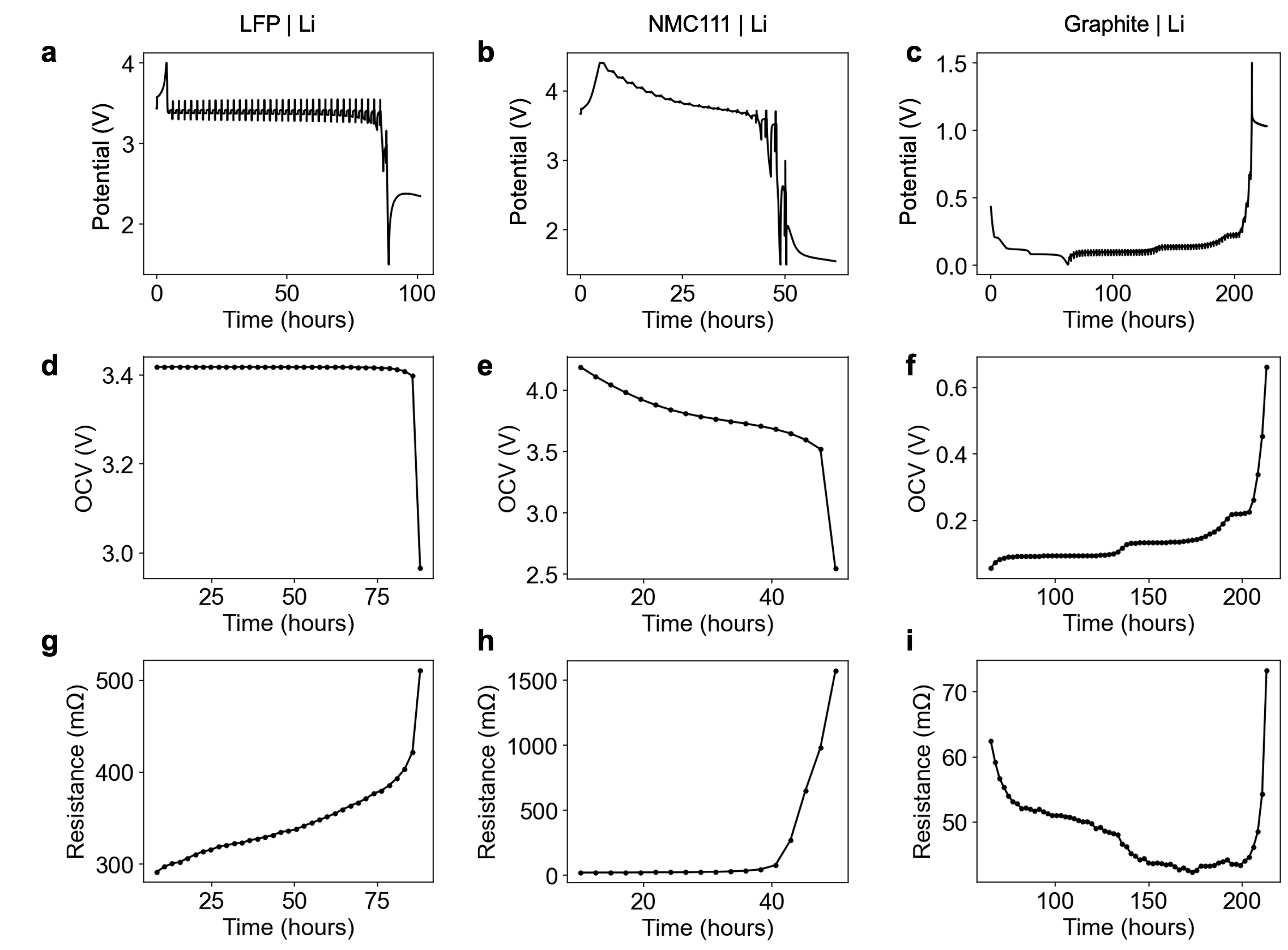}
\caption{\textbf{Hybrid Pulse Power Characterization Tests Performed on Coin Cell Half Cells} \\Potential vs time for (a) LFP/Li, (b) NMC/Li, and (c) Graphite/Li. (d-f) Corresponding pseudo-equilibrium potential curves, where the potential is measured immediately preceding each pulse. (g-h) Calculated 10-second resistances. For the positive electrodes, the resistances are calculated on discharge (i.e. during delithiation). For the graphite negative electrode, the resistances are calculated on charge (i.e. during lithiation).}
\label{fig:coin_cell_hppc}
\end{figure*}

\begin{figure*}[ht]
\centering
\includegraphics[width=0.8\linewidth]{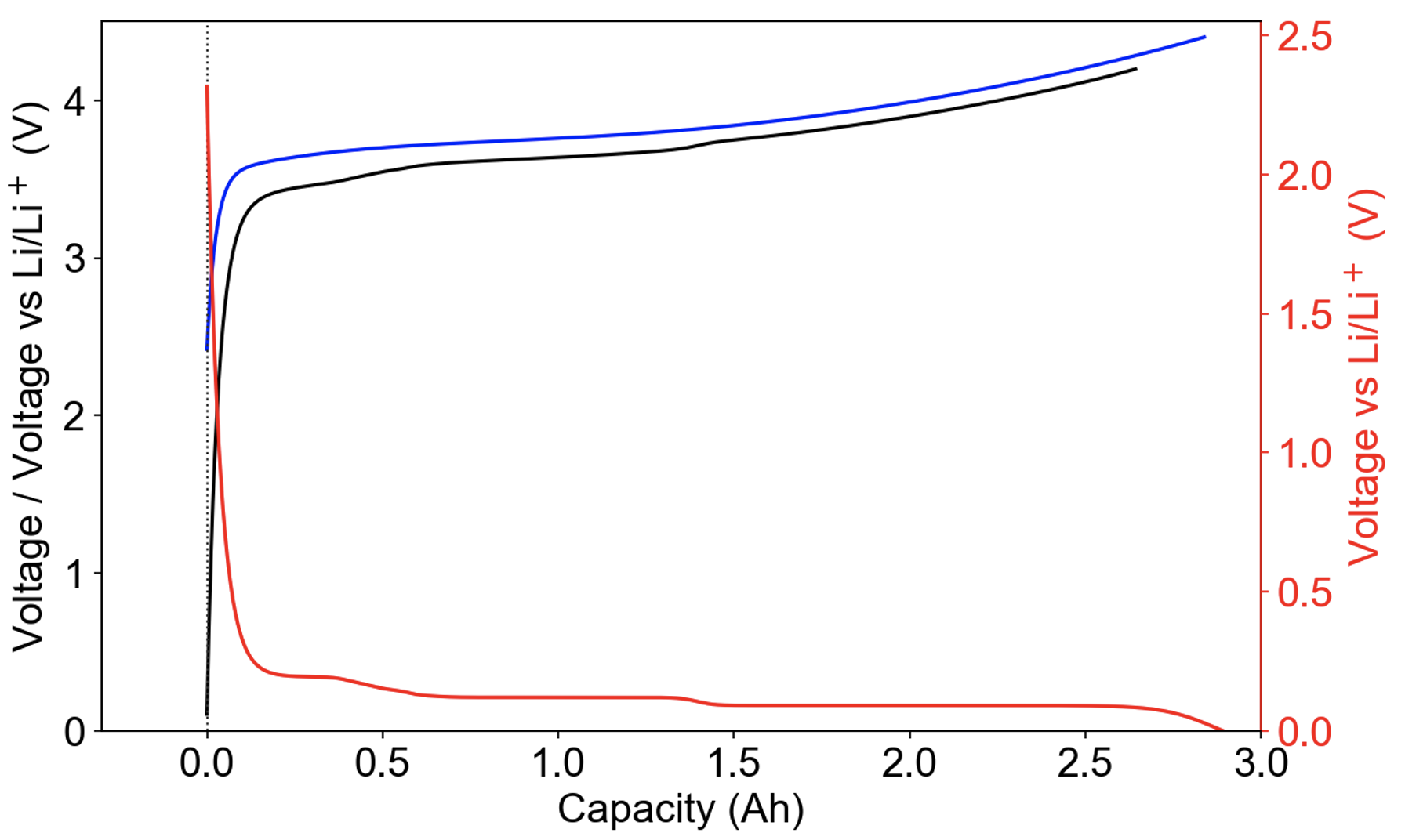}
\caption{\textbf{Initial Cell Equilibrium Potential Curves Before Formation} \\The curves are generated using the electrode stoichiometry model by aligning the point of minimum negative electrode stoichiometry (maximum potential) with the point of maximum positive electrode stoichiometry (minimum potential). Blue: positive electrode. Red: negative electrode. Black: full cell.}
\label{fig:initial-cell-state-before-formation}
\end{figure*}

\begin{figure*}[ht]
\centering
\includegraphics[width=1.0\linewidth]{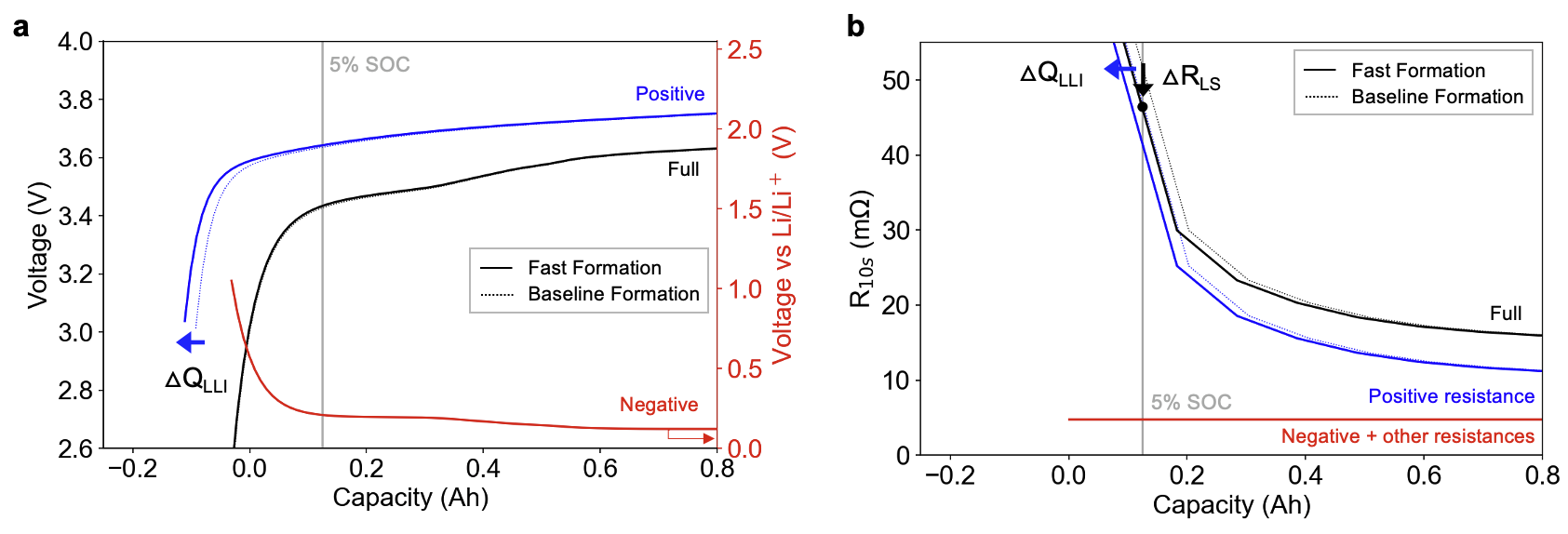}
\caption{\textbf{Electrode Stoichiometry Model Illustration} \\ The model shows the impact of fast formation on the alignment of initial cell equilibrium potential curves (a) and resistance curves (b). In this plot, $\Delta\QLLI$ has been set to $\sim$23 mAh.}
\label{fig:initial-cell-state-mechanism-precise}
\end{figure*}

\begin{figure*}[ht]
\centering
\includegraphics[width=1\linewidth]{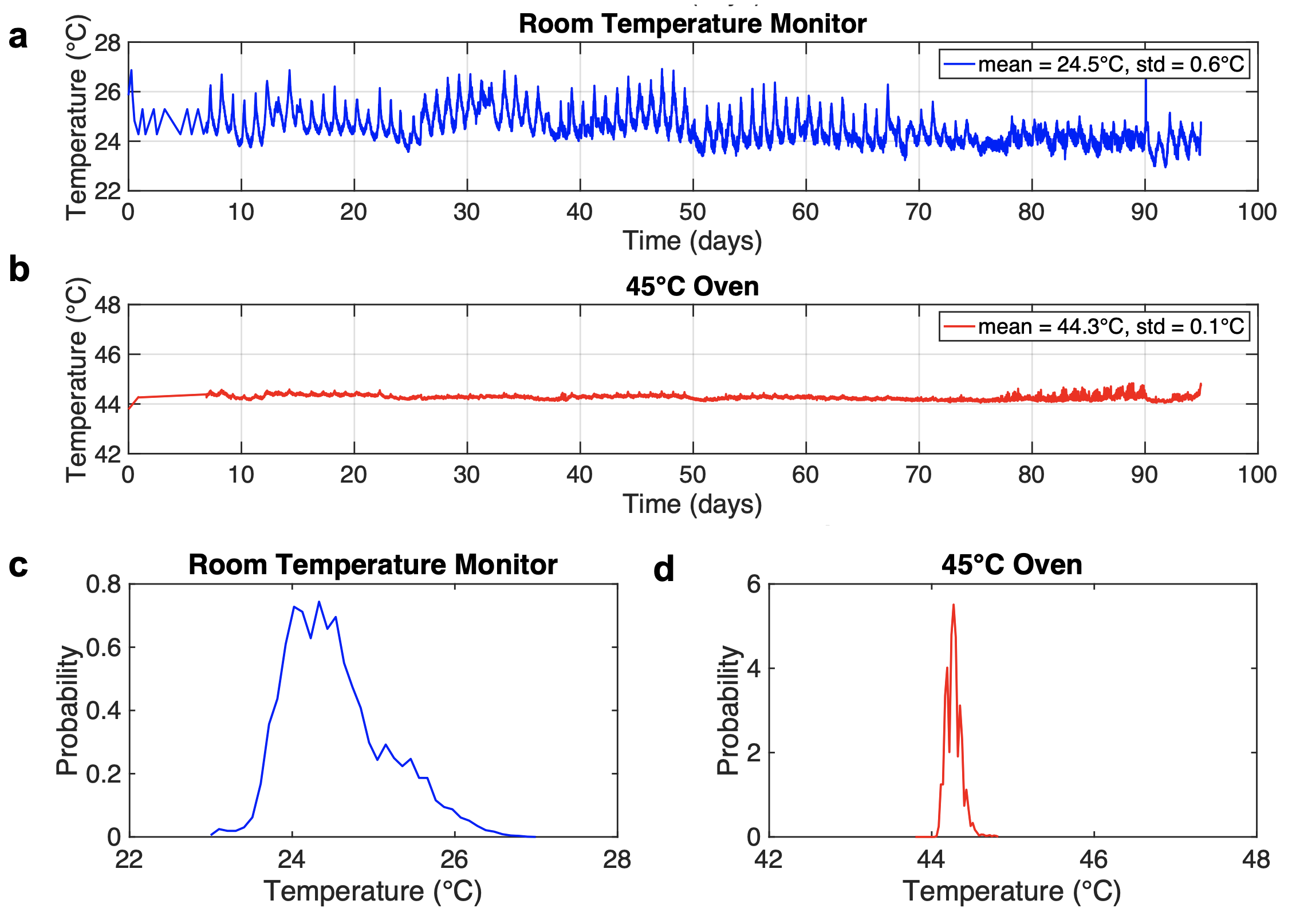}
\caption{\textbf{Temperature Measurement During Cycle Life Testing} \\ (a,b) Time-series data for the room temperature (a) and 45\textdegree C (b) tests. (c,d) Temperature histograms for the room temperature (a) and 45\textdegree C (d) tests.}
\label{fig:aging-test-temperature}
\end{figure*}

%% GENERALIZABILITY STUDIES

\begin{figure*}[ht]
\centering
\includegraphics[width=1\linewidth]{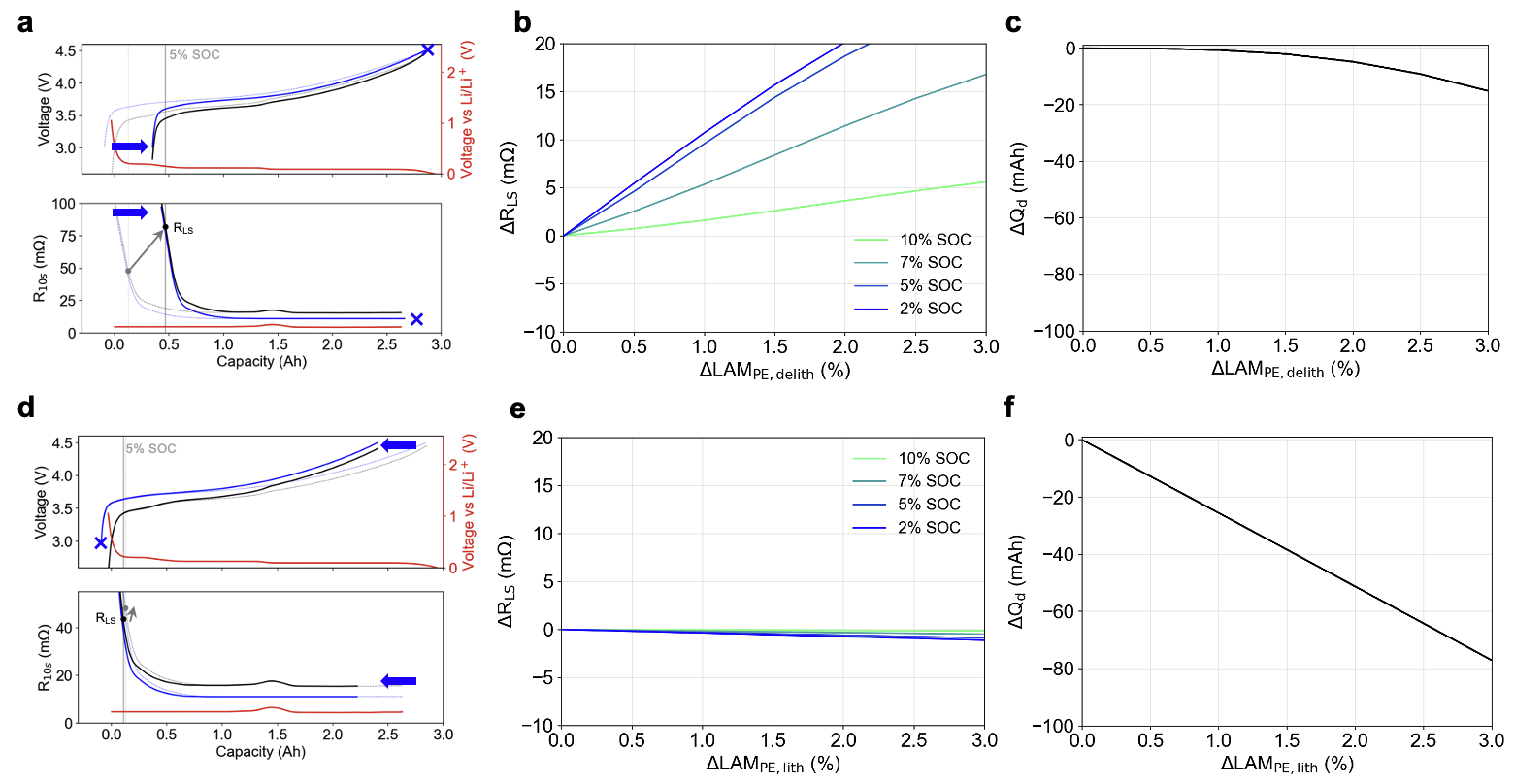}
\caption{\textbf{Sensitivity of Low-SOC Resistance to Loss of Active Material in the Positive Electrode} \\ (a-c) Impact of loss of positive active material in the delithiated state. (d-e) Impact of loss of positive active material in the lithiated state. (c,f) Sensitivity of discharge capacity to loss of active material in the positive electrode. In (a,d), LAM is set to 15\% for graphical clarity.}
\label{fig:generalizability_lam_pe}
\end{figure*}

\begin{figure*}[ht]
\centering
\includegraphics[width=1\linewidth]{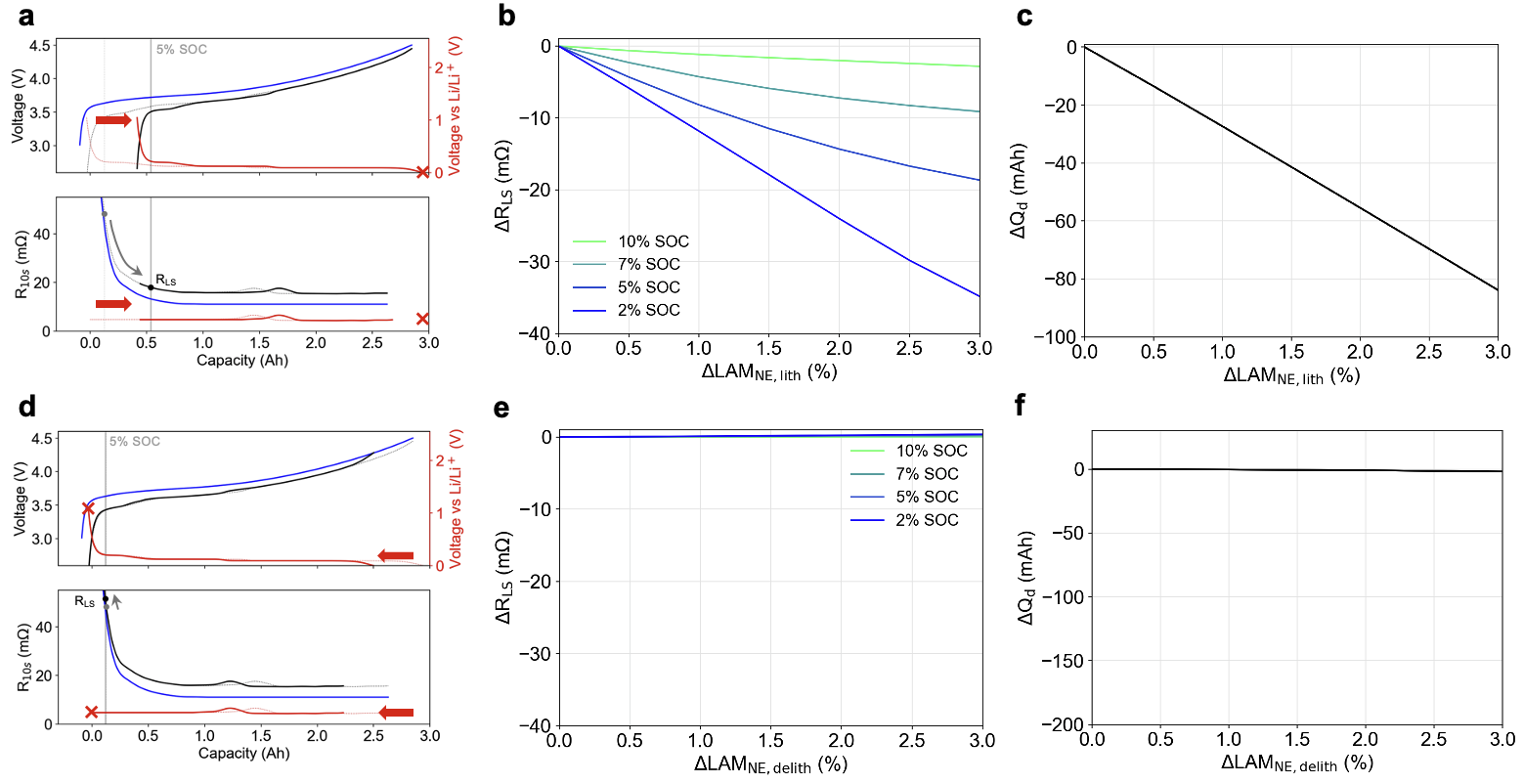}
\caption{\textbf{Sensitivity of Low-SOC Resistance to Loss of Active Material in the Negative Electrode} \\ (a-c) Impact of loss of negative active material in the lithiated state. (d-e) Impact of loss of negative active material in the delithiated state. (c,f) Sensitivity of discharge capacity to loss of active material in the negative electrode. In (a,d), LAM is set to 15\% for graphical clarity.}
\label{fig:generalizability_lam_ne}
\end{figure*}

\begin{figure*}[ht]
\centering
\includegraphics[width=1\linewidth]{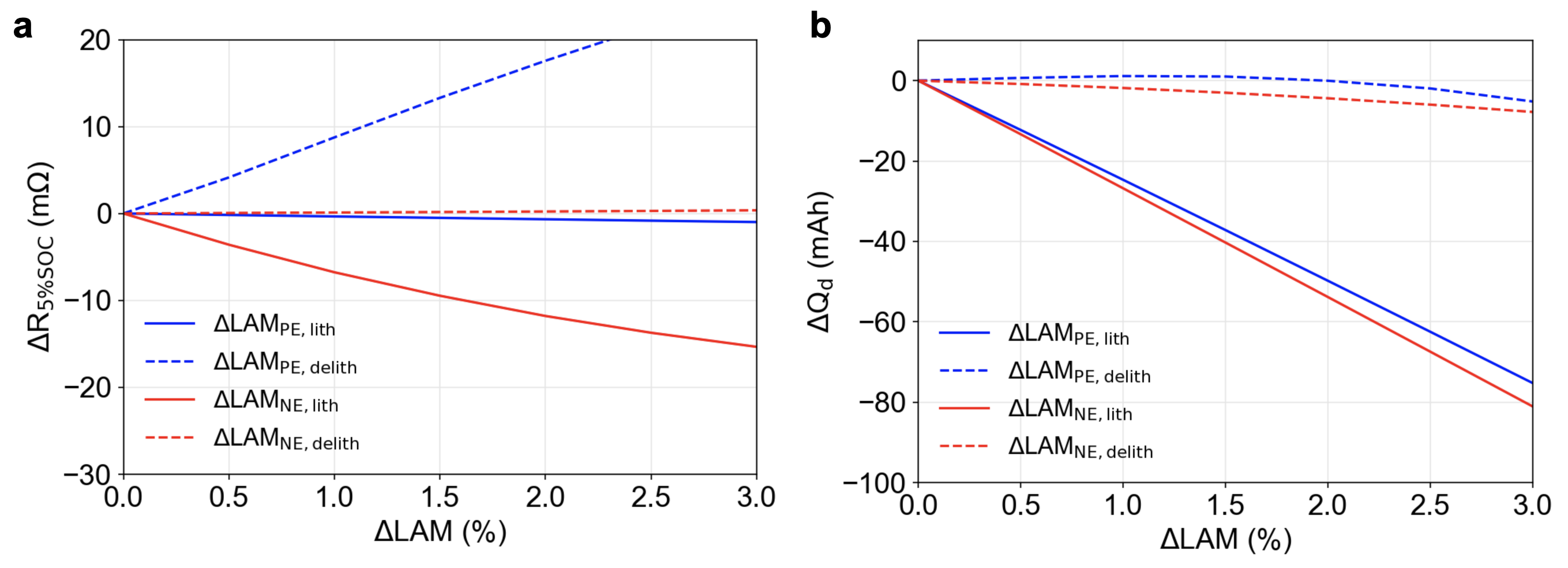}
\caption{\textbf{Sensitivity of Low-SOC Resistance and Discharge Capacity to Active Material Loss: Summary} \\ (a) Sensitivity of low-SOC resistance to different active material loss types. (b) Sensitivity of discharge capacity to different active material loss types. In general, active material can be lost at both the positive and negative electrodes, as well as in the lithiated and delithiated states.}
\label{fig:generalizability_lam_comparison}
\end{figure*}

\begin{figure*}[ht]
\centering
\includegraphics[width=1\linewidth]{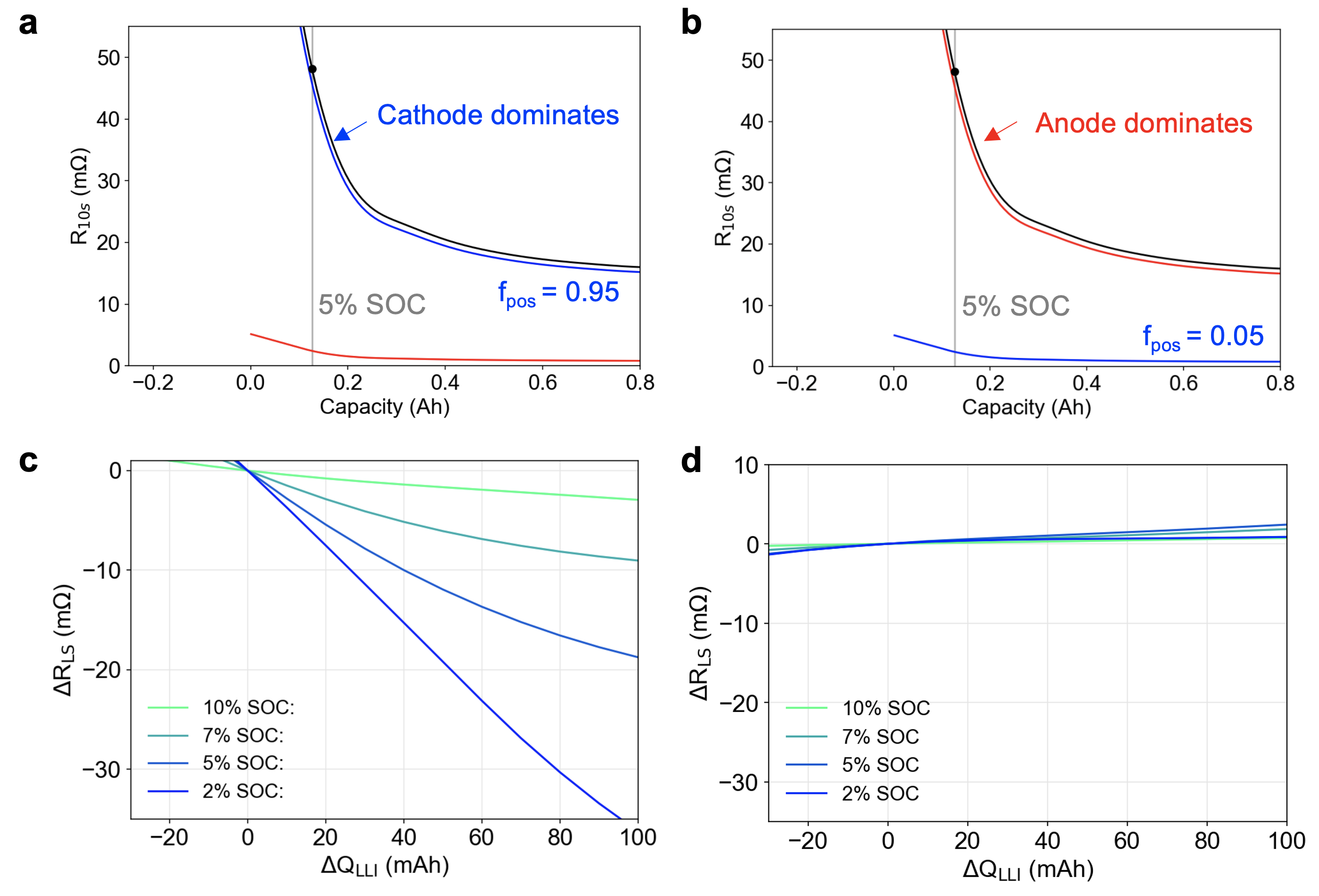}
\caption{\textbf{Sensitivity of Low-SOC Resistance to $\Delta\QLLI$ Depends on Fractional Contribution of the Positive Electrode to the Total Cell Resistance: Example 1} \\ Sensitivity of the low-SOC resistance metric to $\Delta\QLLI$ when (a,c) positive electrode dominates the cell resistance and when (b,d) negative electrode dominates the cell resistance. $f_\mathrm{pos}$ is the fractional contribution of the positive electrode to the total cell resistance.}
\label{fig:generalizability_cathode_dominance_1}
\end{figure*}

\begin{figure*}[ht]
\centering
\includegraphics[width=0.8\linewidth]{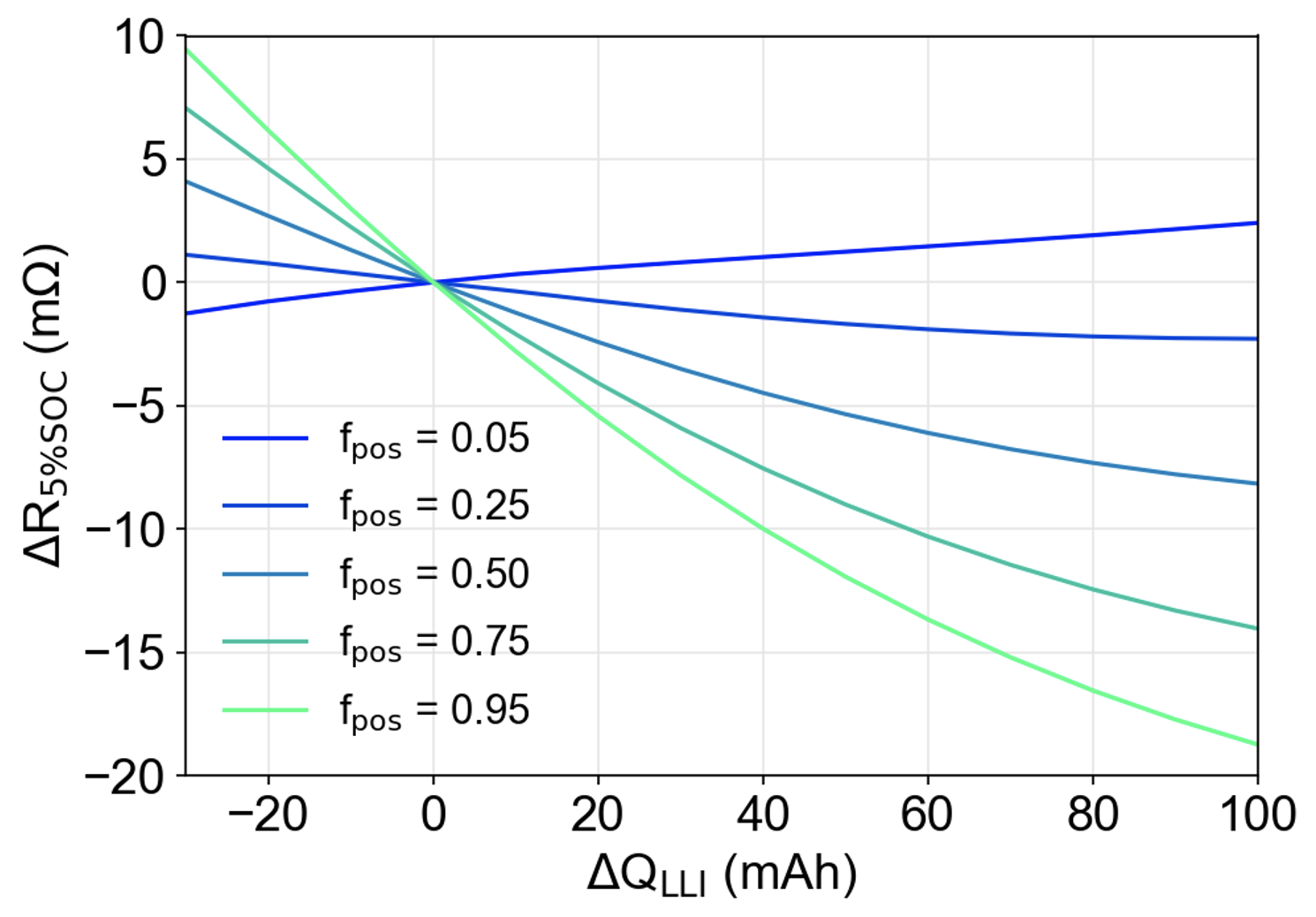}
\caption{\textbf{Sensitivity of Low-SOC Resistance to $\Delta\QLLI$ Depends on Fractional Contribution of the Positive Electrode to the Total Cell Resistance: Example 2} \\ Sensitivity of the resistance at 5\% SOC to $\Delta\QLLI$ for varying positive electrode resistance contributions to full cell resistance. $f_\mathrm{pos}$ is the fractional contribution of the positive electrode to the total cell resistance.}
\label{fig:generalizability_cathode_dominance_2}
\end{figure*}

%% SWELLING AND VARIABILITY

\begin{figure*}[ht]
\centering
\includegraphics[width=1\linewidth]{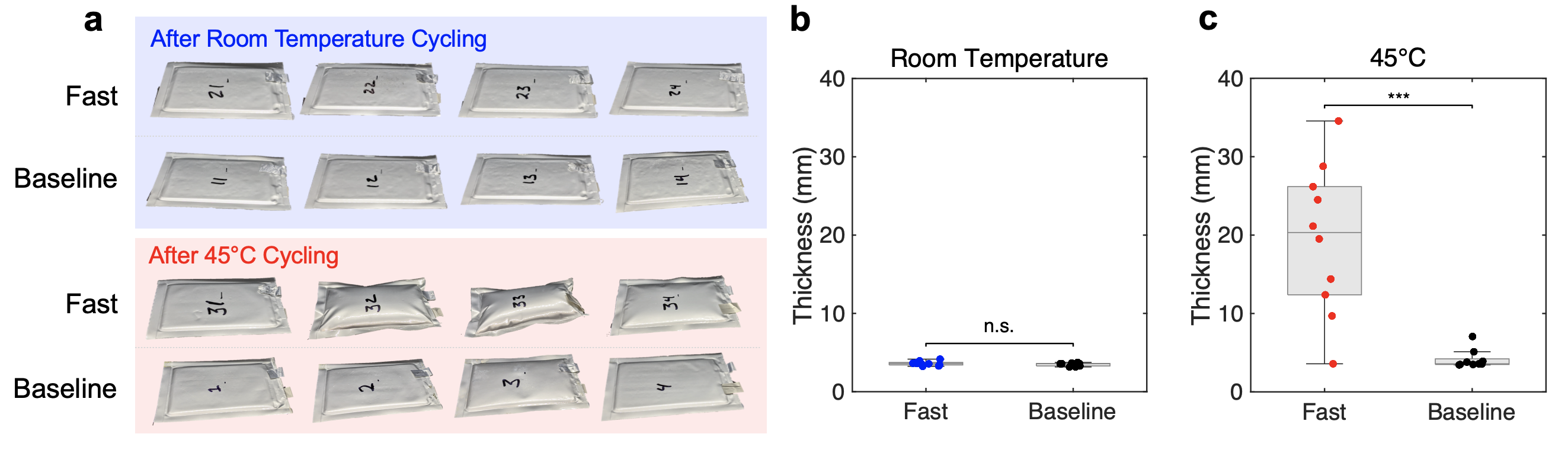}
\caption{\textbf{Pouch Cell Swelling at the End of the Cycle Life Test} \\
(a) Example images of pouch cells taken after aging showing varying degrees of swelling. (b-c) Comparison of pouch cell thicknesses measured at the end of the cycle life test. (b) Cells cycled at room temperature. (c) Cells cycled at 45°C. Cell thickness is measured using a manual caliper, which represents the point of maximum deflection. `***' - statistically significant with $p$-value $< 0.001$. `n.s.' - not statistically significant.}
\label{fig:swelling-comparison}
\end{figure*}

\begin{figure*}[ht]
\centering
\includegraphics[width=1\linewidth]{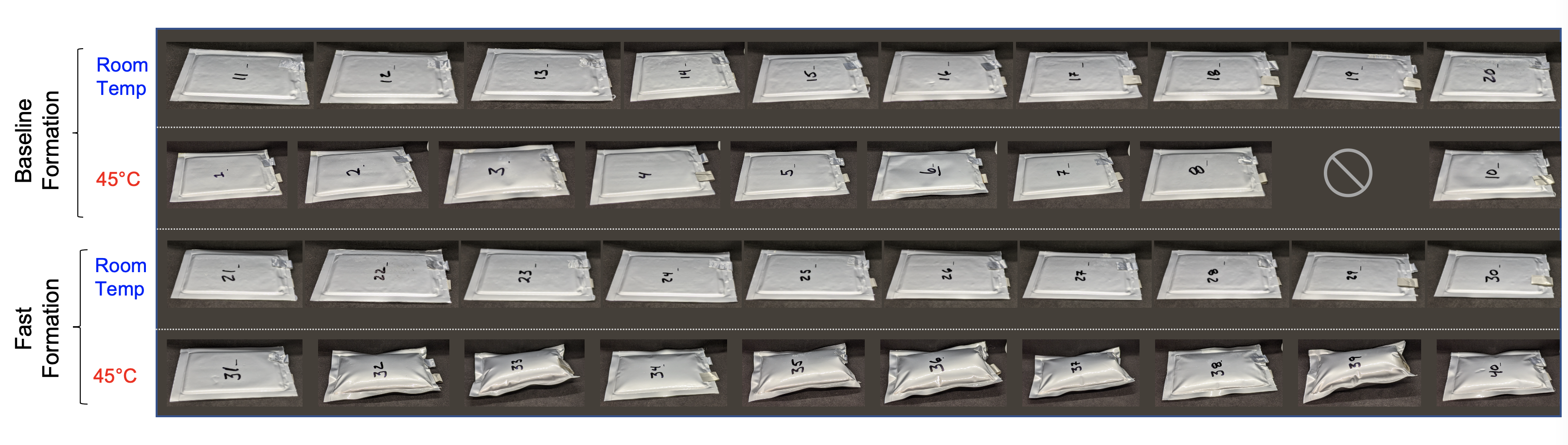}
\caption{\textbf{Images of Pouch Cells Taken After Aging} \\ Cell \#9 has been excluded from the study due to tab weld issues.}
\label{fig:swelling-all}
\end{figure*}

\begin{figure*}[ht]
\centering
\includegraphics[width=1.0\linewidth]{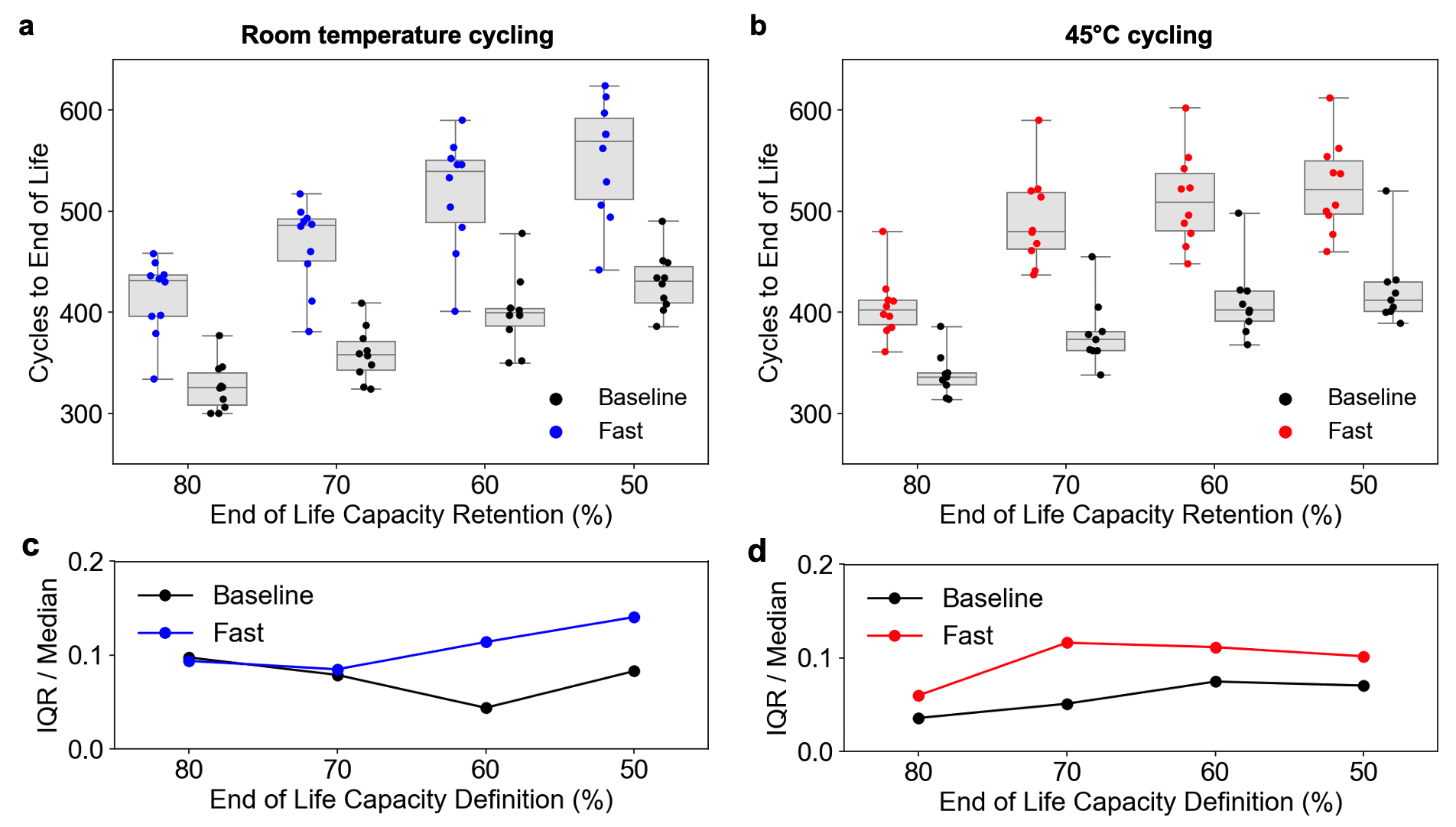}
\caption{\textbf{Aging Variability as a Function of End-of-Life Definition}\\
(a,b) Cycles to end of life under (a) room temperature and (b) 45\textdegree C cycling. Boxes show inter-quartile range (IQR) and whiskers show the min and max values. (c,d) Inter-quartile range (IQR) divided by median plotted as a function of end of life capacity definition for (a) room temperature and (b) 45\textdegree C cycling.}
\label{fig:aging-variability}
\end{figure*}

\clearpage

%%% SIGNAL-TO-NOISE ANALYSIS SECTION
\section*{Note S1: Signal-to-Noise Analysis on Estimates of Lithium Consumed During Formation}

\revb{5}{Here, we perform a simple set of calculations to determine the resolution limit of lithium consumed during formation ($\QLLI$) as a function of the estimation method. We focus on comparing two estimation methods: the low-SOC resistance signal ($\res$) and the cell discharge capacity ($\Qd$). $\Qd$ represents a conventional signal measured using current integration. In this simplified analysis, we consider only the effect of lithium consumption and ignore possible effect from active material losses. For the analysis, we assume hardware specifications from the battery testing equipment used in our experiments (Maccor Series 4000). The hardware specifications and parameters used in the analysis are provided in Table \ref{tab:signal_to_noise_params}.}

\newcommand{\Vp}{\mathrm{V}_\mathrm{p}}
\newcommand{\Ip}{\mathrm{I}_\mathrm{p}}
\newcommand{\IFSR}{\mathrm{I}_\mathrm{FSR}}
\newcommand{\VFSR}{\mathrm{V}_\mathrm{FSR}}
\newcommand{\Iseta}{\mathrm{I}_\mathrm{set,1}}
\newcommand{\Isetb}{\mathrm{I}_\mathrm{set,2}}
\newcommand{\Ierr}{\mathrm{I}_\mathrm{err}}
\newcommand{\Verr}{\mathrm{V}_\mathrm{err}}
\newcommand{\Rlimit}{\mathrm{R}_\mathrm{limit}}
\newcommand{\Rhigh}{\mathrm{R}_\mathrm{high}}
\newcommand{\Rlow}{\mathrm{R}_\mathrm{low}}

\begin{table}[h!]
  \begin{center}
    
    \begin{tabular}{l|c|c|l} 
      \toprule
      \textbf{Parameter} & \textbf{Symbol} & \textbf{Value} & \textbf{Unit} \\
      \midrule
      Voltage Precision & $\Vp$ & 0.02 & \% of full scale range \\
      Voltage Full-Scale Range & $\VFSR$ & 5 & V \\ 
      Current Precision & $\Ip$ & 0.02 & \% of full scale range \\
      Current Full-Scale Range & $\IFSR$ & 5 & A \\
      \midrule
      Current Set-Point for $\res$ Calculations & $\Iseta$ & 2.37 ( = 1C) & A \\
      Current Set-Point for $\Qd$ Calculations & $\Isetb$ & 0.237 ( = C/10) & A \\
      \bottomrule
    \end{tabular}
    \caption{Parameters used for the signal-to-noise analysis}
    \label{tab:signal_to_noise_params}
  \end{center}
\end{table}

The current and voltage errors $\Ierr$ and $\Verr$ can be calculated as

\begin{equation}
    \begin{split}
        \Ierr &= \IFSR \cdot \Ip / 100 \\
                &= 0.001 \mathrm{A}
    \end{split}
\end{equation}

\begin{equation}
    \begin{split}
        \Verr &= \VFSR \cdot \Vp / 100 \\
                &= 0.001 \mathrm{V}.
    \end{split}
\end{equation}

The resolution limit for resistance, $\Rlimit$, can be calculated as

\begin{equation}
    \begin{split}
        \Rlimit &= \Rhigh - \Rlow \\
                &= 0.88 \mathrm{m}\Omega
    \end{split}
\end{equation}

where 

\begin{equation}
    \Rhigh = \frac{\Delta V_\mathrm{meas} + \Verr}{\Iseta - \Ierr}
\end{equation}

\begin{equation}
    \Rlow = \frac{\Delta V_\mathrm{meas} - \Verr}{\Iseta + \Ierr}.
\end{equation}

In these equations, $\Delta\mathrm{V}_\mathrm{meas}$ is the measured voltage drop during the discharge pulse. From Figure \ref{fig:hppc-example}, this value is less than 0.1V at low SOCs. For the remaining calculation, we assume $\Delta \mathrm{V}_\mathrm{meas} = 0.1V$.

The resolution limit for a capacity measurement can be estimated as

\begin{equation}
    \begin{split}
    \mathrm{Q}_\mathrm{limit} & = \mathrm{Q}_\mathrm{high} - \mathrm{Q}_\mathrm{low} \\
                              & = 20 \mathrm{mAh} 
    \end{split} 
\end{equation}

where

\begin{equation}
    \mathrm{Q}_\mathrm{high} = (\Isetb + \Ierr) \cdot 10 \mbox{ hours}
\end{equation}

\begin{equation}
    \mathrm{Q}_\mathrm{low} = (\Isetb - \Ierr) \cdot 10 \mbox{ hours}.
\end{equation}

These capacities correspond to the C/10 constant current discharge steps during the final diagnostic cycle of the formation protocol (Figure \ref{fig:design-of-experiments}b,c). Here, we neglect the effect of voltage inaccuracies which would add an additional error in the voltage termination condition for the discharge cycle. By ignoring this error, the estimate of the resolution limit for $\Qd$ remains optimistic.

To calculate the sensitivity of $\Delta\QLLI$ to $\Delta\res$, we linearize the curve from Figure 4d at 5\% SOC to find that

\begin{equation}
    \begin{split}
    \frac{\Delta\QLLI}{\reslow} &= \frac{40 \mathrm{mAh}}{10 \mathrm{m}\Omega} \\
                                &= 5.00 \mathrm{mAh}/\mathrm{m}\Omega.
    \end{split}
\end{equation}

A similar exercise is done for $\Delta\Qd$ which yields

\begin{equation}
    \begin{split}
    \frac{\Delta\QLLI}{\Delta\Qd} &= \frac{40 \mathrm{mAh}}{37 \mathrm{mAh}} \\
                                  &= 1.08 \mathrm{mAh}/\mathrm{mAh}.
    \end{split}
\end{equation}

Finally, the resolution of $\QLLI$ calculated using the two different methods are computed as

\begin{equation}
    \begin{split}
        \QLLI \mbox{ calculated from } \reslow &= \frac{\Delta\QLLI}{\Delta\reslow} \cdot \Rlimit \\
         &= 3.9 \mathrm{mAh}
    \end{split}
\end{equation}
\begin{equation}
    \begin{split}
        \QLLI \mbox{ calculated from } \Qd &= \frac{\Delta\QLLI}{\Delta\Qd} \cdot \mathrm{Q}_\mathrm{limit} \\
        &= 22.2 \mathrm{mAh}.
    \end{split}
\end{equation}

This simple calculation demonstrates that the low-SOC resistance metric improves the resolution limit of $\QLLI$ by five-fold compared to the same calculation made using $\Qd$. 

We note that Q$_\mathrm{limit} = 20$ mAh corresponds to the magnitude of the difference in lithium consumed between the fast formation and baseline formation protocols (23 mAh), which confirms that it would be difficult to detect such small changes in lithium consumption using ordinary cycler equipment.

The precision of the $\Qd$ measurement can be improved in several ways. First, a higher C-rate discharge would lead to less current integration error and higher accuracy. However, a high-rate discharge would also include a kinetic contribution making it more difficult to correlate to lithium loss. Second, $\Qd$ can also be improved by using higher precision cyclers. For example, if the voltage and current precision both increased by 100-fold, then the resolution limit of $\QLLI$ calculated from $\Qd$ will have improved to 0.2 mAh. However, in this scenario, the resolution limit of $\QLLI$ calculated from $\reslow$ would also have  improved to 0.04 mAh.

\clearpage

%%% ORIGIN OF GAS EVOLUTION SECTION

\section*{Note S2: Origin of Gas Evolution over Cycle Life Due to Fast Formation}

\revc{iii}{At the negative electrode, some common gas-generating reaction pathways include the reduction of ethylene carbonate (EC) to form carbon monoxide (CO) and the reduction of water impurities (e.g. from manufacturing) which can form carbon dioxide (CO$_2$) and hydrogen gas \supercite{Rowden2020}. Since every cell in the experiment was taken from the same manufacturing batch, their water content is expected to be similar. It is therefore unlikely that water reduction pathways is the major source of difference in the amount of gas formed between the two different formation protocols. Next, the reduction of EC is lithium-consuming and thus the amount of CO formed through this reaction pathway is expected to correlate to the loss of lithium inventory (LLI) over life. However, our voltage fitting analysis (Figures \ref{fig:aging_test_signals_high_temp}, \ref{fig:aging_test_signals_low_temp}) suggests that the rate of LLI for fast formation cells is equal to or lower than that of baseline formation. We therefore expect the formation rate of CO from the EC reduction pathway to be comparable between the two different formation protocols. From this brief survey, it is unlikely that a reaction pathway based on the negative electrode can account for differences in measured gas amount over life between the two different formation protocols.}

\revc{iii}{We note that oxidation reactions at the positive electrode (e.g. at high potentials) can also contribute to gas evolution over aging. Common reaction pathways at the positive electrode include the decomposition of lithium carbonate at the surface of the NMC electrodes and the oxidation of EC. The reaction products include the generation of CO$_2$, CO, and oxygen gas \supercite{Rowden2020}. These reactions are known to accelerate at higher positive electrode potentials vs Li/Li$^+$, a phenomenon studied in detail by Jung et al. \supercite{Jung2017}.}

\begin{figure*}[ht]
\centering
\includegraphics[width=1.0\linewidth]{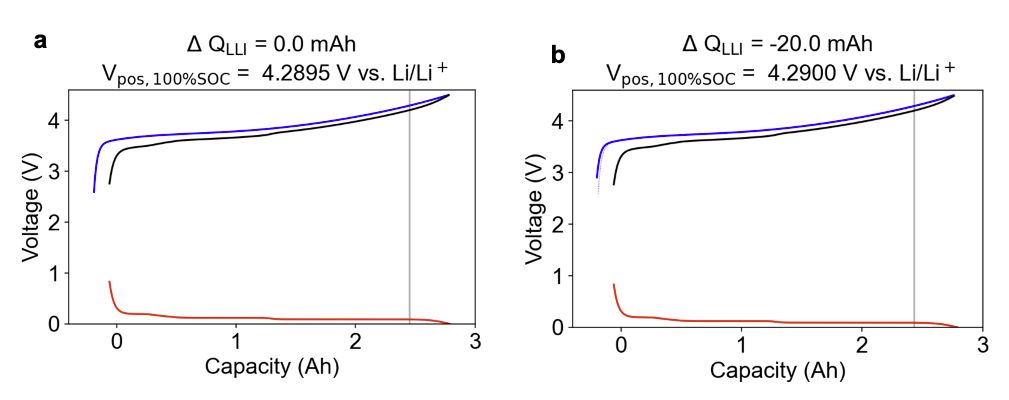}
\caption{Electrode stoichiometry model showing the impact of 20 mAh of additional lithium loss on the positive electrode potential at the top of charge. (a) Model for baseline formation. (b) Model for fast formation.}
\label{fig:cathode_voltage_changing}
\end{figure*}

\revc{iii}{To determine whether fast formation could have caused the positive electrode to experience higher potentials at the top of charge, we used the electrode stoichiometry model to measure the impact of increased lithium consumption during formation on the positive electrode potential vs Li/Li$^+$ at the top of charge (i.e. 4.2V in the full cell). The result, shown in Figure \ref{fig:cathode_voltage_changing}, illustrates how an increase in the magnitude of $\Delta\QLLI$ shifts the positive electrode stoichiometry window towards marginally lower values at both the bottom of discharge and the top of charge. From this graphical analysis, we see that the positive electrode stoichiometry at the top of charge (i.e. 4.2V in the full cell) has increased by 0.5 mV vs Li/Li$^+$. Thus, fast formation may have increased the positive electrode potential at the top of charge, creating a more oxidizing environment and promoting more gas generation. While the difference in the potential is marginal, a small change in oxidation rates could lead to a large difference in accumulated gas generated over the course of hundreds of cycles. We also note that a significant portion of the charge cycles are spent in the CV hold phase where the positive electrode would stay at this higher potential. A similar mechanism has been identified as part of work by Dose et al. \supercite{Dose2020}. Note that Dose et al. observed that the positive electrode impedance increased due to these oxidation reactions, but in our work, this impedance rise was not observed (Figure \ref{fig:aging_test_signals_high_temp}a).}

In general, gas build-up represents the combination of gas both generated and consumed. Xiong et al. \supercite{Xiong2017} demonstrated that gas in NMC-graphite cells can be generated at the positive electrode and subsequently reduced at the negative electrode via a `shuttle' mechanism \supercite{Ellis2017}. At the positive electrode, gas species such as O$_2$, CO, and CO$_2$ can be generated through electrolyte oxidation pathways \supercite{Rowden2020, Jung2017}, and at the negative electrode, gas species can be further reduced into solid products \supercite{Rowden2020}. Fast formation may be exacerbating the gas generation rate, suppressing the gas consumption rate, or both. 

We further hypothesize that the gas build-up over life may have a secondary benefit to cycle life. Krause et al. \supercite{Krause2017} and Chevrier et al. \supercite{Chevrier2018} have reported that the reduction of CO$_2$ at the negative electrode contributes to the SEI growth process and have a stabilizing effect. Since fast formation had increased cycle life, it is possible that more CO$_2$ is being generated at the positive electrode and reduced at the negative electrode, further improving SEI passivation and delaying the knee-point. More experimental work will be needed to confirm this theory.

\clearpage

%%% YMAX AND YMIN SHIFT DISCUSSION
\section*{Note S3: Effect of Lithium Consumption during Formation on Maximum Positive Electrode Stoichiometry}

\revc{ii}{A careful study of the electrode-specific equilibrium potential curves suggests another possible contributor to the improved cycle life seen in fast formation cells. Returning to Figure 4a, we observe that the capacity corresponding to the extra lithium consumed from fast formation, $\Delta\QLLI$, is also associated with a decrease in the maximum positive electrode stoichiometry, $\ymax$. Since fast formation consumed more lithium to create the SEI, the positive electrode becomes less fully lithiated when the cell is fully discharged. By comparison, the positive electrodes of baseline formation cells will be more lithiated at the end of discharge. 

Access to higher positive electrode lithiation states is associated with higher levels of particle-level stress, leading to more likelihood for cracking of the ceramic oxide secondary particles \supercite{Li2020a, Watanabe2014, Zhang2020b}. Stress-induced cracking over life can lead to electrical isolation of particles, resulting in loss of active sites. The cracking may also expose additional surface area which could accelerate the rate of parasitic reactions leading to earlier knee-points \supercite{Ma2019a}. We speculate that this difference in $\ymax$, while small, could protect the fast formation cells against cracking over the course of many cycles, leading to an overall improvement in the overall cycle life. This degradation mechanism is particularly relevant in our testing where every cycle ends on the minimum voltage target of 3.0V. This `hidden degradation mechanism' involving the positive electrode motivates the usage of the low-SOC resistance, $\Delta\res$, to quantify the small changes in $\Delta\ymax$ due to new formation protocols.}

\clearpage

\printbibliography[heading=subbibliography]

\end{refsection}

\end{document}